\documentclass[entropy,review,accept,moreauthors,pdftex,10pt,a4paper]{mdpi}
\pdfoutput=1

\firstpage{1} 
\articlenumber{x}
\doinum{10.3390/------}
\pubvolume{18}
\pubyear{2016}
\copyrightyear{2016}
\externaleditor{Academic Editors: Ignazio Licata and Sauro Succi}
\history{}

\usepackage{graphicx,amsmath,amsthm,amssymb,caption}

\usepackage{subfigure}

\newcommand{\br}{\mathbf{r}}

\Title{Nonlinear Phenomena of Ultracold Atomic Gases in Optical Lattices: Emergence of Novel Features in Extended States}

\Author{Gentaro Watanabe $^{1,2,3,4,5,}$*, B. Prasanna Venkatesh $^{4,6,7,}$* and Raka Dasgupta $^{4,8}$} 

\AuthorNames{Gentaro Watanabe, B. Prasanna Venkatesh and Raka Dasgupta}

\address{%
$^1$ \quad Center for Theoretical Physics of Complex Systems, Institute for Basic Science (IBS), 34051 Daejeon, Korea\\
$^2$ \quad Department of Physics, Zhejiang University, 38 Zheda Road, 310027 Hangzhou, China\\
$^3$ \quad  University of Science and Technology (UST), 217 Gajeong-ro, Yuseong-gu, 34113 Daejeon, Korea\\
$^4$ \quad Asia Pacific Center for Theoretical Physics (APCTP), Pohang, 37673 Gyeongbuk, Korea\\
$^5$ \quad Department of Physics, Pohang University of Science and Technology (POSTECH), Pohang, 37673~Gyeongbuk, Korea\\
$^6$ \quad Institute for Quantum Optics and Quantum Information, Austrian Academy of Sciences, Technikerstra\ss e 21a, A-6020 Innsbruck, Austria\\
$^7$ \quad Institute for Theoretical Physics, University of Innsbruck, A-6020 Innsbruck, Austria\\
$^8$ \quad Department of Physics, University of Calcutta, 92 Acharya Prafulla Chandra Road, 700009 Kolkata, India

}

\corres[2]{Correspondence: gentaro@pcs.ibs.re.kr, gentaro.wtnb@gmail.com (G.W.); prasanna.venkatesh@oeaw.ac.at (B.P.V.)}

\abstract{
The system of a cold atomic gas in an optical lattice is governed by two factors: nonlinearity originating from the interparticle interaction, and the periodicity of the system set by the lattice. The high level of controllability associated with such an arrangement allows for the study of the competition and interplay between these two, and gives rise to a whole range of interesting and rich nonlinear effects. This review covers the basic idea and overview of such nonlinear phenomena, especially those corresponding to extended states. This includes ``swallowtail'' loop structures of the energy band, Bloch states with multiple periodicity, and those in ``nonlinear lattices'', \textit{i.e.}, systems with the nonlinear interaction term itself being a periodic function in space.
}

\keyword{optical lattices; Bose--Einstein condensates; cold Fermi gases; superfluidity}

\PACS{67.85.Hj; 03.75.Lm; 67.85.Lm; 03.75.Ss}

\begin{document}

\section{Introduction}

Following a long series of developments in the experimental techniques of atomic and optical physics, the Bose--Einstein condensation (BEC) of cold alkali atomic gases was realized in 1995\linebreak (see, e.g., \cite{pethick_smith,pitaevskii_stringari,leggett_review} and references therein). The creation of this new state of quantum matter has opened up a new research field, the physics of ultracold atomic gases. The novelty of this system lies in its high controllability: various system parameters such as the dimensionality, the configuration of the external potentials, and the strength and the sign of the inter-atomic interaction can be manipulated dynamically as well as statically. In addition, this system has high measurability: since both the spatial and temporal microscopic scales of this system are relatively large, real time observation and direct imaging are possible. With these unique features, ultracold atomic gases serve as an unprecedented playground of the quantum world.

Due to the emergence of the superfluid order parameter, BECs acquire a nonlinear character originating from the interparticle interaction. Here, nonlinearity means that the basic equation governing the state of the system depends on the state itself. A variety of phenomena caused by nonlinearity such as solitons \cite{burger99,denschlag00,strecker02,khaykovich02} and matter-wave mixing \cite{deng99}, \textit{etc}. have been predicted and realized in BECs.
Remarkably, the strength of the nonlinearity in BECs of cold atomic gases is controllable. This is because here $s$-wave scattering length $a_s$, the parameter characterizing the interatomic interaction, can be tuned using the Feshbach resonance. Therefore, the realization of BECs of cold atomic gases has opened up a new horizon for the study of nonlinear phenomena\linebreak (see, e.g., \cite{kevrekidis}).

With further development of technology and tools, superfluidity has been realized using cold fermionic atoms as well. It has been shown that, by increasing the interatomic attraction using Feshbach resonances, the state of an atomic Fermi gas can be varied in a controlled manner from a Bardeen--Cooper--Schrieffer (BCS) superfluid of delocalized Cooper pairs to a BEC of tightly-bound dimers \cite{zwierlein05}. Furthermore, these two limits are smoothly connected without a phase transition: the so-called BCS-BEC crossover \cite{eagles69,leggett80,giorgini_review,bloch_review}. The experimental confirmation of the BCS-BEC crossover is one of the prime achievements in the field of cold atomic gases. Using BCS-BEC crossover, we can understand both the Bose and Fermi superfluids from a unified perspective.

Another important development in the field of cold atomic gases is the realization of the external periodic potential called an ``optical lattice'': Pairs of counter-propagating laser fields detuned from atomic transition frequencies, act as a free-of-defect, conservative potential for atoms via the optical dipole force. The realization of optical lattices has opened up the connection between the physics of cold atomic gases and solid state/condensed matter physics (see, e.g., \cite{bloch_review,lattice,yukalov_review,jkps_review} for reviews), and especially enables the simulation of theoretical models of solid state physics using cold atoms.

As a consequence of the competition and interplay between the effects of the periodic potential and nonlinearity, rich phenomena are expected to emerge in cold atomic gases in optical lattices. 
Especially, equipped with Feshbach resonance, a knob for controlling the strength of the nonlinearity, cold atomic gases in optical lattices allow us to enter a regime in which the effect of the nonlinearity is comparable to (or even dominates over) that of the periodic potential.
Such a strongly nonlinear regime beyond the tight-binding approximation has not been well-explored in conventional solid state physics. For example, a loop structure called ``swallowtail'' in the Bloch energy band \cite{wu_st,diakonov02} is a representative novel phenomenon emerging in this regime. In addition, using cold atomic gases, direct observations of the resulting nonlinear phenomena are possible, which is also a difficult task using solids.

In this short review article, we discuss nonlinear phenomena of superfluid cold atomic gases in optical lattices. Especially, we consider extended states and focus on the following phenomena: the swallowtail band structure, Bloch states with multiple periods of the applied optical lattice potential called multiple period states, and those in nonlinear lattices, \textit{i.e.}, systems with a periodically modulated interaction strength in space. This article is complementary to the existing review article on nonlinear phenomena in lattices \cite{malomed_review}, which focuses mainly on localized states. Superfluidity is the most important macroscopic quantum phenomena and superfluid flow in a periodic potential is ubiquitous in many other systems, such as superconducting electrons in superconductors and even in astrophysical environments such as superfluid neutrons in ``pasta'' phases in neutron star crusts (see, e.g., \cite{pasta_review,qmd_review} and references therein). Through the study of cold atomic gases in optical lattices, one may also expect to get deeper insights into these other systems.

This article is organized as follows. In Section \ref{sec:framework}, we explain the setup of our system and basic theoretical formalism employed in the later discussions. In Sections \ref{sec:swallowtail}--\ref{sec:nonlinlat}, we provide a comprehensive overview of the selected nonlinear phenomena in optical lattices starting with a simple physical explanation for each topic: swallowtail loops in Section \ref{sec:swallowtail}, multiple period states in Section \ref{sec:multiperiod}, and nonlinear lattices in Section \ref{sec:nonlinlat}. Finally, summary and prospects are given in Section \ref{sec:conclusion}.


\section{Theoretical Framework \label{sec:framework}}

\subsection{Setup of the System}

In the present article, we discuss superfluid flows of either fermionic or bosonic atoms in the presence of the externally imposed periodic potentials.
For the external periodicity, we mainly consider one of the most typical cases: one-dimensional (1D) sinusoidal potential of the form,
\begin{equation}
V({\bf r}) = V(x) = s E_R \sin^2{q_Bx} \equiv V_0 \sin^2{q_Bx}\, ,
\label{eq:lat}
\end{equation}
either in quasi-1D or 3D systems.
Here, $E_R = \hbar^2q_B^2/2m$ is the recoil energy, $m$ is the mass of atoms, $q_B = \pi/d$ is the Bragg wave number (note that $q_B$ is different from the fundamental vector of a 1D reciprocal lattice, $2\pi/d$, by a factor of $2$), and $d$ is the lattice constant, $V_0 \equiv sE_R$ is the lattice height, and $s$ is the dimensionless parameter characterizing the lattice intensity in units of $E_R$. For simplicity, we also assume that the superflow is in the same direction as the periodic potential (\textit{i.e.}, $x$ direction).
Throughout the present article, we set the temperature $T = 0$.

The systems which we discuss in this article consist of a large number of particles (the number of particles per site is also large) at temperatures close to absolute zero. One of the most convenient ways to deal with such many-body systems is to use the mean-field approximation. In this formalism, one focuses on a particular single particle, and the interactions produced by all the other particles are replaced by an averaged interaction described by the ``mean field''. Thus the complicated many-body problem is effectively reduced to a far simpler one-body problem.

The mean-field theory provides a minimal framework to study the nonlinear phenomena emerging from the presence of the superfluid order parameter. The mean-field theory enables us to predict novel nonlinear phenomena and obtain qualitative understanding of them although its validity is not always guaranteed. 
We resort to a mean-field description throughout as it readily fits our motivation in the present review---to provide a physical explanation of some selected, novel nonlinear phenomena of superfluids in periodic systems.

In the rest of this section, we provide a brief explanation of the theoretical framework used in the discussions in the remaining part of this article. The main purpose of this section is to provide a minimal explanation and define the notation. Therefore, interested readers are encouraged to refer to other references (e.g., \cite{pethick_smith,pitaevskii_stringari,dalfovo_review,giorgini_review}) for further details.

\subsection{Bosons}

The mean-field theory describing Bose--Einstein Condensates (BECs) at zero temperature is given by the Gross--Pitaevskii (GP) equation \cite{pitaevskii61,gross61,gross63,pethick_smith,pitaevskii_stringari,dalfovo_review}:
\begin{equation}
\label{GP1} 
  i\hbar \dfrac{\partial \psi({\bf r},t)}{\partial t}=\left[-\dfrac{\hbar^2}{2m}\nabla^2+ V({\bf r})+g|\psi({\bf r},t)|^2\right]\psi({\bf r},t)\, .
\end{equation} 
Here $\psi({\bf r},t)$ is the superfluid order parameter (or the condensate wave function) 
and $g$ is the effective coupling constant between two interacting bosons given by
\begin{equation}
g=\dfrac{4 \pi \hbar^2 a_s}{m}\, ,
\end{equation}
where $a_s$ is the $s$-wave scattering length.
The average number density $n$ is 
\begin{equation}
  n= \frac{N}{\mathcal V} = \frac{1}{\mathcal V}\int|\psi({\bf r})|^2\, d{\bf r}\, ,
\end{equation}
where $N$ is the total number of particles and ${\mathcal V}$ is the volume of the system.
Note that the GP equation can be viewed as the dynamical equation that results from a governing Hamiltonian known as the GP energy functional given by:
\begin{equation}
E[\psi] = \int d \br \left(\frac{\hbar^2}{2m} \vert \nabla \psi \vert^2+ V(\br) \vert \psi \vert^2 + \frac{g}{2} \vert \psi \vert^4 \right) \label{eq:GPerg}.
\end{equation}

The stationary solution of Equation (\ref{GP1}) is given by
\begin{equation}
\label{GP2} 
\mu \psi({\bf r})=\left[-\dfrac{\hbar^2}{2m}\nabla^2 + V({\bf r})+g|\psi({\bf r})|^2\right] \psi({\bf r})\, ,
\end{equation}
where $\mu$ is the chemical potential.

Nonlinearity of the GP equation (the third term in the rhs of Equations (\ref{GP1}) and (\ref{GP2})) originates from the interaction between bosonic atoms.
Many previous studies have shown that GP equation describes BECs of dilute, weakly interacting bosons at zero temperature quite successfully\linebreak (see, e.g., \cite{dalfovo_review} and references therein).

\subsection{Fermions}

A useful method for treating superfluid Fermi gases is the standard BCS mean-field theory of superconductivity.
Such a mean-field theory for inhomogeneous systems is given by the Bogoliubov-de Gennes (BdG) equations \cite{degennes,giorgini_review}:
\begin{equation}
\begin{pmatrix}
H'({\bf r}) & \Delta({\bf r})\\ \Delta^*({\bf r})& -{H'}({\bf r})
\end{pmatrix}
\begin{pmatrix}
u_i({\bf r})\\ v_i({\bf r})
\end{pmatrix}
=\epsilon_i
\begin{pmatrix}
u_i({\bf r})\\ v_i({\bf r})
\end{pmatrix}\, .
\label{eq:bdg}
\end{equation}
Here $H'({\bf r}) = -\dfrac{\hbar^2}{2m}\nabla^2+V({\bf r})-\mu$.  Also,  $v_i({\bf r})$ and $u_i({\bf r})$ are the quasiparticle amplitudes, associated with the probability of occupation and unoccupation of a paired state denoted by an index $i$, while $\epsilon_i$ is the corresponding eigen-energy. The quasiparticle amplitudes $v_i({\bf r})$ and $u_i({\bf r})$ satisfy the normalization condition $\int d{\bf r}\, [u_i^*({\bf r})u_j({\bf r}) + v_i^*({\bf r})v_j({\bf r})] = \delta_{i,j}$. $\Delta$ is the order parameter (or the pairing field), which reduces to the pairing gap in the single quasiparticle spectrum in the region of $\mu>0$ for the uniform system. The pairing field $\Delta({\bf r})$ and the chemical potential $\mu$ in Equation (\ref{eq:bdg}) are self-consistently determined from the gap equation,
\begin{equation}
\Delta({\bf r}) = -g \sum_i u_i({\bf r})v^*_i({\bf r})\, ,\label{eq:gap}
\end{equation}
and the average number density
\begin{equation}
n= \frac{N}{\mathcal V} = \frac{1}{\mathcal V} \int n({\bf r})\, d{\bf r} = \frac{2}{\mathcal V}\sum_i\int|v_i({\bf r})|^2 d{\bf r}\, .
\end{equation}
Since $\Delta$ depends on $\{u_i\}$ and $\{v_i\}$, the BdG equations (\ref{eq:bdg}) are nonlinear for nonzero interatomic interaction parameter $g$.

The superfluid Fermi systems bear a direct analogy with traditional superconducting systems, and likewise $g$, the contact interaction, plays similar role as the weakly attractive interaction term in the BCS-model. Only, now $g$ can be both small or large, and its value can be externally tuned using Feshbach resonances by applying a magnetic or an optical field. This controllability leads to a crossover between two ends: a weakly attractive BCS-like superfluid and a condensate of tightly bound bosonic molecules of a pair of fermionic atoms; popularly called the BCS-BEC\linebreak crossover \cite{eagles69,leggett80}. 

For contact interactions, the right-hand side of Equation (\ref{eq:gap}) has an ultraviolet divergence, which has to be regularized by replacing the bare coupling constant $g$ with the two-body T-matrix related to the $s$-wave scattering length \cite{randeria}. A standard scheme \cite{randeria} is to introduce a cutoff energy $E_c \equiv \hbar^2k_c^2/2m$ in the sum over the BdG eigenstates and to replace $g$ by the following relation:
\begin{equation}
\dfrac{1}{g} = \dfrac{m}{4 \pi \hbar^2 a_s}-\sum_{k<k_c}\dfrac{1}{2\epsilon_k^{(0)}}\, ,
\end{equation}
with $\epsilon_k^{(0)}\equiv \hbar^2k^2/2m$.

\subsection{Discrete and Continuum Models}

The systems of cold atomic gases can be studied by solving the GP equation (bosons) or the BdG equations (fermions) for the full continuum model.
Let us, for the sake of simplicity, consider quasi-1D bosonic systems. So instead of $V({\bf r})$, we think of a potential in $x$ direction only: $V(x)$.\linebreak  If there is a periodicity in the form of $V(x)$, or, if $g$ itself is a periodic function of $x$, one approach is to try the Bloch solutions $\psi(x)=e^{ikx}\phi(x)$, where $\hbar k \equiv P$ is the quasimomentum of the superflow and $\phi(x)$ is a periodic function with the same periodicity as the externally imposed periodicity by $V(x)$ or $g(x)$. One can expand $\phi(x)$ in terms of plane waves 
to give the following form for the order parameter:
\begin{equation}
\psi(x) = e^{ikx} \phi(x) = e^{ikx}\sum_{l=-l_{\rm max}}^{l_{\rm max}} a_l e^{i 2\pi l x/d}\, , \label{eq:Blochansatz}
\end{equation}
to find the Bloch solutions. 
The normalization condition yields $\sum_l |a_l|^2 =1$.

Instead of going for the full solution, one easier approach is to map the system to a discrete model, borrowed from the idea of tight-binding model in solid-state physics. 
In this approach the density of bosonic/fermionic atoms is assumed to be concentrated around the minima of the optical lattice potential.
For example, in a quasi-1D periodic potential, the condensate wave function can be approximated by a superposition of wave functions $\phi_j(x)$ localized at the lattice sites, denoted by $j$, which are normalized as $\int |\phi_j(x)|^2 dx =1$. Thus, $\psi(x,t)= \sum_j\psi_j(t)\phi_j(x)$. The coefficient $\psi_j$ is dependent on the site index $j$.

The Hamiltonian for such a discrete model is 
\begin{equation}
\label{hamilt}
H = -K \sum_j  (\psi_j^* \psi_{j+1}+ \psi_{j+1}^* \psi_j)
+\dfrac{U}{2}\sum_{j}|\psi_j|^4\, .
\end{equation}
In the case of the periodic solution with the same periodicity as that of the lattice (lattice constant $d$), the normalization condition is given by 
\begin{equation}
\int_{-d/2}^{d/2} |\psi(x)|^2 dx = \nu\, ,
\end{equation}
where $\nu$ is the filling factor (number of particles per site) with
\begin{equation}
  \nu = |\psi_j|^2\, .\label{eq:normdiscrete2}
\end{equation}
In evaluating the normalization condition, one neglects the overlap between $\phi_j$'s localized at different~sites.

The first term in the Hamiltonian describes hopping between the nearest-neighbor sites.\linebreak The hopping parameter $K$ is given by $K = -\int \phi_j[-\frac{\hbar^2}{2m}\nabla^2 + V(x)] \phi_{j+1} dx$. The on-site interaction parameter $U$ characterizes the interaction energy between two atoms on the same site, and gives the nonlinear term. This $U$ is connected to the interatomic interaction parameter $g$ by $U=g\int|\phi_j(x)|^4 dx$.  {In a manner similar to the continuum model one can also derive a dynamical equation for the amplitudes $\psi_j$ from an extermisation of the energy functional corresponding to Equation (\ref{hamilt}). Such a dynamical equation is known in literature as the discrete nonlinear Schr\"{o}dinger equation (analogous to Equation (\ref{GP1}) for the continuous case) \cite{DNLS1,DNLS2} and has served as an important tool to study ultracold atoms in optical~lattices.}

\subsection{Energetic and Dynamical Stability\label{subsec:thframeergstab}}

Once one can solve for the system, either from the full continuum model or the discrete version, the next step is to study the energetic and dynamical stability of the stationary solutions. 
Energetic stability guarantees that the stationary states are a local energy minimum of the energy functional [Equation (\ref{eq:GPerg})] and dynamical stability means that the time evolution of the system is stable with respect to small perturbations (this issue will be discussed in detail later in Figure \ref{fig:niustab1}a in Section \ref{sec:loopstability}). The standard approach in this context is the linear stability analysis described below \cite{pethick_smith,wu01,machholm03,machholm04}.

Let $\delta\phi_{q}(x)$ be the deviation from the stationary Bloch wave solution $\phi(x)$ for a given quasimomentum $\hbar k$ of the superflow. This can be written in the following form:
\begin{equation}
\delta\phi_{q} =u(x,q) e^{i q x}+ {v}^*(x,q)e^{-i q x}\, .
\end{equation}
Here $\hbar q$ is the quasimomentum of the perturbation. The energy deviation from the stationary states can be written in the following form
\begin{equation}
\delta E=\int dx 
\begin{pmatrix}
u^{*}& v^{*}
\end{pmatrix}
M(q)
\begin{pmatrix}
u\\
v\\
\end{pmatrix}\,.
\label{eq:ergstabcondn}
\end{equation}
The matrix $M(q)$ is Hermitian and gives the curvature of the energy landscape around the stationary solution. 
The system is energetically stable as long as all of the eigenvalues of $M(q)$ are positive. When even one of the eigenvalues is non-positive, the solution is no longer a local minimum and the system is energetically unstable. This is often termed as ``Landau instability''.

For the same perturbation $\delta\phi_q$, the time-evolution of the system for each $k$ is found to be:
\begin{equation}
i \dfrac{\partial}{\partial t}\begin{pmatrix}
u\\
v\\
\end{pmatrix}=
\sigma_z M(q)
\begin{pmatrix}
u\\
v\\
\end{pmatrix},
\label{dynmat}
\end{equation}
with
\begin{equation}
  \sigma_z =
\begin{pmatrix}
1 & 0\\
0 & -1
\end{pmatrix}.
\end{equation}
Unlike $M(q)$, the matrix $\sigma_z M(q)$ is not Hermitian. If the eigenvalue of $\sigma_z M(q)$ is complex, the perturbation corresponding to an eigenvalue with positive imaginary part grows exponentially in time: in this case the stationary solution is dynamically unstable. If the imaginary part is zero, the stationary state remains stable (\textit{i.e.}, dynamically stable). Therefore, by noting the eigenvalues of $M(q)$ and $\sigma_z M(q)$, one can learn whether the system belongs to an energetically stable region to start with, and how it evolves during the course of time. This is important because for BECs and superfluid Fermi gases, this stability translates to the sustenance of superfluidity in the system.


\section{Swallowtail Loops in Band Structure \label{sec:swallowtail}}

In this section we consider one unique manifestation of the nonlinearity (see Equation (\ref{GP1})) governing the dynamics of BECs in optical lattices---the so-called swallowtail loops in band structures. In the first Section \ref{subs31} we provide the basic physical idea and the context in which swallowtail structures arise in the energy dispersions. {Moreover, we discuss the key implication of having such swallowtail dispersions---breakdown of adiabaticity even at very slow driving.} The~purpose of this subsection will be to communicate the central physical picture succinctly, hence we shall sacrifice chronology and use the most clear presentation (in our opinion) \linebreak following~\cite{niu00,diakonov02,machholm03,mueller02}. In the second Section \ref{subs32} we present a more detailed account of the various theoretical results on swallowtail band structures including their energetic and dynamical stability. In the third Section \ref{subsec:nonlinexp} we present some experimental results that consider the effects of such nonlinear structures. In~the following Section \ref{subs31}, we present a brief account of more recent developments that extend the swallowtail phenomena to situations beyond the standard $s$-wave interacting Bosons in optical lattices including dipole-dipole interactions, superfluid Fermi gases, \textit{etc.} We conclude the section with some brief remarks indicating future prospects.

\subsection{Basic Physical Idea: The Nonlinear Landau--Zener Model and Variational Ansatz for Condensate Wavefunction in Optical Lattices}
\label{subs31}

The GP equation (\ref{GP1}) for the order parameter describing a BEC at zero temperature differs from the Schr\"odinger equation for a single particle in one key aspect --- the presence of the interaction term $g \vert \psi (\mathbf{r}, t)\vert^2$ in addition to the externally applied potential term $V(\br)$. When the order parameter $\psi(\br)$ is expanded in terms of a given complete basis set of single particle wavefunctions, the nonlinear term in the GP equation essentially leads to an effective potential that is dependent on the occupation probability of the different single particle states. One may anticipate that in the limit that the nonlinearity is comparable to the external potential, \textit{i.e.}, $g n \sim V(\br)$, the resulting dynamics as well as the stationary solutions of the GP equation (\ref{GP2}) can be very different from the equivalent ones for the non-interacting $g=0$ case which are simply governed by the eigen-energies of the Schr\"{o}dinger equation.

\begin{figure}[H]
\center{
\begin{tabular}{c  c}
\includegraphics[scale=0.31]{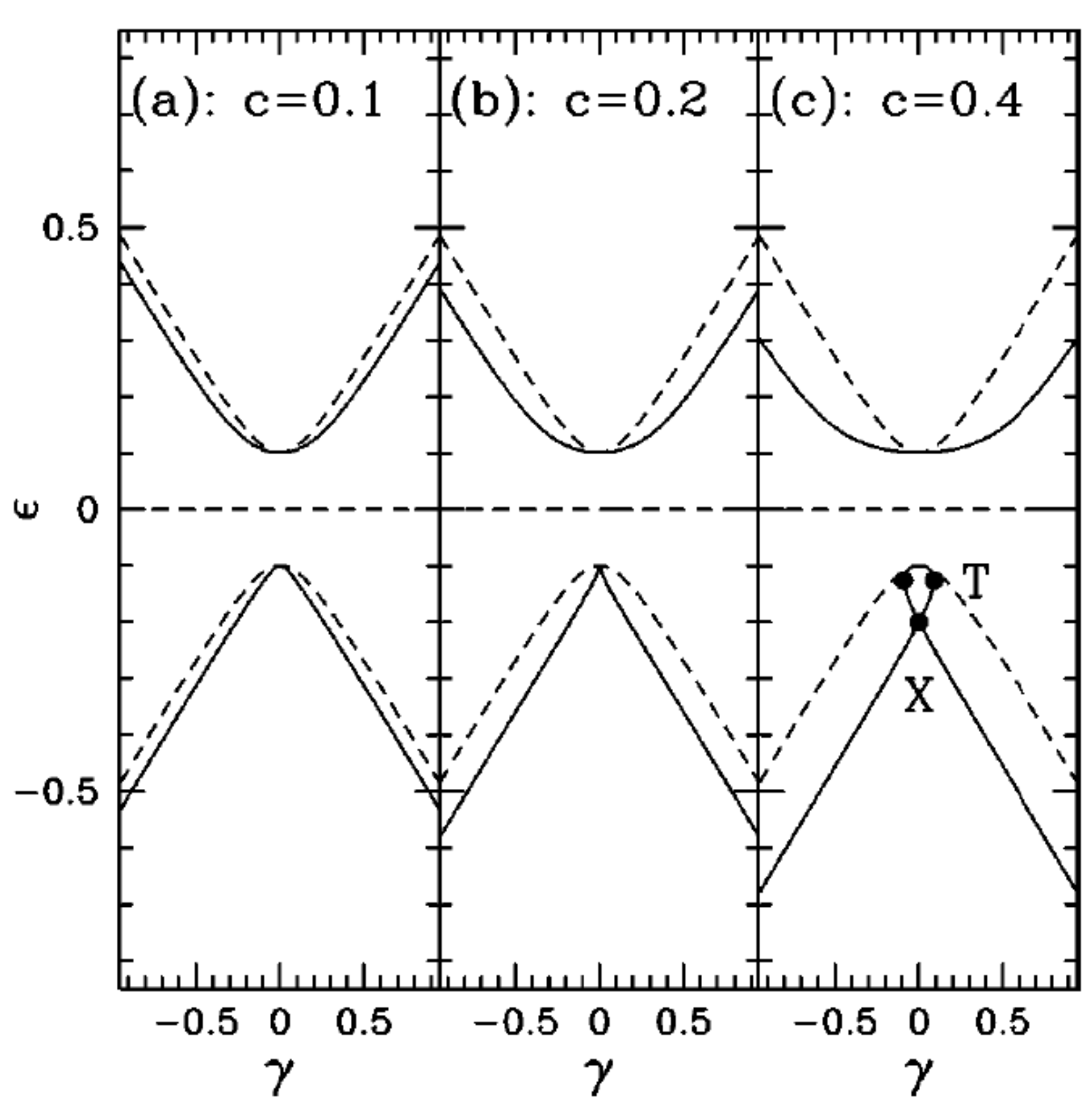}&
\includegraphics[scale=0.315]{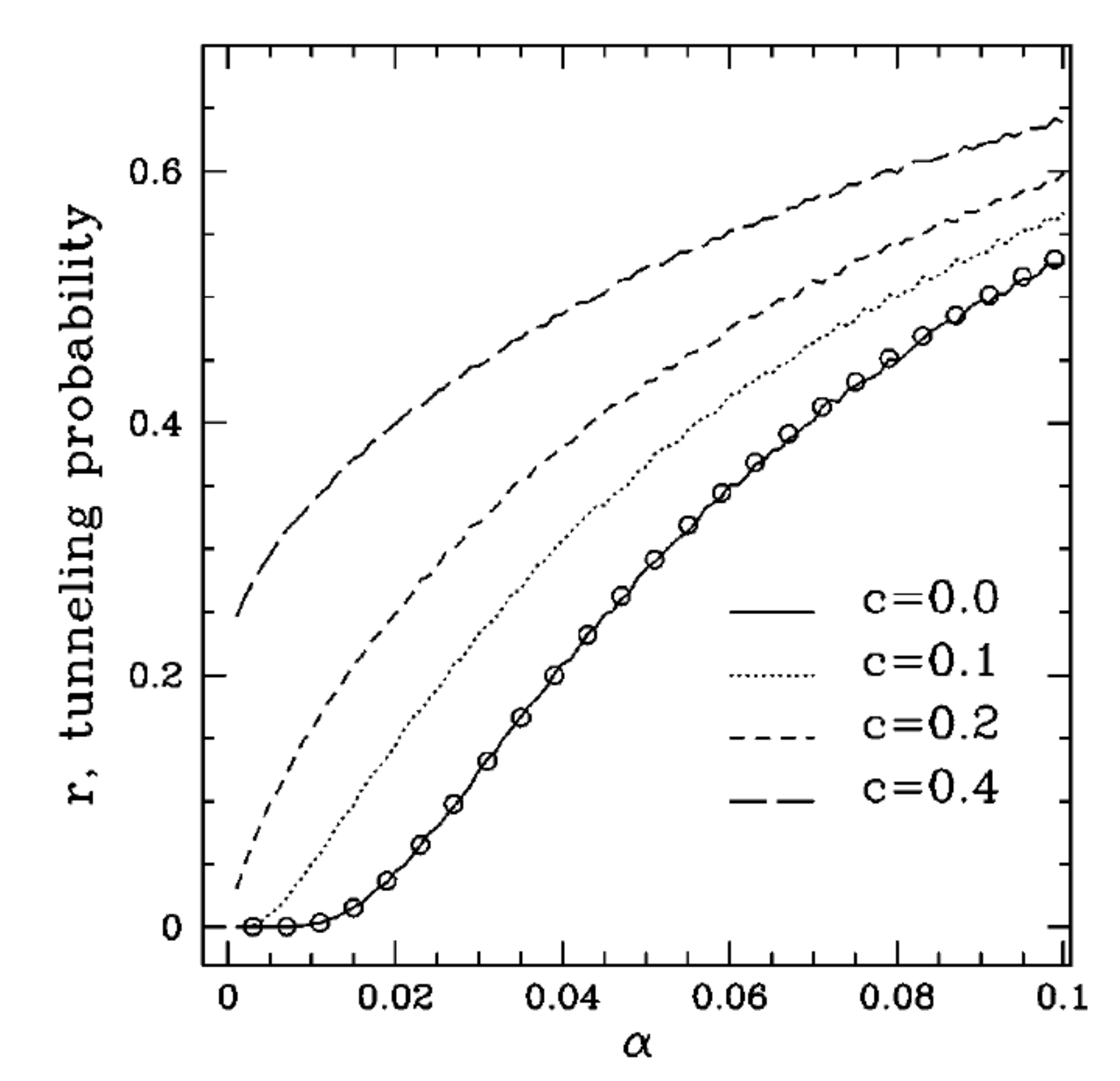}\\
       (\textbf{Left}) Energy levels.  & \qquad (\textbf{Right}) Transition probability.
       \end{tabular}
}
\caption{(\textbf{Left}) Adiabatic energy levels (or the chemical potential $\mu$) as a function of the nonlinearity $c$ for the nonlinear LZ problem showing the emergence of loops for $c>v$. The dashed lines represent the adiabatic energy levels for the $c=0$ case. (\textbf{Right}) Tunneling probability as a function of drive speed $\alpha$ at different values of $c$. For the linear case the result of the classic formula $r_0 = \exp(-\pi v^2/2\alpha)$ is displayed by the open circles. The figures are taken from \cite{niu00}.}
\label{fig:nLZ12}
\end{figure}

The simplest model to see the structures that emerge in the limit of large nonlinearities is the nonlinear two-level system introduced in \cite{niu00}:
\begin{align}
i \frac{\partial}{\partial t} \begin{pmatrix}
a \\ b
\end{pmatrix} &= H(\gamma) \begin{pmatrix}
a \\ b
\end{pmatrix}\, ,\label{eq:nonlinLZ}\\
H(\gamma) &= \begin{pmatrix}
\frac{\gamma}{2}+\frac{c}{2} \left(|b|^2-|a|^2 \right) & v/2 \\
v/2 & -\frac{\gamma}{2}-\frac{c}{2} \left(|b|^2-|a|^2 \right)
\end{pmatrix}. \label{eq:nLZmat}
\end{align}
Here $\gamma$ gives the level separation and $v/2$ is the coupling between the levels. \mbox{With $\gamma=\alpha t$}, and $g=0$, the above Hamiltonian is the well known Landau--Zener (LZ) model \cite{lz32,zn32}.\linebreak Equation (\ref{eq:nonlinLZ}) represents an extension of the LZ model representing the dynamics of a BEC with an order parameter whose overlap with the two relevant single particle levels has the amplitude $a$ and~$b$. Let us now consider the main results of such a model. In the limit of the small $c=0.1$, as shown in\linebreak Figure \ref{fig:nLZ12} left, 
the adiabatic energy levels (or the chemical potential $\mu$ in the language of our theoretical framework) is qualitatively similar to the linear case with the characteristic avoided crossing. When~the nonlinearity exceeds the coupling strength $v$, the adiabatic energy levels are drastically changed with the development of the characteristic looped structures. Note that the shape of the loop for the chemical potential is not a swallowtail, the energy function has such a swallowtail structure.

{While the modification of the energy level structure due to the strong nonlinearity is novel, what makes the looped energy structures particularly interesting from a physical point of view is the crucial implication of the looped structure for the transition probability between adiabatic energy levels.\linebreak In the linear case ($c=0$), the transition probability for the LZ model is given by $r_0 = \exp (-\pi v^2/ 2 \alpha)$. In this case in the so-called adiabatic limit of $\alpha \rightarrow 0$, the transition probability vanishes. On the other hand in the nonlinear case, as shown in Figure \ref{fig:nLZ12} right, for strong enough nonlinearity such that $c>v$ there is a finite transition probability even in the adiabatic limit. A simple explanation for this behavior is discernible from the looped energy structure in the third panel of Figure \ref{fig:nLZ12} left---starting from the lower energy state initially, in the adiabatic limit there is very little tunneling as the system passes the point ``X'' and continues upwards along the loop to reach the final point ``T''. At this point the system has to make a non-adiabatic transition to either the upper or lower level irrespective of how slow the sweep rate $\alpha$ is. This breakdown of adiabaticity is the key implication as well as an indication of the loop structure arising from the nonlinearity.}

Having captured the basic setting in which loops arise via the nonlinear LZ model, we move on now to the system central to this review---BEC in an optical lattice. We are interested in the stationary states of the GP equation that satisfy Equation (\ref{GP2}). In analogy to the LZ model, when the nonlinearity $g n$ is comparable to the applied lattice potential strength $V(x)$ we can expect interesting energy dispersions to arise. To be specific, henceforth we follow the treatment in \cite{machholm03,diakonov02} and consider a one-dimensional potential of the form:
\begin{align}
V(x) = V_0 \cos(2\pi x/d), \label{eq:potl}
\end{align}
and look for stationary states of the Bloch form, \textit{i.e.}, $\psi(x) = e^{ik x} f(x)$ with $f(x+d) = f(x)$.\linebreak Here $k$ denotes the quasimomentum of the condensate and is the proper quantum number in a periodic system. A key quantity of interest is the energy per unit volume given by
\begin{align}
\mathcal{E} = \frac{1}{d} \int_{-d/2}^{d/2} dx \, \left[ \frac{\hbar^2}{2m} \vert \nabla \psi \vert^2+ V_0 \cos(2\pi x/d) \vert \psi \vert^2 + \frac{g}{2} \vert \psi \vert^4 \right]. \label{eq:ergpervol}
\end{align}
In the absence of interactions, \textit{i.e.}, $g=0$, the energy per unit volume is arranged into the usual band structure which repeats after every reciprocal lattice momentum $k = 2 \pi/d$ and has gaps at the zone edges $k = r \pi/d$ with $r$ an odd integer in the extended zone representation of the bands (or at $k=\pm \pi/d$ in the reduced zone schemes).

\begin{figure}[H]
\center{
\begin{tabular}{c }
\includegraphics[scale=0.35]{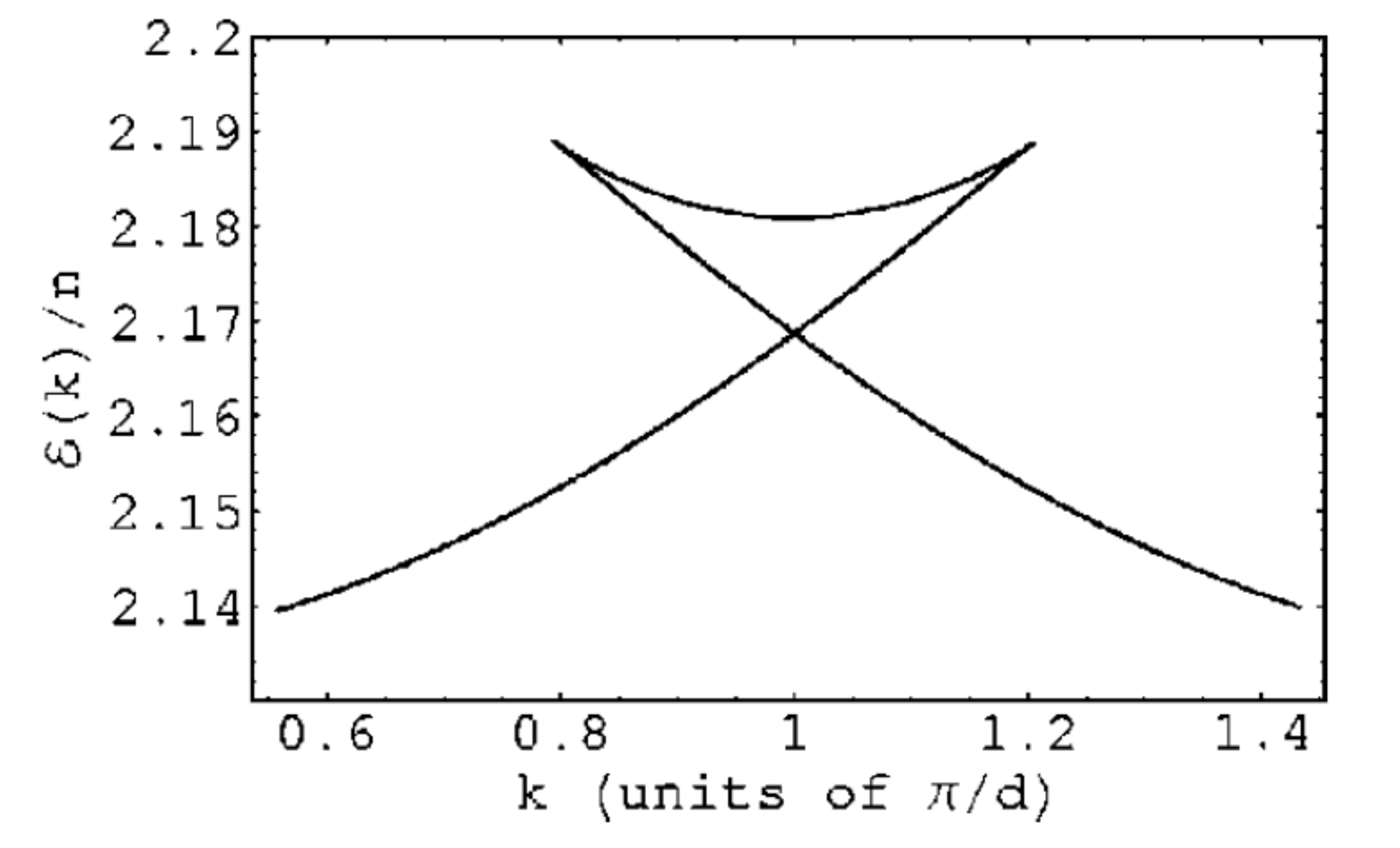}\\
(\textbf{Top})  Zone-edge loop.\vspace{6pt}\\
\includegraphics[width=.55\textwidth]{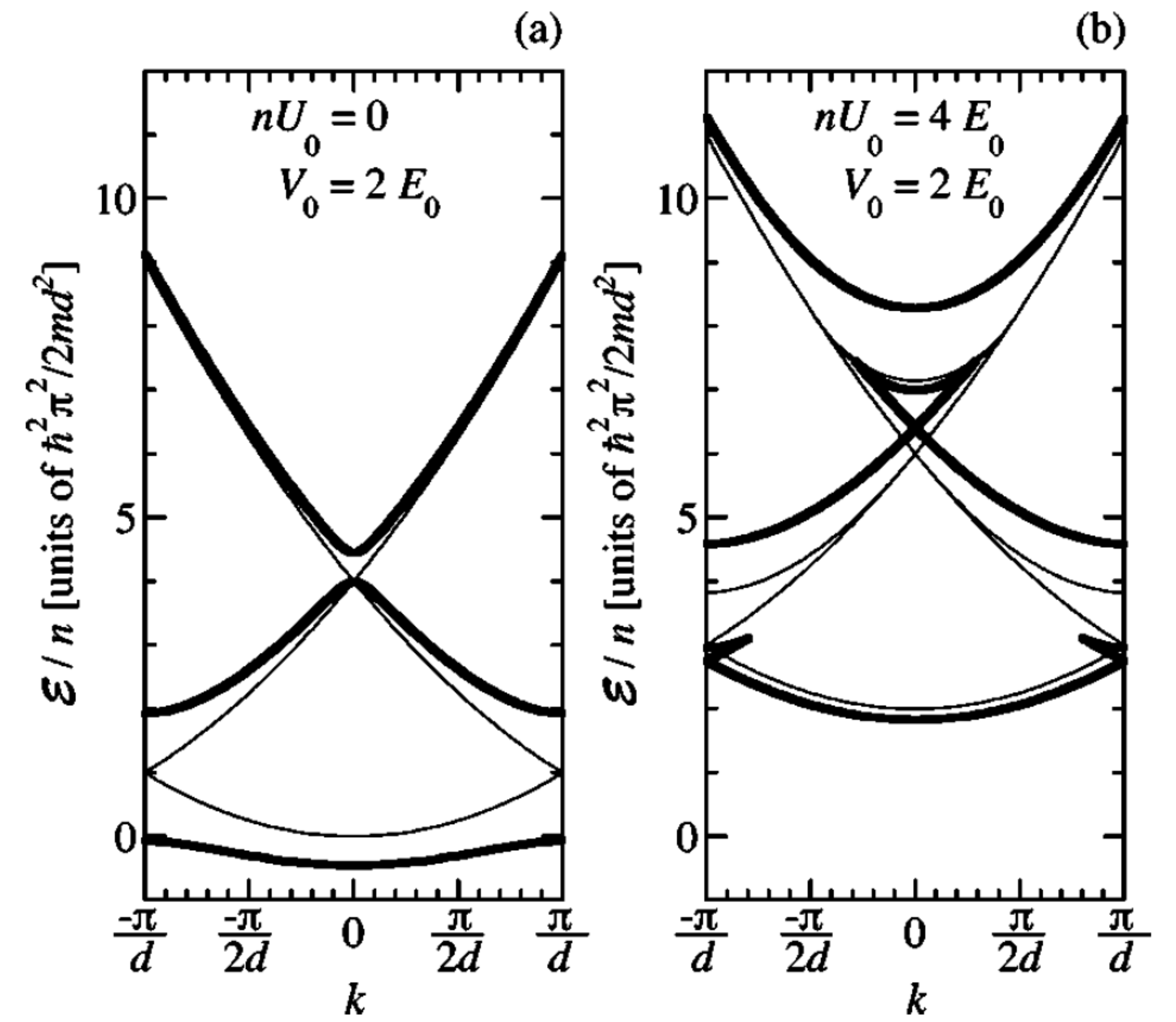}\\
(\textbf{Bottom}) Zone-center \& zone-edge loops.
\end{tabular}
}
\caption{(\textbf{Top}) Swallowtail loop structure of the energy per particle as function of $k$ computed from the variational ansatz (\ref{eq:bandedgeansatz}) for $V_0 = 3\hbar (4 V_0/m)^{1/2} \pi/d$ and $n = 1.2 V_0/U_0$. Figure was taken from \cite{diakonov02}. (\textbf{Bottom}) {Energy per particle} obtained using the variational ansatz (\ref{eq:varansatz}) demonstrating the presence of swallowtail loops at both the zone edge and zone center for sufficiently strong nonlinearity. $U_0$ in the figure is $g$ in our notation. Moreover, the loops persist and extend over the entire band in the limit $V_0 \rightarrow 0$. Taken from \cite{machholm03}.}
\label{fig:diakedgeloops}
\end{figure}

Equation (\ref{eq:Blochansatz}) gives the full plane wave expansion for a Bloch state but a variational ansatz restricting to just three plane wave states, \textit{i.e.}, $l_{\rm max} =1$ can already capture much of the physics as shown in \cite{machholm03}. Since the relative phases of the plane wave amplitudes $(a_0,a_1,a_{-1})$ do not change the energy, they can be chosen as real and using their normalization property restricted to the form:
\begin{align}
a_0 = \cos{\theta},\quad a_1 = \sin{\theta}\, \cos{\phi},\quad a_{-1} = \sin{\theta}\, \sin{\phi}\, . \label{eq:varansatz}
\end{align}
The trial wavefunction (\ref{eq:Blochansatz}) with the coefficients of the form (\ref{eq:varansatz}) is inserted into the energy per unit volume expression (\ref{eq:ergpervol}) and extremized with respect to the parameters $\theta$ and $\phi$. Also the recoil energy $E_0 = \hbar \pi^2/2md^2$ ($E_0=E_R$ in the notation of this review article) serves as a convenient unit for different energies in the problem.
Let us first consider the situation close to the Brillouin zone edge $k= \pm \pi/d$. Using intuition from the nearly free particle models for periodic potentials, the ansatz (\ref{eq:varansatz}) may further be simplified by taking $\phi = \{0 ,\pi/2\}$ \cite{diakonov02} giving:
\begin{align}
\psi(x) = \sqrt{n}\, e^{ikx} \left(\cos{\theta} + \sin{\theta}\, e^{-i2\pi x/d} \right) \label{eq:bandedgeansatz}
\end{align} 
with $n$ the average particle density. In fact it was shown in \cite{niu00} that such an ansatz can indeed be used to map the problem to that of the nonlinear LZ discussed earlier.  At the zone edge $k=\pi/d$, upon extremization of energy density functional (\ref{eq:ergpervol}) yields the solutions $\cos{2 \theta} = 0$ or $\sin{2 \theta} = V_0/(g n)$. When $g n < V_0$, the~only possible solutions are $\theta = \pi/4$ or $\theta = 3 \pi/4$ representing the zone edge solutions of the lowest and first excited band and are qualitatively similar to their counterparts in the linear, $g=0$, limit. The condensate current density $J$, given by the derivative of energy $\mathcal{E}$ with respect to $k$, is zero for such solutions. When the interaction is strong enough such that $n g > V_0$, the other solution with $\theta = \sin^{-1}{\left[ V_0/(gn) \right]}/2$ is also allowed. Moreover this solution has no linear counterpart and has non-zero current even at the zone edge with:
\begin{align}
J = \pm \frac{\hbar \pi}{md}\sqrt{n^2-\frac{V_0^2}{g^2}}\, . \label{eq:loopcurrent}
\end{align}
For values of $k$ away from the zone edge with $g>V_0/n$, the two solutions originally located at $\theta=\pi/4$ and $\sin^{-1}{\left[ V_0/(gn) \right]}/2$ approach each other and finally merge giving rise to the typical loop structure depicted in Figure \ref{fig:diakedgeloops} top. Also note that the band-gap at $k=\pi/d$ in the weak lattice limit $V_0 \ll E_0$ is given by $V_0$ and the condition to have looped energy dispersion is that the interaction energy per particle $gn$ be greater than this band gap.

Remarkably, the variational trial wavefunction (\ref{eq:bandedgeansatz}), with $\theta$ values such that $\mathcal{E}$ is extremized, is identical with an exact analytical solution found in \cite{bronski01,wu_st}. While this initially leads one to suspect that loops are somehow a restricted phenomenon subject to the availability of such exact solutions, further work in \cite{machholm03} dispelled this notion by demonstrating the possibility of loops even at the zone center, \textit{i.e.}, with $k=0$, without any known exact solutions (see Figure \ref{fig:diakedgeloops} bottom). Such solutions are also captured by the variational ansatz (\ref{eq:varansatz}). In this context the key solution in the linear limit is the one corresponding to the top of the second band with $k=0$ and energy $\mathcal{E}/n =4 E_0$ per particle in the absence of interactions $g=0$ and $V_0 \ll E_0$. This solution with an equal admixture of $a_1$ and $a_{-1}$, and very small contribution from $a_0$ has $\theta = \pi/2$ and $\phi = 3 \pi/4$ (the solution with $\phi = \pi/4$ corresponds to the third band now) persists even when $g \neq 0$. However, it was shown in \cite{machholm03} that, for sufficiently large values of $g$, this solution becomes unstable and two new solutions emerge signaling the lower and upper point of the loop at $k=0$ depicted in Figure \ref{fig:diakedgeloops} bottom. The condition for emergence of loops at the zone center is given by:
\begin{align}
g n > (16 E_0^2+V_0^2)-4E_0\, .
\end{align} 
In the weak lattice limit, the above condition simplifies to $g n > V_0^2/(8 E_0)$ which is precisely analogous to the condition that interactions exceed the band gap similar to the criterion to have loops at the zone edge. Figure \ref{fig:diakedgeloops} bottom illustrates the zone-edge and zone-center loops computed using the ansatz (\ref{eq:varansatz}). Furthermore it was found in \cite{machholm03} that the size of the loops, \textit{i.e.}, their extent in $k$ increases monotonically as the ratio $gn/V_0$ is increased. Most interestingly in the limit of vanishing lattice potential $V_0 \rightarrow 0$, the swallowtail loops extend over the entire Brillouin zone and the upper edge of the swallowtail becomes degenerate with the states in the bands above. As we discuss more in detail in the next subsection, this behavior stems from the fact that the states in the upper edge of the swallowtail correspond to periodic soliton solutions of the GP equation in free space whose degeneracy is lifted by the application of a periodic potential.

\begin{figure}[H]
\centering
  \subfigure[]{\label{fig:muefig1}\includegraphics[width=.45\textwidth]{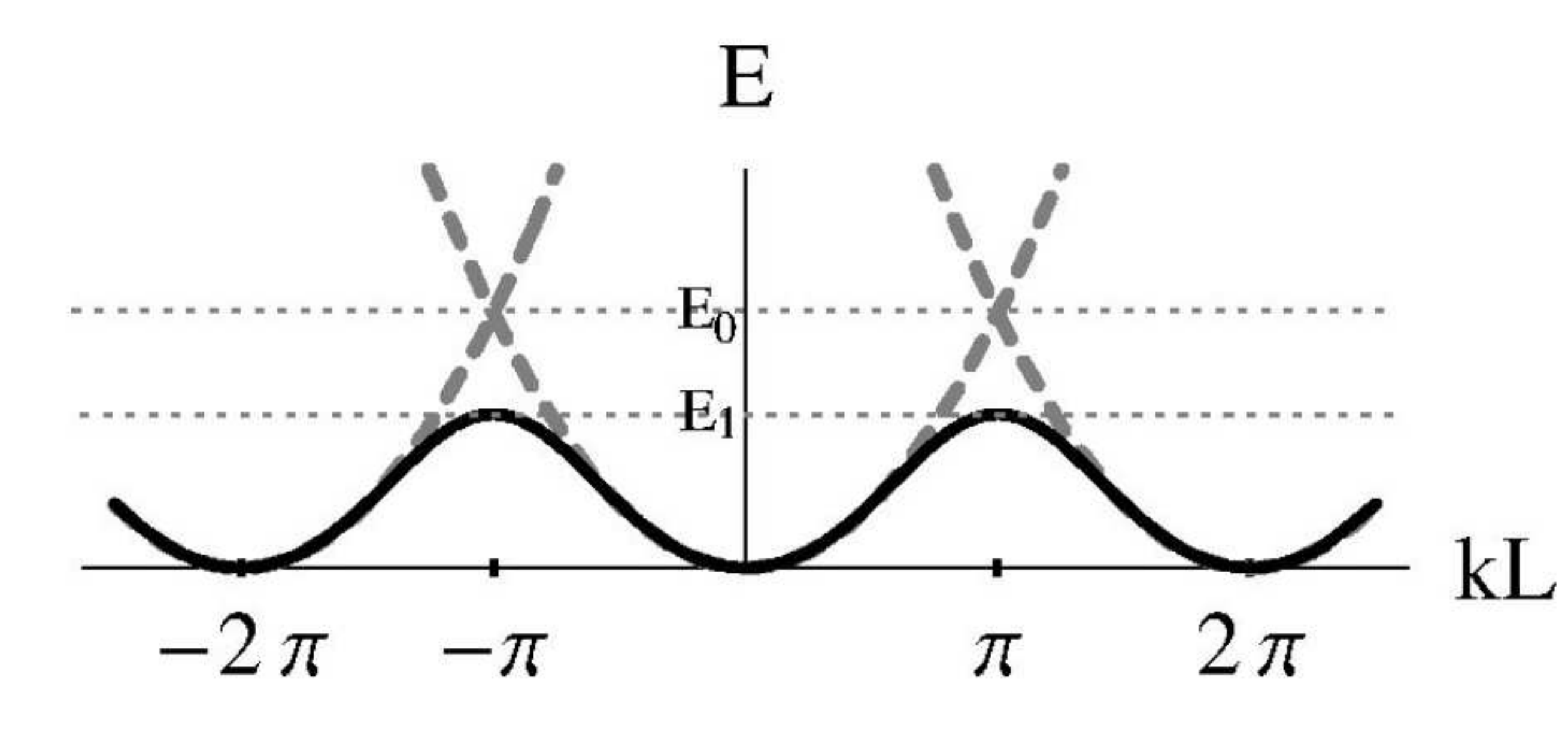}}
  \hfill
  \subfigure[]{\label{fig:muefig2}\includegraphics[width=.45\textwidth]{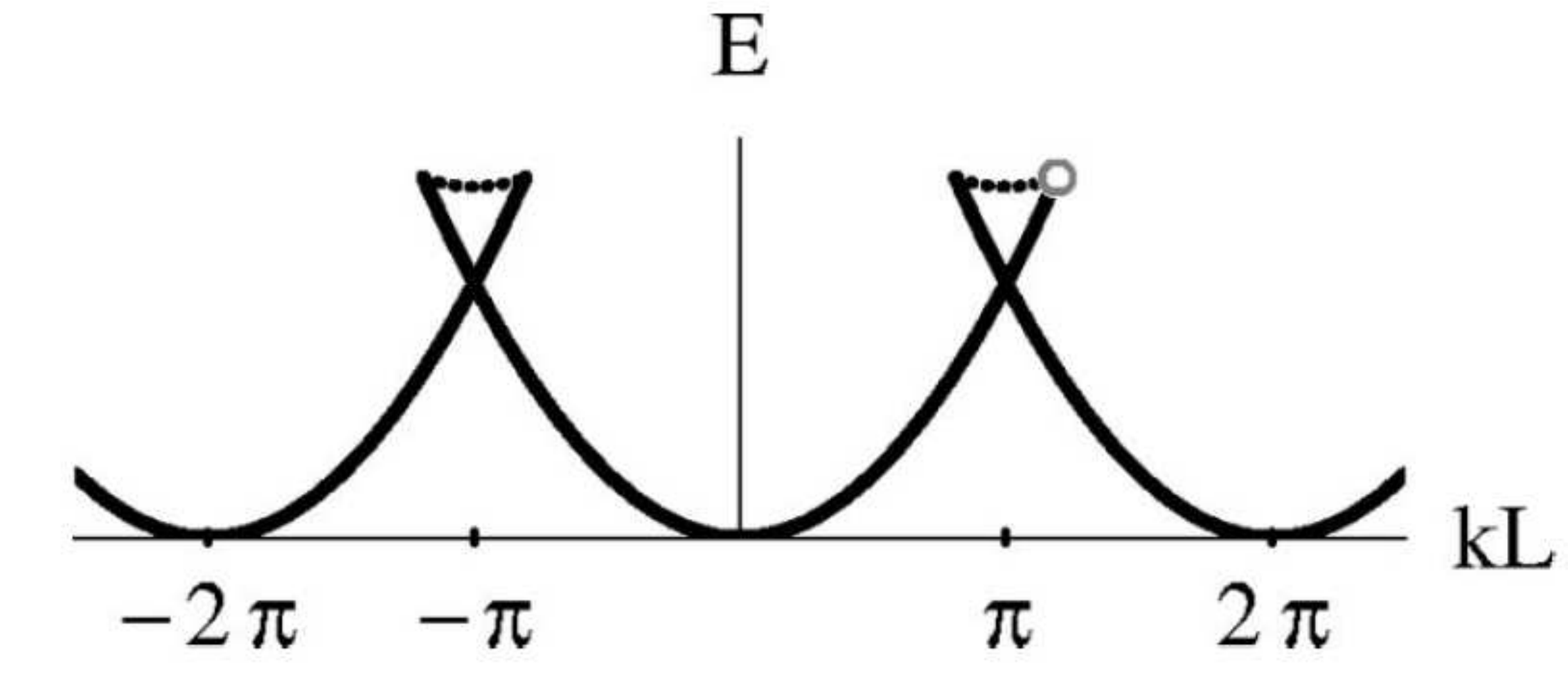}}
\caption{(\textbf{a}) The dashed line schematically shows the energy dispersion for a free particle which is modified to the solid line showing the (lowest) band structure on the application of a periodic potential of period~$L$. (\textbf{b}) Schematic plot of energy bands for a superfluid in an optical lattice that screens out the lattice to maintain non-zero flow velocity at the zone edge. Plots taken from \cite{mueller02}.}
\label{fig:muefig12}
\end{figure}

Finally, the swallowtail loop structures discussed so far may be intuitively viewed as generic features that arise in hysteretic systems as clarified in the work of \cite{mueller02}. Moreover the specific case of loops in the energy band structure of a BEC in an optical lattice can also be understood as a manifestation of superfluidity  and extended to other analogous systems such as superfluids in annular rings that have been realized experimentally \cite{eckel14}. The key insight of \cite{mueller02} can be explained from Figure \ref{fig:muefig12}. In Figure \ref{fig:muefig12}a, the approximately sinusoidal energy dispersion (lowest band) of a quantum particle in a periodic potential is shown. The corresponding Bloch eigenstates at different quasimomentum $\hbar k$ have periodic probability distributions commensurate with applied lattice potential. When a uniform external force is applied, the particle adiabatically follows this energy dispersion and performs periodic Bloch oscillations. In contrast for a BEC with interactions obeying the GP equation, \textit{i.e.}, a superfluid in a periodic potential, the interaction term tends to prefer uniform density distributions. Hence for strong enough interactions, as shown in Figure~\ref{fig:muefig12}b, the adiabatic band structure tends to be more similar to the quadratic free particle dispersion.\linebreak This behavior has been referred to as the screening of the lattice potential by the superfluid in \cite{mueller02}.\linebreak  As a result, the velocity of the superfluid does not go to zero at the zone edge leading to non-zero current as expressed by Equation (\ref{eq:loopcurrent}). As shown in Figure \ref{fig:muefig12}b, the velocity cannot increase indefinitely and the dispersion terminates (gray circle in Figure \ref{fig:muefig12}b) when the velocity becomes comparable to the Landau critical velocity of the superfluid giving rise to the swallowtail structure. Clearly adiabatic evolution along such a trajectory has to breakdown beyond this terminal point and forcing the system across this point from left to right and back will not restore the initial state---which is a clear sign of hysteresis. Also looking at the vicinity of the zone edge, it is clear that there are two possible minima for the energy which naturally leads to a saddle point separating the two minima given by the upper branch of the loop (the dotted lines in Figure \ref{fig:muefig12}b). 

The presence of multiple minima is a general characteristic of hysteretic systems. The number of minima for the superfluid in a periodic potential can be ``controlled'' by varying the quasimomentum $\hbar k$ or the interaction strength $g$ that sets the size of the swallowtail loop (remember that the swallowtail loops disappear if the $g$ is below some critical value). In the purview of catastrophe theory~\cite{thombook}, which is the study of singularities of gradient maps (any physical theory where a generating function is minimized to identify the stationary states of the system such as Fermat's principle in optics or the extremization of the GP energy functional here are examples of gradient maps), the swallowtail structures of energy bands represents a cusp catastrophe where two control parameters ($g$ and $k$) may be varied to change the number of extrema of the energy by $2$. 

\subsection{Swallowtail Loops Structures for Bosons in Optical Lattices}
\label{subs32}

Having provided a physical picture of swallowtail loops in the previous subsection we proceed now to a comprehensive overview of the main results. We will divide this subsection into two parts. In the first part we provide an account of the different results obtained regarding the occurrence of swallowtail loop structures for BECs in optical lattices and in the second part focus on the energetic and dynamical stability of the solutions.

\subsubsection{Occurrence of Loop Solutions}

The phenomenon of swallowtail loops in energy band structures arising from the GP equation were first investigated by Wu and Niu in \cite{niu00} for the nonlinear LZ introduced in detail in the previous subsection. In this work it was also pointed out that the nonlinear LZ model naturally arises near the zone edge in the dispersion for a BEC in a periodic optical lattice potential. The key results from this pioneering work are noted in Figure \ref{fig:nLZ12}. Following this work, an exact solution for the GP equation in a periodic potential of the form $V(x) = -V_0 \mathrm{sn}^2(x,\kappa)$ with $\mathrm{sn}(x,\kappa)$ denoting the Jacobian elliptic sine function (with elliptic modulus $0\le \kappa\le 1$) was discovered in \cite{bronski01}. These solutions for $\kappa=0$ (Equation (2.1) of \cite{wu_st} or Equation (10) of \cite{bronski01} with elliptic modulus $k=0$ in their notation) take the following form in our notation:
\begin{align}
\psi_{\rm exact}(x) =  \frac{\sqrt{c-v} + \sqrt{c+v}}{2\sqrt{c}} e^{i \pi x/d} +  \frac{\sqrt{c-v} - \sqrt{c+v}}{2\sqrt{c}} e^{-i \pi x/d}  \label{eq:carr}
\end{align}
with $c = gn/(8 E_0)$ and $v = V_0/(8 E_0)$, which is of the Bloch wave form for the zone edge $k = k_L = \pi/d$. When $c>v$ (which is also the condition for loops to appear at the zone edge), Equation (\ref{eq:carr}) gives the solution with finite current (see Equation (\ref{eq:loopcurrent})). 

\begin{figure}[H]
\centering
  \subfigure[~Zone-edge loops.]{\label{fig:machloopwd1}\includegraphics[width=.48\textwidth]{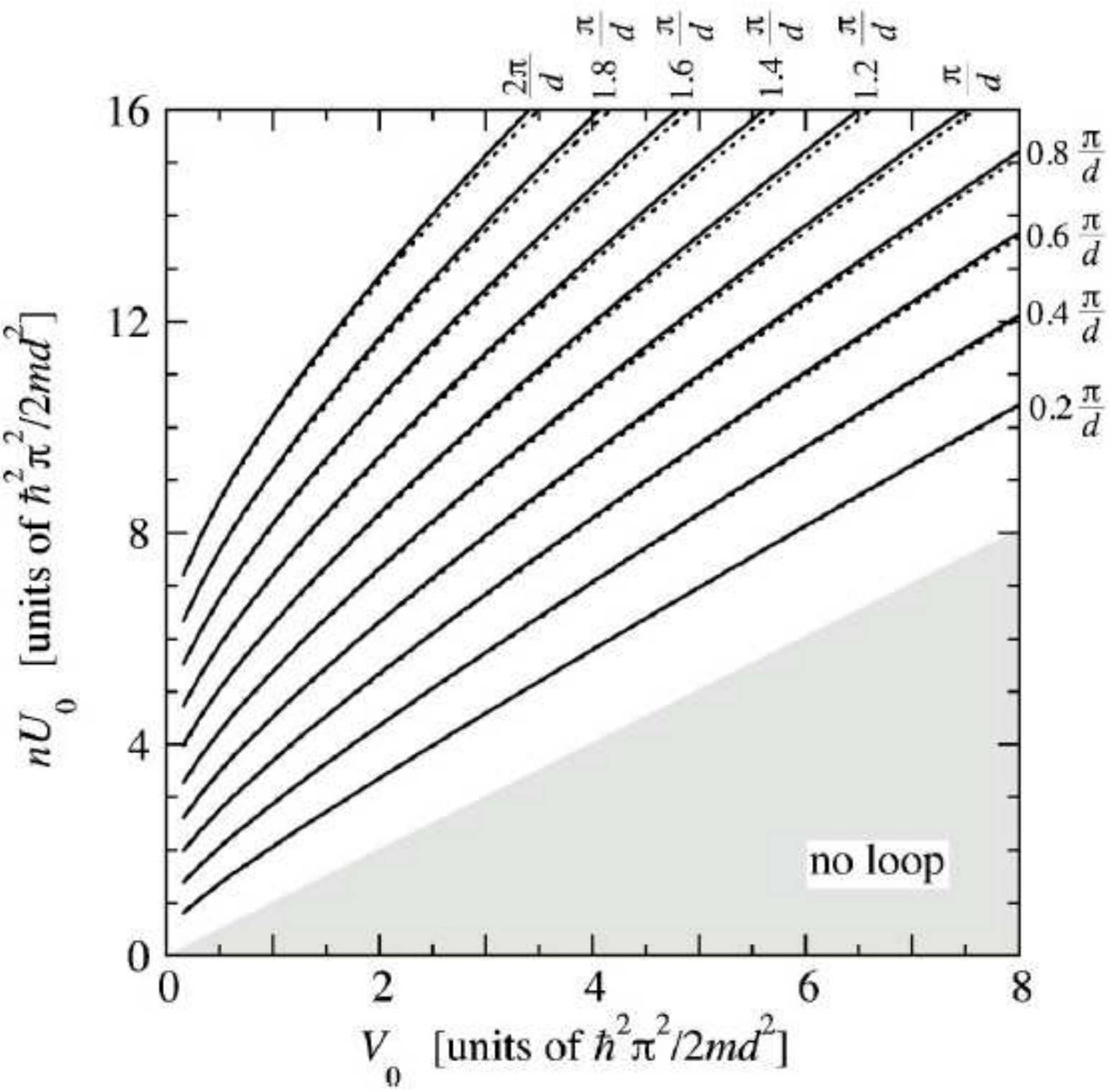}}
\quad
  \subfigure[~Zone-center loops.]{\label{fig:machloopwd2}\includegraphics[width=.48\textwidth]{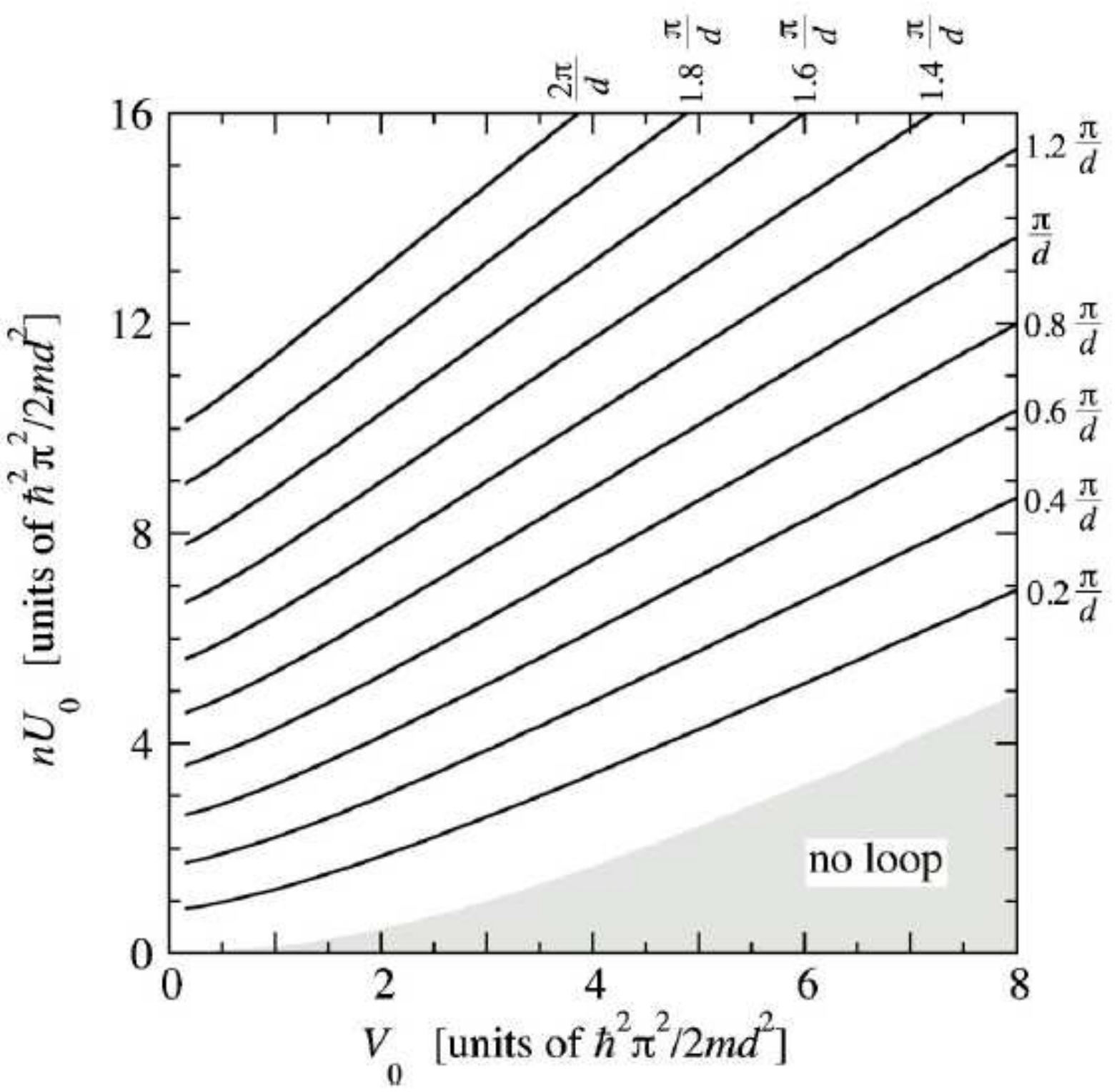}}
\caption{Contour plot of width of swallowtail loops from a numerical minimization using the variational wavefunction (\ref{eq:Blochansatz}) as a function of interaction ($nU_0$ is $ng$ in the notation of this review) and optical lattice depth $V_0$ for zone-edge loops (\textbf{a}) and zone-center loops (\textbf{b}). Dashed lines in (\textbf{a}) are for truncation with $l_{\rm max} =2$ and solid lines for $l_{\rm max} = 3$. Taken from \cite{machholm03}.}
\label{fig:machloopwd12}
\end{figure}

While on the one hand the exact solution lent more credibility \cite{wu_st} to the looped dispersions found from numerical solutions in \cite{niu00}, they also led to the suspicion that such solutions may not be a general feature. This was quickly dispelled by a simple variational calculation for the loops at zone edge by Diakonov \textit{et al.} \cite{diakonov02}, followed by a more comprehensive analysis by \mbox{Machholm \textit{et al.} \cite{machholm03}} demonstrating loops at the band center which we described in the previous subsection. \mbox{Machholm \textit{et al.} \cite{machholm03}} also provided a thorough analysis of the width of the swallowtail loops, defined as the extent in quasimomentum space in an extended zone scheme, as the key parameters $gn$ and $V_0$ are varied for both the zone-edge loops (Figure \ref{fig:machloopwd12}a) and zone-center loops (Figure \ref{fig:machloopwd12}b). 

The results in Figure~\ref{fig:machloopwd12} were calculated by a numerical minimization of the GP energy functional (\ref{eq:GPerg}) with the ansatz (\ref{eq:Blochansatz}). An unexpected feature of the results in Figure \ref{fig:machloopwd12} is that the width of the swallowtail loops remain non-zero even in the limit $V_0 \rightarrow 0$. In this limit the swallowtail solutions correspond to periodic soliton solutions of the GP equation. The zone edge solution with $k=\pi/d$ represents an equally spaced array of dark solitons (in one dimension, a dark soliton's wave function vanishes at some point in space whereas for a gray soliton there is a density dip of non-zero value at some point in space) with one soliton per lattice spacing $d$. The wave number of the solution $k = \pi/d$ can then be justified as the average phase change per unit length giving the right change in phase of $\pi$ across the dark solitons in each period. When $V_0 = 0$, soliton array solutions with density dips located at $x = r d$ and $x=(r+1/2)d$ (with $r$ integer) are degenerate and correspond to the highest energy state in the first band and the one immediately above in the second band, respectively. The lattice potential breaks this degeneracy as the solution with soliton centers at potential maxima $x=r d$ have lower energy giving rise to the energy gap at the center of the swallowtail loop. The loops at $k=0$ can also be understood with an analogous argument. The solutions with $k \neq 0$ correspond to arrays of gray solitons. The phase change across a gray soliton is less than $\pi$ and they move with some finite velocity $v_{\mathrm{soliton}}$ in the absence of the potential. In order to create a stationary state from such solutions at finite $V_0$, one has to imagine boosting the condensate velocity by $-v_{\mathrm{\rm soliton}}$ giving a spatial dependent phase to the condensate wave function. The wave vector corresponding to such solutions now depends both on the density and the phase shift implied by the finite velocity boost. Moving away from $k =0$ or $\pi/d$, the minimum density and $v_{\mathrm{soliton}}$ increases eventually going to zero and the sound speed $(gn/m)^{1/2}$ respectively leading to the loop branch merging with the free particle dispersion as shown in Figure \ref{fig:diakedgeloops} bottom. This manner of understanding the emergence of swallowtail loop structures from the soliton solutions provides a complementary physical picture of the phenomenon.

\begin{figure}[H]
\centering
\includegraphics[width=0.45 \textwidth]{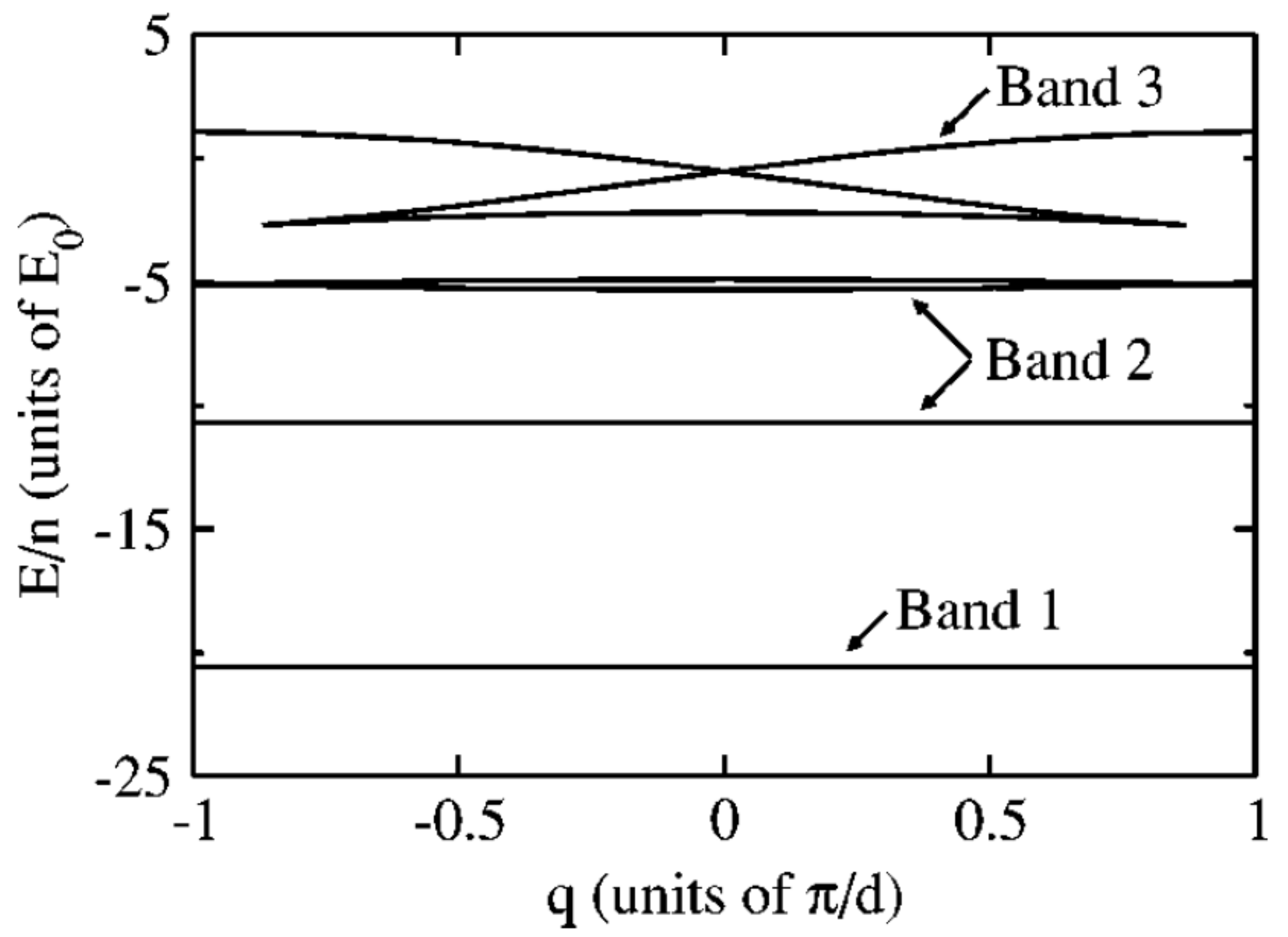}
\caption{Energy bands with swallowtail loops in the excited bands for a Kronig--Penney potential [Equation (\ref{eq:kronigpenney})] with $V_0 = E_0$ and attractive interaction $g n = -10 E_0$. Taken from \cite{seaman05a}.}
\label{fig:attloop}
\end{figure}

The work by Mueller in \cite{mueller02} provides a general way to understand swallowtail loops from the point of view of hysteresis and superfluid response. A corollary of such an approach is that it is possible to extend the phenomena of loops to systems beyond the standard system in this review (BEC in a periodic potential). Amongst the various examples discussed in \cite{mueller02}, the case of a BEC in an annular trap is of particular interest in the context of the experiment \cite{eckel14}. The Hamiltonian in the rotating frame describing a bosonic superfluid (mass $m$) in a 1D ring of length $L$, rotating at a frequency $\Omega$ is
\begin{align}
\frac{H}{\hbar^2/2mL^2} = \sum_{j} (2 \pi j + \Phi)^2 c_j^{\dagger} c_j + \frac{\tilde{g}}{2} \sum_{j+k=l+m} c_j^{\dagger}c_k^{\dagger}c_l c_m + \lambda \sum_{j} \left( c_j^{\dagger} c_{j-1}+c_j^{\dagger}c_{j+1} \right)\, . \label{eq:SFringham}
\end{align}
In the above, $c_j$ stands for the annihilation operator for bosons with angular momentum $j \hbar$ in a quantized picture or the amplitude of occupation of the same mode within a mean-field GP-like picture, $\Phi = 2 mL^2 \Omega/\hbar$ is the rotation speed in a dimensionless form, $\tilde{g} = 4 \pi a_s L/d_{\perp}^2$ is the effective interaction for a trap perpendicular to the ring with harmonic oscillator length $d_{\perp}$, and $\lambda$ is an impurity term that breaks the rotational symmetry coupling different angular momenta. This can arise naturally due to imperfections in the container or be generated, for instance, by applying a laser potential externally. The key point is that there is a one-to-one correspondence between the Hamiltonian (\ref{eq:SFringham}) and the GP energy functional (\ref{eq:GPerg}) after substituting the Bloch ansatz (\ref{eq:Blochansatz}) with $\Phi$ playing the role of quasi-wave number $k$ and $\lambda$ playing the role of the periodic potential strength leading to swallowtail loops as shown in Figure 11 in \cite{mueller02}. We will revisit Equation~(\ref{eq:SFringham}) in Section \ref{subsec:nonlinexp}.
{Seaman \textit{et al.} \cite{seaman05a} and Dong and Wu \cite{dong07}} provided an interesting insight into the swallowtail loop structures for both repulsively ($g>0$) and attractively ($g<0$) interacting BECs for the special case of a Kronig--Penney periodic potential which is of the form,
\begin{align}
V(x) = V_0 \sum_{j=-\infty}^{\infty} \delta( x - j d). \label{eq:kronigpenney}
\end{align}
The Bloch states in such a potential can be solved analytically. For the repulsive interactions case, the results in \cite{seaman05a} agreed qualitatively with the earlier numerical and approximate calculations but had unique features such as the fact that the critical interaction strength to have loops $g n > 2 V_0$ for \emph{all} bands of the energy spectrum unlike the sinusoidal band case discussed in \cite{machholm03}. In the case of attractive interactions with $g<0$, they found the loop structures occurred in the upper branch at the band gaps, starting from the second band as shown in Figure \ref{fig:attloop}. Moreover for the strongly attractive case $gn = -10 E_0$ shown in Figure \ref{fig:attloop} the loop in the second band spans the entire Brillouin zone and splits from the original band. 

\subsubsection{Stability of Loop Solutions\label{sec:loopstability}}

\begin{figure}[H]
\centering
\hspace{2cm}
  \subfigure[~Schematic.]{\label{fig:niustabschem}\includegraphics[scale=0.5]{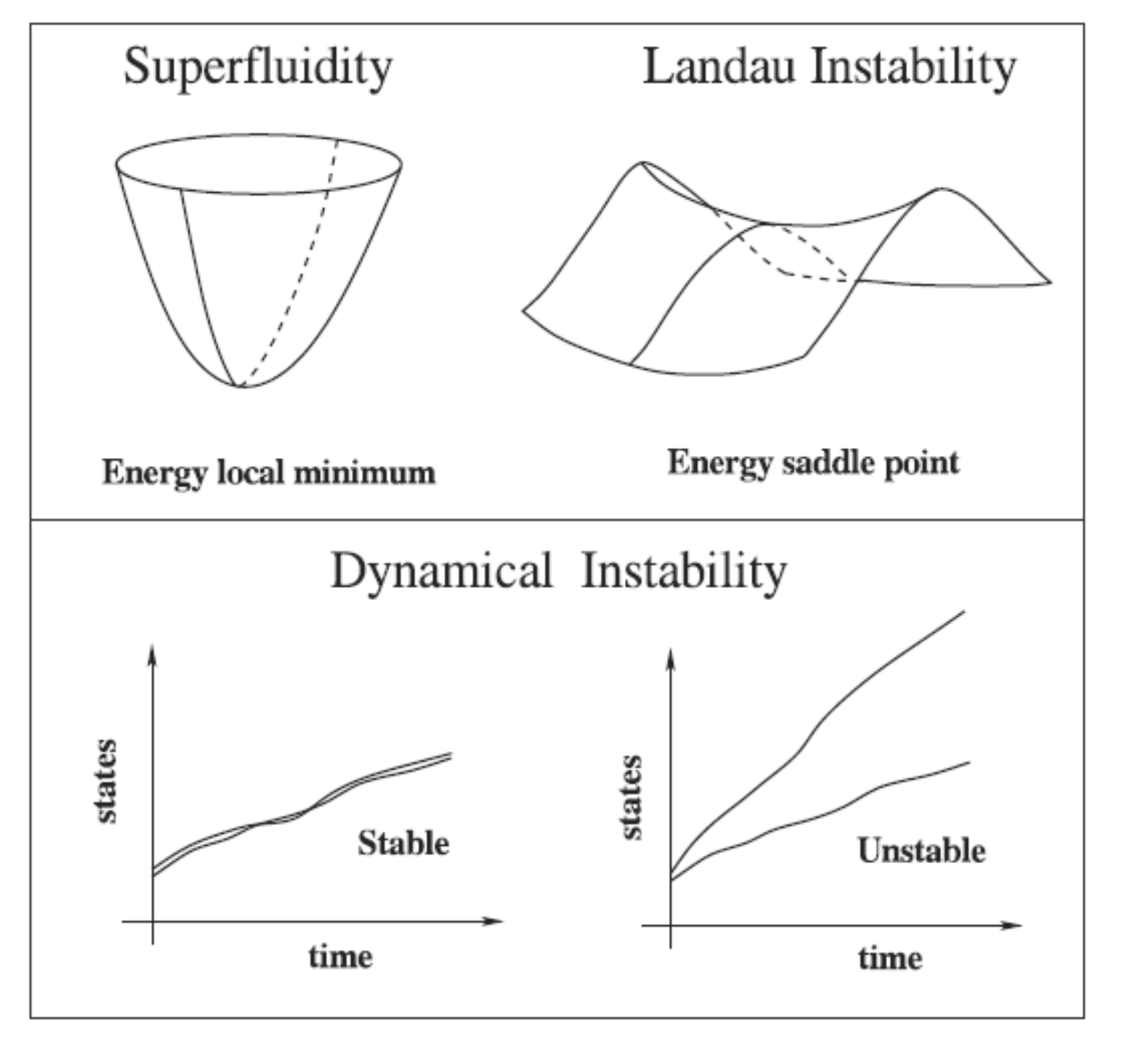}}
  \newline
  \subfigure[~Stability of Bloch solutions.]{\label{fig:niustabfull}\includegraphics[scale=0.4]{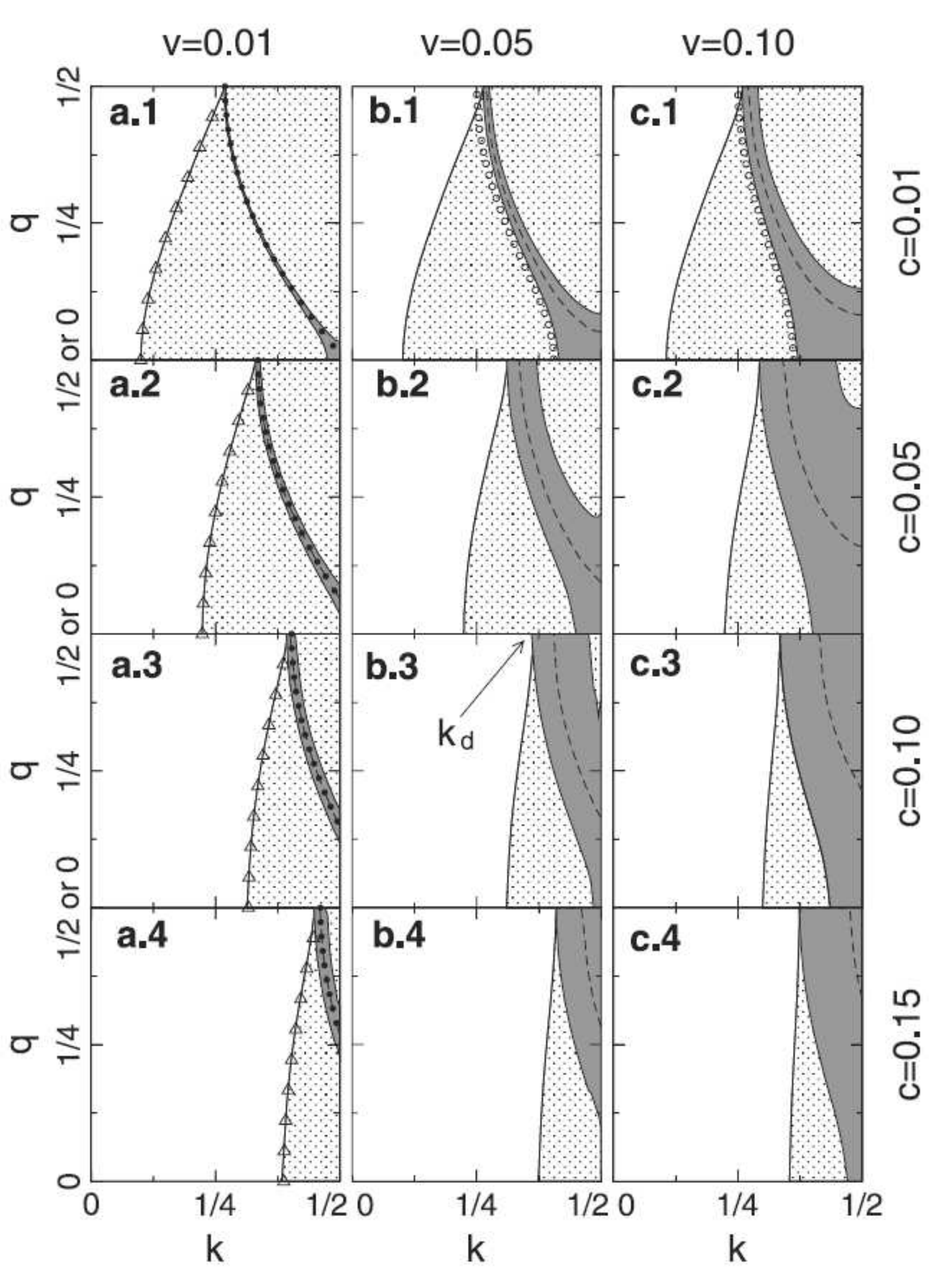}}
\caption{(\textbf{a}) Schematic illustration of the physical principle behind the definitions of energetic and dynamical stability. A dynamically unstable state is always energetically unstable whereas the vice-versa is not true. (\textbf{b}) Stability phase diagram at different values of lattice depth and nonlinearity parameter for a BEC in an optical lattice. The relation between the notation in the picture and this review article is scaled potential $v = V_0/(8 E_0)$ and nonlinearity parameter $c = gn/(8E_0)$. $q$ is the wave number of the perturbation modes and $k$ the quasimomentum in units of $2 K = 4 \pi/d$. In the shaded dark and light areas the system is energetically unstable while in the shaded dark area it is dynamically unstable. Triangles in (a.1--a.4) represent the boundary $q^4+c=k^2$ expected for $v=0$ and full dots in first column are from the results of an analytical calculation using Equation (\ref{eq:stabcondnfig}) valid for $v\ll 1$. Broken curves indicate the most unstable modes for each quasimomentum. The open circles in (b.1) and (c.1) represent analytical result valid for $c\ll v$ (see Equation (5.14) of \cite{wu03}). Figure taken from \cite{wu03}.}
\label{fig:niustab1}
\end{figure}

In the discussion of swallowtail loops so far, we have completely ignored the stability properties of the solutions. In what follows, we discuss results regarding the energetic and dynamical stability of steady state solutions of the Bloch form (of which swallowtail loops are special cases) for BECs in periodic potentials. The stability of the solutions we find by extremizing the GP energy functionals (\ref{eq:GPerg}) are crucial as they will provide us clues as to whether in an experiment the system can reach such equilibrium solutions and if they do how long can they be stable. As a part of the the theoretical framework Section \ref{subsec:thframeergstab}, we provided the conditions for energetic stability but a physical picture of the two kinds of stability as exemplified in Figure \ref{fig:niustab1}a is helpful---for energetic stability the equilibrium solution has to be a local minimum of the energy functional (\ref{eq:GPerg}) whereas for dynamical stability perturbations about the equilibrium state should not grow with time when evolved according to the time-dependent GP equation (\ref{GP1}).

The stability of Bloch states in the lowest band excluding loops was discussed by Wu and Niu in \cite{wu01} and \cite{wu03}. Figure \ref{fig:niustab1}b represents the results from a numerical calculation mapping out the stability of Bloch states with quasimomentum $\hbar k$ under perturbations of the Bloch wave form with quasimomentum $\hbar q$ for different values of the potential $v = V_0/(8 E_0)$ and nonlinearity $c = gn/(8E_0)$. Let us denote the stability matrix, Equation (\ref{eq:ergstabcondn}), for a state with quasimomentum $\hbar k$ under a perturbation of quasimomentum $\hbar q$ as $M_k(q)$. For the special case of a free BEC with no lattice, \textit{i.e.}, $v=0$, the eigenvalues of the matrix can be computed analytically and the requirement for positivity of eigenvalues leads to the well-known Landau criterion given by
\begin{align}
\vert k \vert \geq \sqrt{q^2/4+c} \label{eq:freeLandaucrit}.
\end{align}
The shaded light and dark regions in Figure \ref{fig:niustab1}b represent regions of energetic instability with $M_k(q)$ having negative eigenvalues. In the limit of small $v$ the equality in the expression (\ref{eq:freeLandaucrit}) accurately reproduces the energetic stability region shown by triangles in the plot. Another key feature to note in   Figure~\ref{fig:niustab1}b is that even at $v$ comparable to $c$, as the nonlinearity $c$ is increased, the BEC is energetically stable over an increasing area in the $k$-$q$ space, which can be anticipated from expression (\ref{eq:freeLandaucrit}).

In Figure \ref{fig:niustab1}b, the system is dynamically unstable in the dark shaded regions. The first thing to notice is that energetic instability is a pre-requisite for dynamical instability and this can also be shown in general (see appendix of \cite{wu03}). Further insight into dynamical stability can be obtained by considering the structure of the eigenvalues of the matrix $\sigma_z M_k(q)$ since the requirement for dynamical stability is to have real eigenvalues for this matrix. The states represented by the eigenvectors of $\sigma_z M_k(q)$ can also be viewed as quasiparticle excitations in the BEC, \textit{i.e.}, phonons \cite{wu03}, with the positive eigenvalues giving the phonon spectrum. In general the eigenvalues of $\sigma_z M_k(q)$ can be complex but always occur in complex conjugate pairs \cite{machholm03,wu03} owing to the real nature of the matrix in momentum representation. In the case of $v=0$, the eigenvalues of $\sigma_z M_k(q)$ are always real and given by $\epsilon_{\pm}(q) = kq \pm \sqrt{q^2 c + q^4/4}$ where $k$ and $q$ are measured in units of $4 \pi/d$. This implies that in free space BEC, flows are always dynamically stable. However, Figure \ref{fig:niustab1}b shows that situations change significantly when the lattice potential is introduced as also evidenced in experiments \cite{burger01,burger01comment,burger01comment1}. In general for all parameter regimes there is a critical wave number $k_d$ beyond which Bloch waves are dynamically unstable. At $k=k_d$, dynamical stability always sets in for $q=\pi/d$, which interestingly also corresponds to a period doubling revealing a link further explored in Section \ref{sec:multiperiod}. At the point where dynamical instability sets in, the eigenvalues of $\sigma_z M_k(q)$ change character from real to pairs of complex conjugates, \textit{i.e.}, with equal real parts. Hence as explained in \cite{wu01,wu03,machholm03}, dynamical instability can be viewed as arising from a lattice induced resonance between a pair of excitations that are degenerate in the $v=0$ limit. Thus when the instability sets in, two phonons with the sum of their momenta given by the primitive reciprocal lattice vectors $\pm G= \pm 2\pi/d$ are created from zero energy, \textit{i.e.}, they satisfy 
\begin{align}
\epsilon_{+}(q) + \epsilon_{+}(2 \pi/d-q) = 0\, . \label{eq:stabcondnfig}
\end{align}
This can be used to clearly justify the observation (valid at small lattice depths $V_0$) that instability at the critical wave number $k_d$ always sets in with $q=G/2=\pi/d$ and the critical vector satisfies $\vert k_d \vert = (\pi/d)(gn/2E_0+1/4)^{1/2}$ agreeing with the numerical results in Figure \ref{fig:niustab1}b.

\begin{figure}[H]
\centering
  \subfigure[~Energetic stability.]{\label{fig:pethergstab}\includegraphics[width=.45\textwidth]{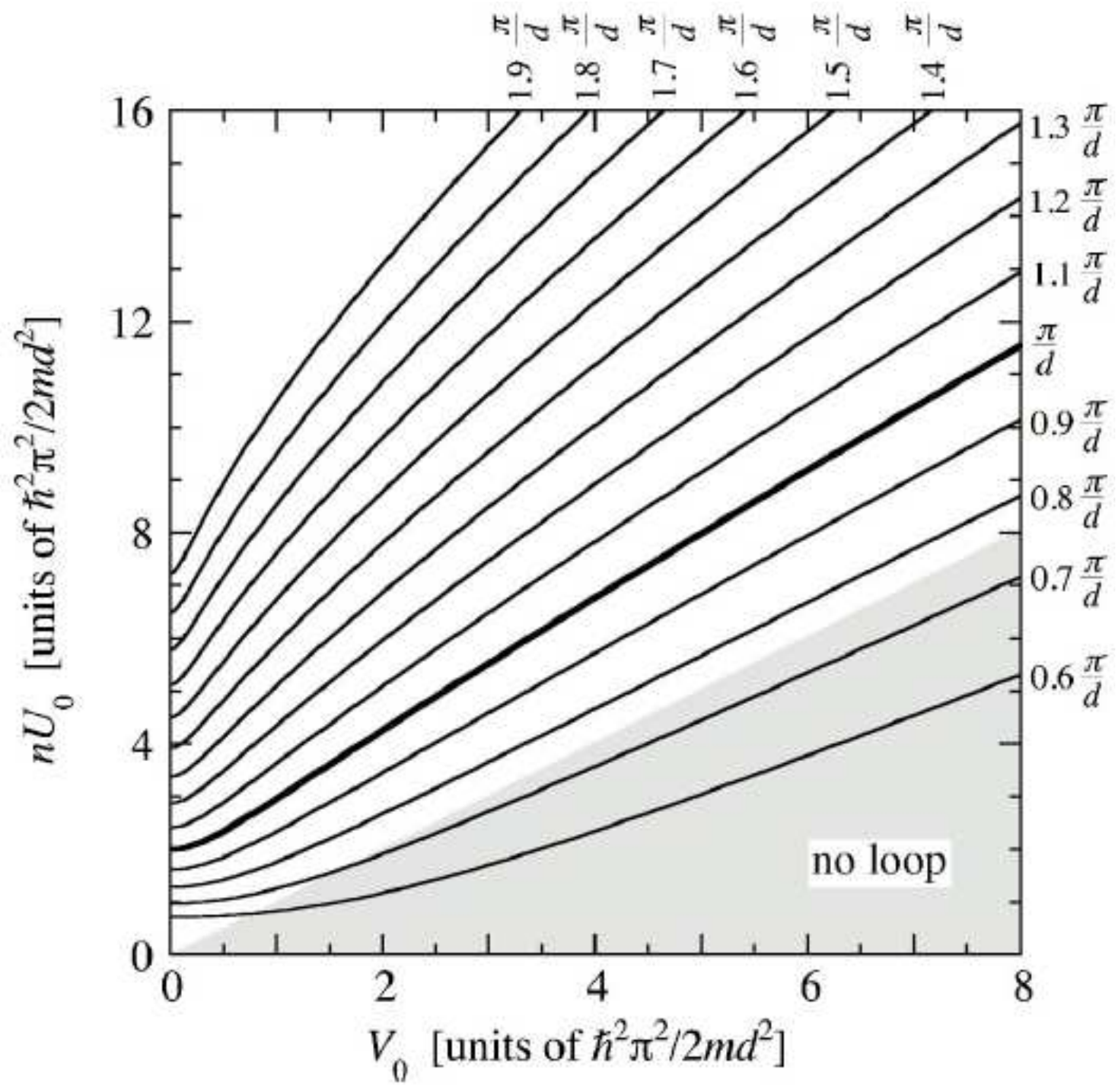}}
  \hspace{6pt}
  \subfigure[~Dynamical stability.]{\label{fig:pethdynstab}\includegraphics[width=.45\textwidth]{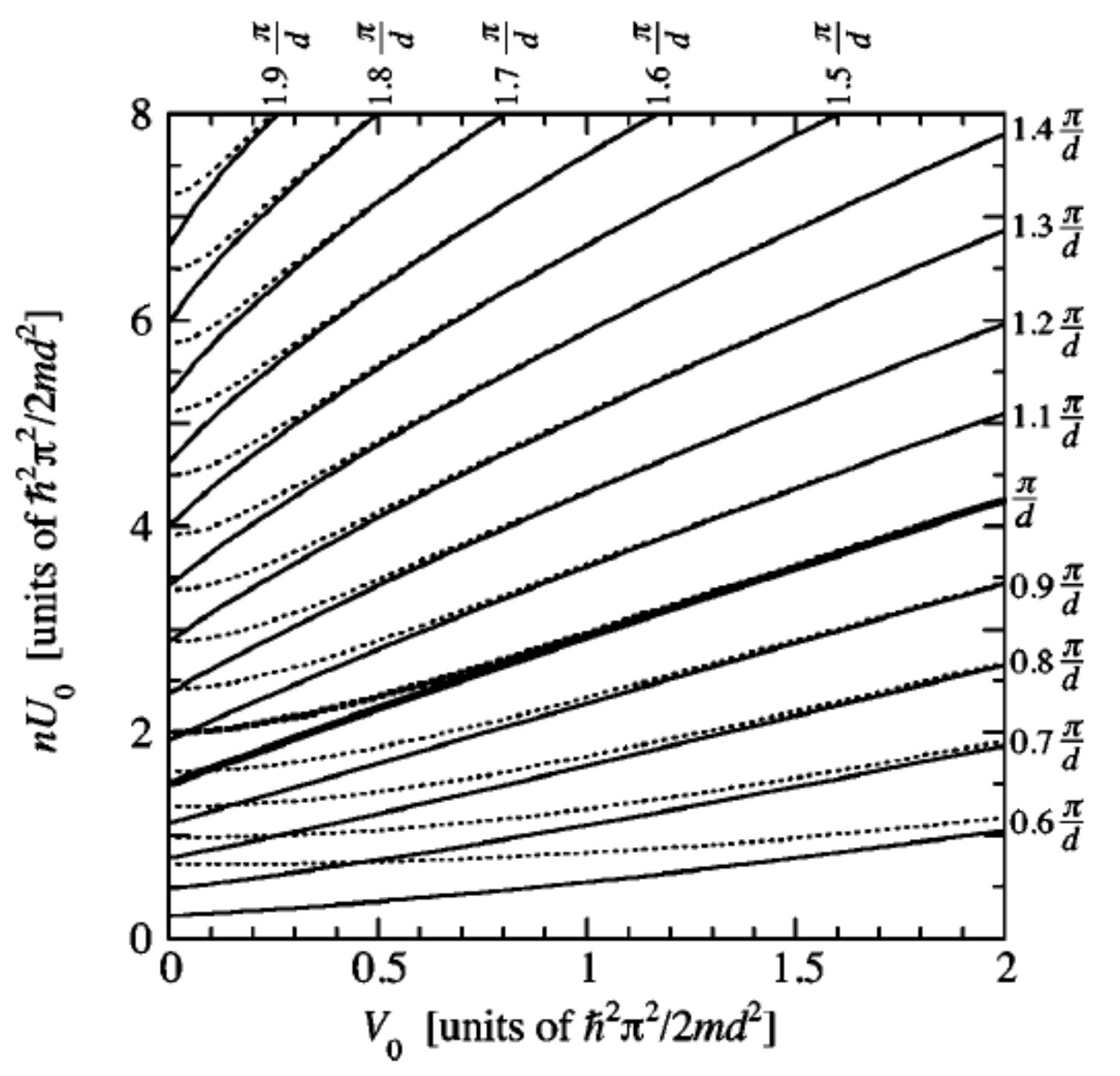}}
\caption{Contour plot of maximum quasimomentum $k$ for energetic stability (\textbf{a}) and dynamical stability (\textbf{b}) as a function of interaction and lattice depth for zone-edge loops. $k\leq \pi/d$ represents states on the lowest branch of the ground band and $k > \pi/d$ the states on the lower edge of the swallowtail loop. The dashed lines in  (\textbf{b}) are energetic stability curves for comparison. \mbox{Taken from~\cite{machholm03}.}}
\label{fig:pethergdynstab}
\end{figure}

In the stability analysis of \cite{machholm03}, in addition to standard Bloch states, also the ones corresponding to the loop solutions (lower branch) were considered. In Figure \ref{fig:pethergdynstab} the results of this analysis is shown by plotting the largest Bloch wave vector $k$ for which the states are energetically stable as a function of $ng$ and $V_0$. States with $0\leq k \leq \pi/d$ correspond to states with the lowest energy and $k>\pi/d$ represents lower edge-loop states. In general it was found that the range of quasimomentum values at which the system is energetically stable increases with the interaction strength $gn$. They also found that the wave vector of the tip of the swallowtail sets a natural limit for the wave vector at which instability sets in for parameter regimes where swallowtail loops occur. Moreover they found, in agreement with \cite{wu01}, with increasing $k$ the long wavelength perturbations with $q \rightarrow 0$ become energetically unstable first.
Hence a hydrodynamic description can be constructed \cite{mamaladze66,hakim97,vc} and it also gives analytical predictions for the instability contour as a function of $V_0$ and $g n$ for the zone edge with $k=\pi/d$ which compares favorably with the numerical calculations \cite{vc}. Note that as expected the states on the upper edge of the loop are always energetically and dynamically unstable.

Finally, the discussion so far was limited to only the linear stability of equilibrium solutions but the full response of the solutions of time dependent GP equation (\ref{GP1}) to perturbations was also studied numerically in \cite{seaman05a}. Here the stable lifetime of a given initial equilibrium state was defined as the time taken for the variance of the Fourier spectrum of the time-dependent order parameter from the initial Fourier spectrum (normalized to the initial spectrum) to exceed the value $1/2$, \textit{i.e.}, the time at which the order parameter becomes very different from the initial solution and taken as an indicative time for onset of dynamical instability. They found that for weakly attractive condensates the zero quasimomentum Bloch state in the first band is long-lived under white noise perturbations but highly unstable for time periodic perturbations. For both weak and strong attractive interactions the higher band Bloch states are unstable but the first band Bloch states with non-zero quasimomentum can be stable even to harmonic perturbations owing to their negative effective mass (defined as the inverse of the Band curvature). For weak repulsive interactions, Bloch states in the lowest band are stable as long as the quasi-wave number $k<\pi/d$, as at larger quasi-wave numbers the effective mass becomes negative, for the particular choice of Kronig--Penney potential (\ref{eq:kronigpenney}) used in \cite{seaman05a}. However, in agreement with the linear stability analysis, as the interaction strength grows, larger parts of the energy band including the lower branch of the loops are stable. In a later publication the same authors \cite{danshita07,sc07} showed that there is always a small part of the loop in the repulsive case that has negative effective mass but this area's extent monotonically decreases as $gn$ is increased. {A clear discussion of the dynamical stability of attractive BECs in an optical lattice is provided in \cite{barontini07}.\linebreak A recent review \cite{zhu15} also provides detailed treatment of stability of BECs in optical lattices.}

\subsection{Experimental Realization} \label{subsec:nonlinexp}

In the past decade and half there has been tremendous progress in the field of ultracold atoms in optical lattices \cite{bloch_review,lattice} with a range of experiments tackling interesting many-body physics. In this light it must be said that specific experiments dealing with nonlinear energy dispersions in optical lattices have been few and far in between. 

The earliest relevant experiment concerns the instability of superfluids in optical lattices by the group at LENS, Florence reported in Burger {\it et al.} \cite{burger01,burger01comment,burger01comment1}. In this experiment a cigar shaped quasi-1D BEC was localized in an harmonic trap and an optical lattice potential was also turned on. Following this, the center of the trap was suddenly shifted, corresponding to a sudden change of the quasimomentum in our language. For small shifts $\Delta x$ the dynamics of the BEC was coherent but for shifts greater than a critical value $\Delta x > \Delta x_c$, the oscillations are disrupted and the dynamics became dissipative. The authors of the experiment attributed this to the energetic Landau instability of the condensate. But theoretical work from Wu and Niu \cite{burger01comment,burger01comment1,wu03} showed that it may be more appropriate to associate this behavior with dynamical instability especially considering that the experimental parameters fall in a regime ($c\sim0.02$, $v\sim0.2$) where dynamical instability is rampant (see Figure \ref{fig:niustab1}b) and the critical displacement $\Delta x_c$ increases with decrease of lattice depth in the experiment (the energetic stability is essentially independent of lattice depth in the linearized stability treatment). {A follow-up comment from the authors of the experiment \cite{burger01comment,burger01comment1} was essentially inconclusive but hinted that even the GP equation may not be a valid description for some of their experimental results and beyond mean-field effects may have to be included. Following this, in~\cite{modugno04} a thorough theoretical treatment of the problem including 3-dimensional GP equations and effective 1-dimensional  GP equations taking into account the transverse degrees of freedom was undertaken. This theoretical treatment adapted to the experiment \cite{burger01} clearly showed that the onset of instability observed in the experiment was due to dynamical instability.  There are two main reasons that led to some uncertainty regarding the conclusions in the experiments \cite{burger01}---the inability to distinguish experimentally between dynamical and energetic stability as both processes manifest as an enhanced loss of atom number from the condensate and the inability to accurately set the initial quasimomentum of the condensate as a mean displacement from the harmonic trap center can in general lead to a mixture of quasimomenta and band eigenstates.

In the follow-up experiment from the LENS group \cite{dynstabexp,dynergstabexp} both these issues were succesfully resolved. In \cite{dynstabexp}, a moving optical lattice was implemented by frequency detuning the two laser beams creating the lattice. The ground state in a moving lattice is simply a state with a finite quasimomentum. By varying the amount of detuning, they were able to load the BEC adiabatically into states with a given quasimomentum both in the ground and excited bands. A subsequent measurement of the loss rates as a function of quasimomentum revealed a threshold for the onset of dynamical instability in very good agreement with the theoretical expectation for the lattice depth used. In \cite{dynergstabexp} they were able to also distinguish between the onset of energetic and dynamical instability by an ingenious use of a radio-frequency (RF) shield to selectively control the thermal fraction of the atomic cloud. The presence of the thermal cloud can effectively trigger energetic instability, which has generally a lower threshold in quasimomentum, by providing a dissipation channel. Hence when a large thermal fraction is present, the onset of dynamical instability is marred by energetic instability. On the other hand when the RF shield is turned on to remove the thermal fraction and experiments are performed with nearly pure condensate, the onset of dynamical instability stands out via a dramatic loss of atom number. It is important to emphasize at this point that experiments such as \cite{burger01} are performed at small lattice depths $v$. In this regime, as evident from Figure \ref{fig:niustab1}b, there are significant regions of quasimomentum space that are energetically unstable but dynamically stable. On the other hand at large lattice depths (see Figure 1 of \cite{dynergstabexp}), the region of quasimomentum space that exhibits only energetic instability is tiny. Thus in this so called tight-binding lattice limit, there was unambiguous agreement between experiments \cite{dynstabexpdeep} and the theoretical \cite{wu03,modinstabsmerzi} and numerical \cite{dynstabnumdeep,dynstabnumdeep1} treatments that predicted dynamical instability.\linebreak It was also shown in \cite{modinstabsmerzi} that the dynamical instability in this regime can also be viewed as a kind of modulational instability which is a general feature of nonlinear wave equations where a small perturbations of a carrier wave can exponentially grow as a result of interplay of dispersion and~nonlinearity.

In \cite{dynstabMISF}, the behaviour of the critical quasimomentum upto which superfluidity persisits across the superfluid-Mott insulator (SF-MI) quantum phase transition was studied. Within a mean-field GP picuture, as we have discussed in Section \ref{sec:loopstability}, the stability of the superfluid is in general enhanced with increasing interaction and lattice depth. On the other hand, within the full quantum model, there is a critical interaction strength to tunneling ratio beyond which the system is no longer superfluid and goes into the Mott insulating phase where the critical quasimomentum is trivially equal to $0$. This study maps the behaviour of the critical quasimomentum as it goes from a finite value in the SF phase to zero in the MI phase giving an accurate determination of the phase boundary. In \cite{ferris08} beyond dynamical instability at the zone edge is investigated experimentally and theoretically within the truncated Wigner approach which can account for beyond GP effects including the thermal depletion of the condensate. Finally, in recent experiments \cite{dynstabSOC} dynamical instability of spin-orbit coupled (SOC) BECs in moving optical lattices was investigated and a manifestation of the breakdown of Galilean invariance predicted for SOC systems was evidenced by the difference in the strength of the dynamical instability (measured by atom loss rate) depending upon the direction of motion of the~lattice.}

In the seminal experiment of the Bloch group \cite{chen11}, some aspects of the nonlinear LZ tunneling phenomena originally considered in the theoretical work of Wu and Niu \cite{niu00} was explored.\linebreak The experimental system consisted of an array of tubes of BECs in a 2D optical lattice potential.\linebreak A superlattice potential along $x$ direction allows pair-wise coupling between tubes giving many copies of coupled double wells. In the experiment they effectively realize the nonlinear LZ energy function of the the form:
\begin{align}
E[\psi_R,\psi_L] \approx \frac{\Delta}{2} \left(\vert \psi_R \vert^2 -\vert\psi_L \vert^2 \right) - J (\psi_R^{*}\psi_L + \mbox{c.c.}) + \frac{U}{2} \langle \delta \hat{n}^2\rangle  \left(\vert\psi_R \vert^4 + \vert \psi_L \vert^4 \right) \label{eq:blochexperg},
\end{align}
where $\psi_L$ and $\psi_R$ represent the amplitude to occupy either the left or the right tube. $\Delta$, the energy detuning between the tubes, and $J$, the tunnel coupling between the tubes can be controlled by varying the relative phase and lattice depth respectively of the superlattice potential along $x$ direction.

\begin{figure}[H]
\center{
\begin{tabular}{c  c }
\includegraphics[width=.45\textwidth,height=10cm]{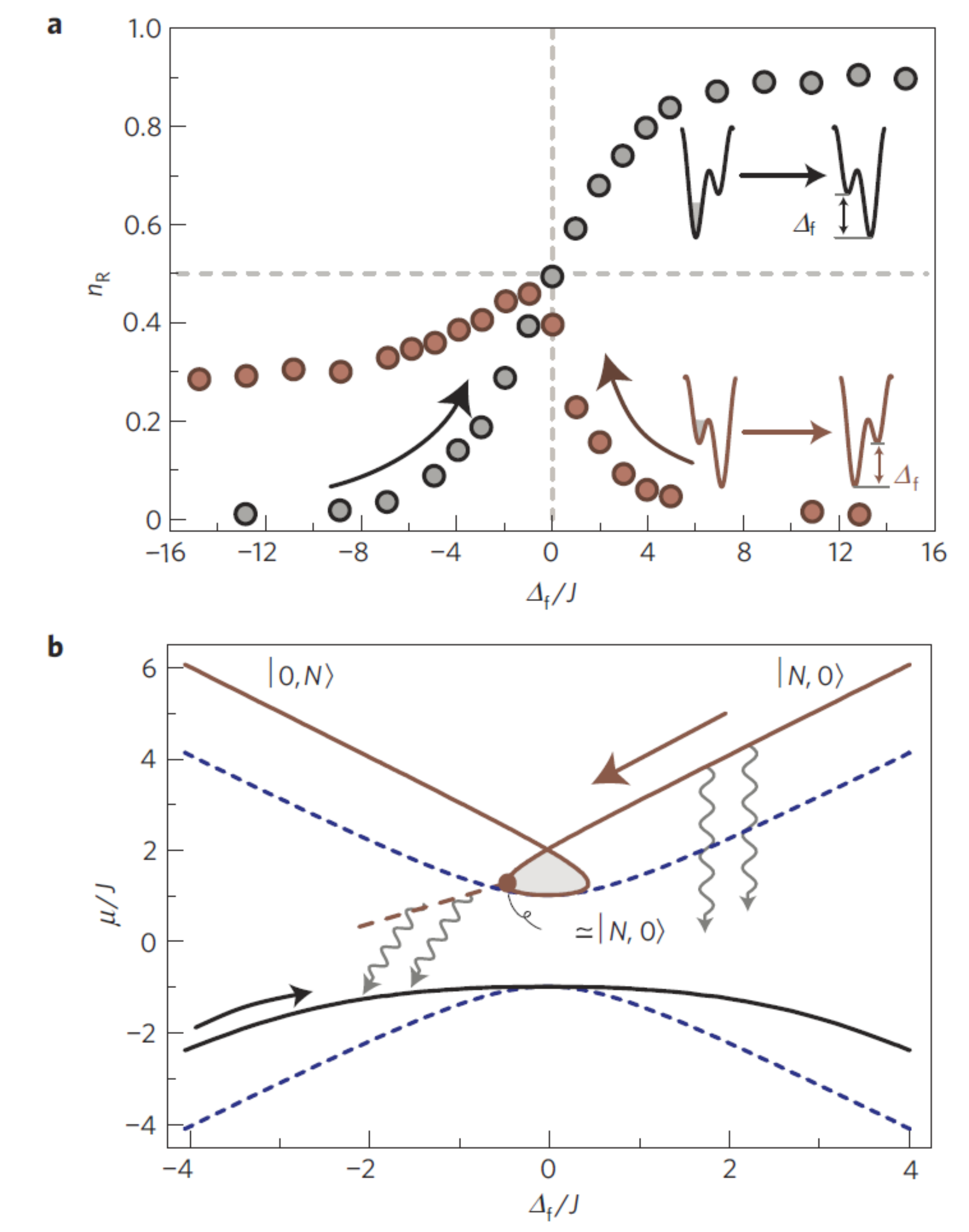}&
\includegraphics[width=.45\textwidth,height=10cm]{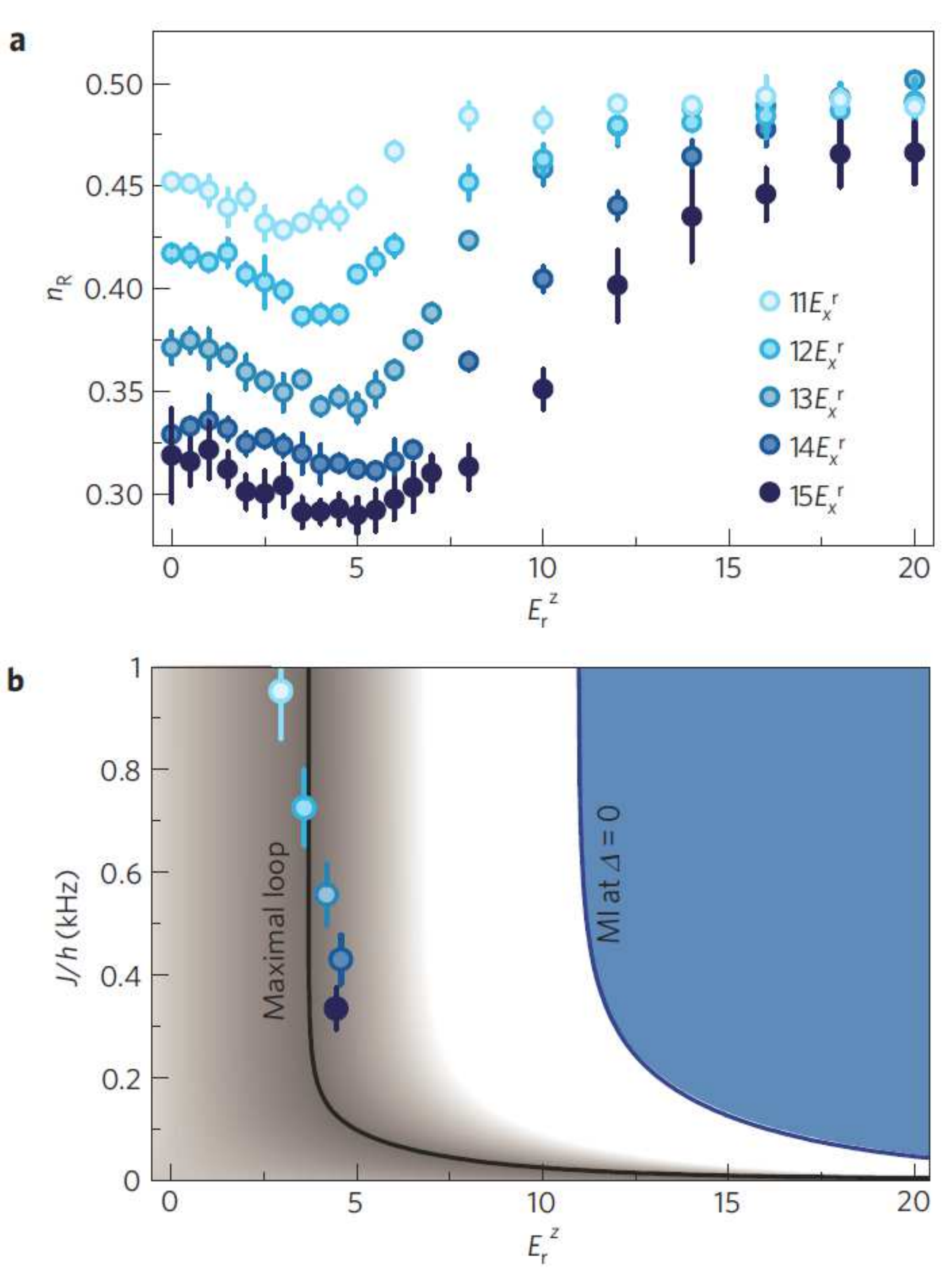}\\
(\textbf{Left})  & (\textbf{Right}) 
\end{tabular}
}
\caption{(\textbf{Left}) {({\bf a})} Transfer efficiency $n_R$ from filled to unfilled well (tube) as a function of the energy detuning for ground state sweep (gray dots) and excited state sweep (red dots) for constant sweep rate of $ 2 \pi J^2/\hbar |\alpha| = 2.1(1)$; {({\bf b})} Adiabatic energy levels for the excited (with loops) and ground state levels including strong repulsive interaction (solid lines) and excluding interactions (dashed lines). The curved lines indicate the relaxation process enabling non-adiabatic transitions. $\vert n_R, n_L \rangle$ is a shorthand denoting the left- and right-well (tube) population respectively. (\textbf{Right}) {({\bf a})} Transfer efficiency as a function of $z$-lattice depth at constant sweep rate $0.53(3)$; {({\bf b})} Phase diagram of the metastable excited condensate branch. In the gray shaded area there is a loop in the adiabatic energy level. Data points represent minima in the measured transfer efficiency and agree well with the solid line depicted for the calculated maximum loop size. Taken from \cite{chen11}.}
\label{fig:blochexp35}
\end{figure}

The experimental protocol consists of preparing all the atoms initially in the left tube with the initial detuning $\Delta_i$ either chosen to be negative (ground state) or positive (excited state) and sweeping the detuning at the linear rate $\alpha$ and finally measuring the number of atoms in the left and right tube. As already anticipated by Wu and Niu \cite{niu00}, the model [Equation (\ref{eq:blochexperg})] is exactly the same as the one introduced in Equation (\ref{eq:nonlinLZ}) except for the interaction term's sign is switched. As a result once the ratio of interaction to tunneling $\eta = U \langle \delta \hat{n}^2\rangle / J$  ($c/v$ in \cite{niu00}) is large, there is a loop in the upper branch as shown in lower panel of Figure~\ref{fig:blochexp35} left. In the experiment, for sweeps along the ground state branch (gray dots in Figure~ \ref{fig:blochexp35} left) there is no adiabaticity breakdown observed. In the sweep starting with the excited state (left tube at higher energy), there is a complete breakdown of adiabaticity even for reasonably small sweep rates $\alpha$ (red dots in Figure~\ref{fig:blochexp35} left). Moreover, for small sweep rates the transfer efficiency, given by number of atoms $n_R$ in the final state, decreases with decreasing sweep rate completely opposing the expected LZ behavior. The presence of the loop in the upper branch contributes to this adiabaticity breakdown for small sweep rates, as the atoms follow the upper branch and are \emph{self-trapped} in the middle branch of the loop, (see lower panel of Figure \ref{fig:blochexp35} left) which is a local maxima. They finally make a diabatic transition to the adiabatic ground state at the loop edge leading to the sharp breakdown of transfer efficiency near zero detuning seen in Figure \ref{fig:blochexp35} left upper panel. A confirmation of the effect of loops on the adiabaticity breakdown is provided by the non-monotonic behavior of the transfer efficiency (number of atoms in the initially empty right tube at the end) at a given sweep rate and tunnel coupling as a function of the $z$-lattice depth shown in top panel of Figure \ref{fig:blochexp35} right.\linebreak At small $z$-lattice depths, an increase in lattice depth effectively increases the on-site interaction $U \langle \delta \hat{n}^2\rangle$, leading to larger loop sizes which leads to lower transfer efficiency but eventually beyond a certain lattice depth the suppression of on-site fluctuations $\langle \delta \hat{n}^2 \rangle$ dominates leading to a decrease in the effective interaction and increase of transfer efficiency restoring standard LZ behavior. Moreover as shown in the lower panel in Figure \ref{fig:blochexp35} right, the experimentally determined minimum transfer efficiency agrees with a theoretical calculation for the position of the maximum loops size as a function of the $z$-lattice depth and tunnel coupling. 

In the experiment by Eckel {\it et al.} \cite{eckel14}, a physical situation approximately~corresponding to the Hamiltonian in Equation (\ref{eq:SFringham}) was realized for a BEC of $^{23}$Na atoms confined in a ring shaped trap \cite{amico05}. The goal of this experiment was to observe hysteresis between quantized states of circulation of the superfluid BEC caused by the presence of swallowtail loops in the energy landscape of such a system \cite{mueller02,baharian13}. In the experiment \cite{eckel14}, they were concerned with the quantized circulation states with winding numbers $n=0$ and $n=1$ with frequency $n$ times the rotational quanta $\Omega_0 = \hbar/mR^2$ and drive transitions between these states by tuning the relative angular velocity between the trap and the superfluid $\Omega$ which can be controlled by applying a repulsive rotating laser potential.

\begin{figure}[H]
\center{
\begin{tabular}{c c}
\includegraphics[width=.45\textwidth]{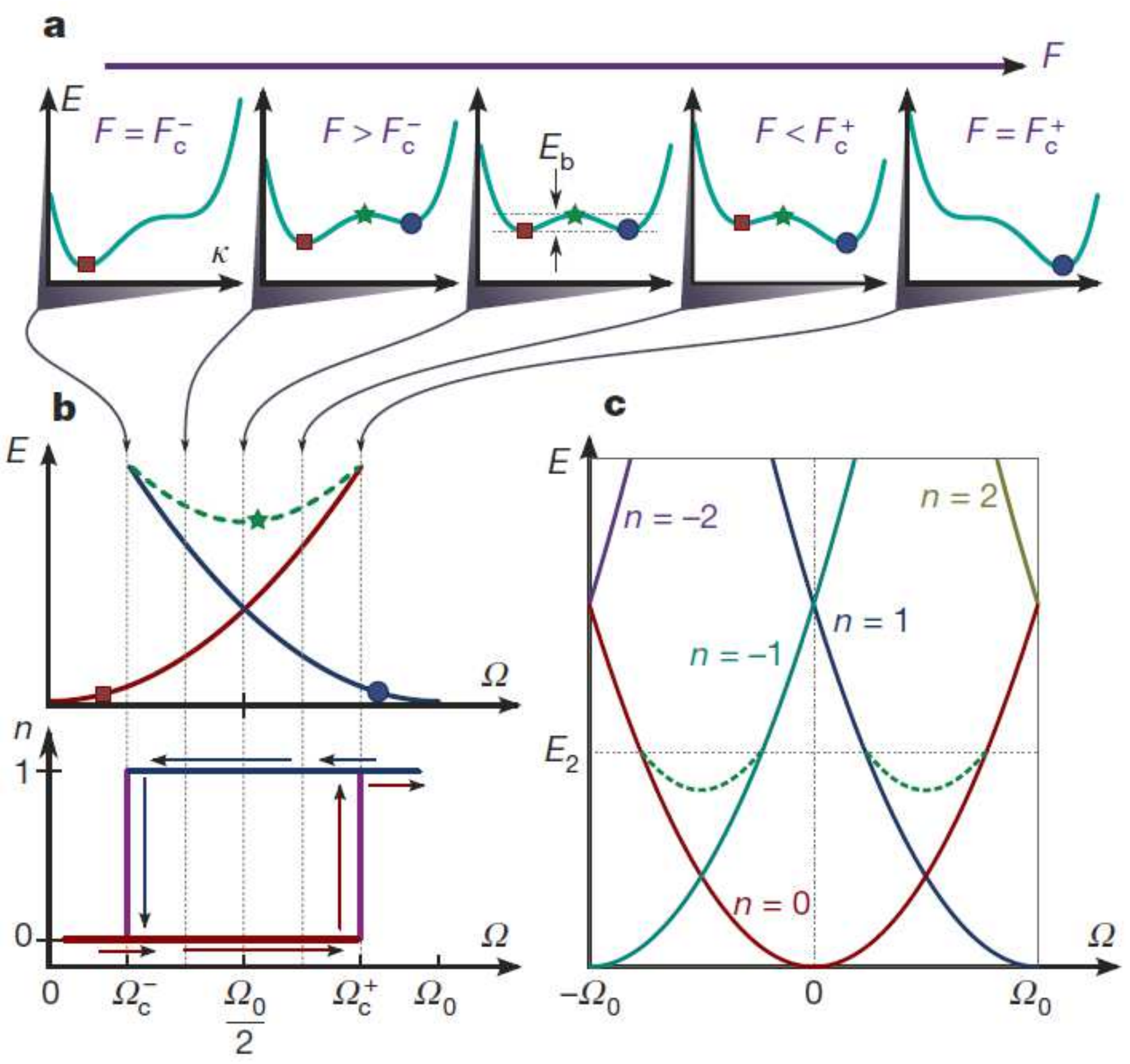}&
\includegraphics[width=0.45\textwidth]{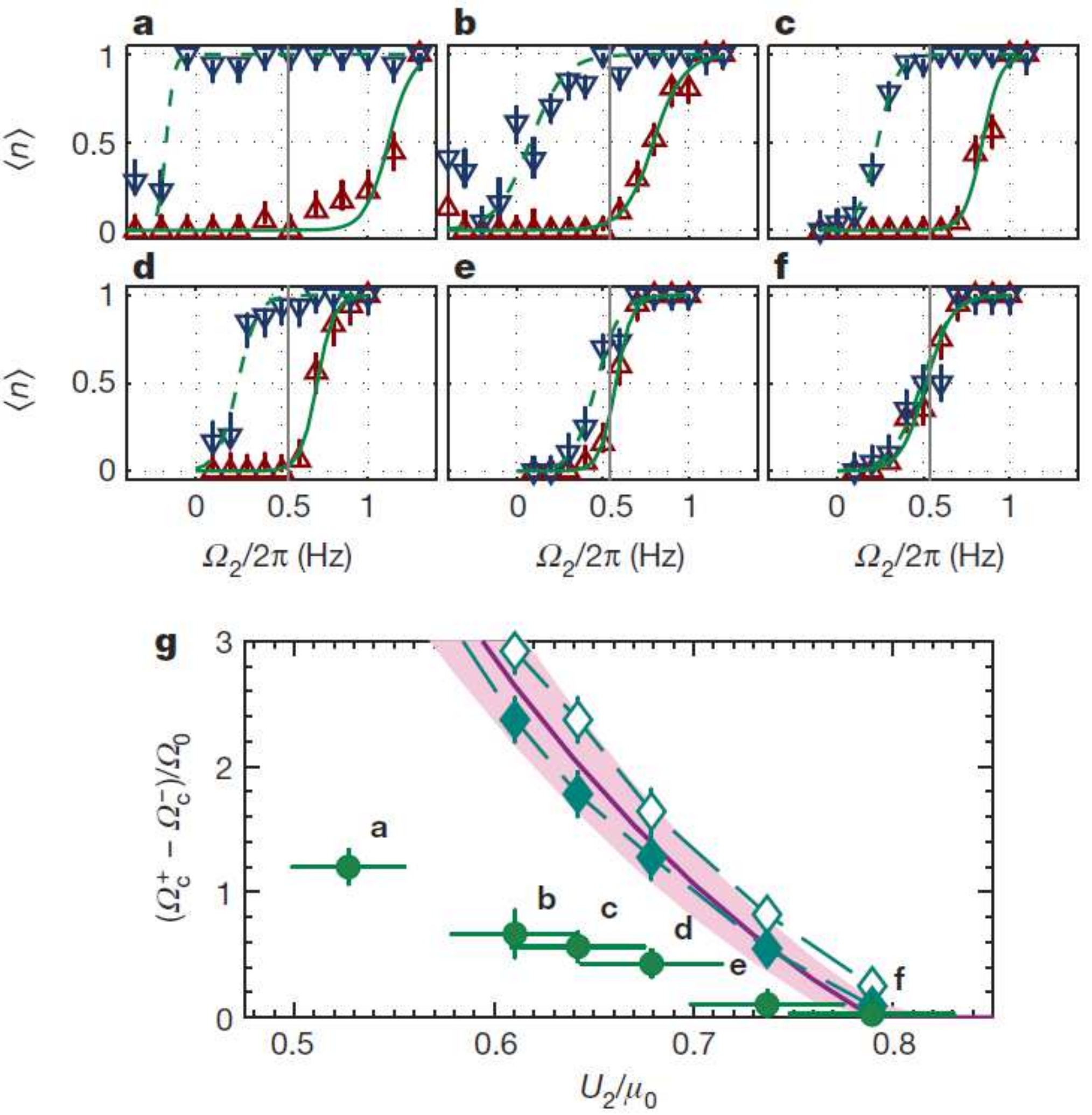}\\
(\textbf{Left}) & (\textbf{Right})
\end{tabular}
}
\caption{(\textbf{Left}) {({\bf a})} Schematic plot of the energy landscape of a hysteretic system as a function of the applied field $F$ (for example $\Omega$ the rotation rate of the superfluid) showing stable states of different energies separated by a barrier; {({\bf b})} Energy diagram for a superfluid as a function of $\Omega$ showing a swallowtail loop and the related hysteresis loop; {({\bf c})} The swallowtail structure is periodic in the rotation quanta $\Omega_0$. (\textbf{Right}) {({\bf a})--({\bf f})} Measured hysteresis loop with sigmoid fits at different values of potential strength $U_2$ shown in {({\bf g})}. The red up-triangles (blue down-triangles) are results from $20$ shot averages while starting from $n=0$ ($n=1$). {({\bf g})} Experimentally determined size of the hysteresis loop (green dots) as a function of $U_2$ and open and filled cyan diamonds are results of numerical GP calculation of dynamics with differing amounts of phenomenological dissipation. Taken from \cite{eckel14}.}
\label{fig:eckelexp13}
\end{figure}

In the absence of coupling between the different circulation states, the energy landscape of the interacting superfluid forms a swallowtail loop shape as a function of the relative angular velocity $\Omega$. At a fixed value of $\Omega$ in the swallowtail region, $n=0$ (red line in Figure \ref{fig:eckelexp13} left) and $n=1$ (blue line in Figure \ref{fig:eckelexp13} left) states form the minima of a double-well energy landscape with a barrier state (green dashed line) separating them. If the system begins in the $n=0$ state and its angular velocity is increased, the flow will be stable as long as $\Omega < \Omega_{c+}$ when it reaches the edge of the swallowtail and after this it will make a transition to the lower energy $n=1$ state. Beginning with the $n=1$ state, a similar stable flow can exist as long as $\Omega > \Omega_{c-}$, leading to the hysteresis loop shown in lower panel of Figure~\ref{fig:eckelexp13} left. The rate at which the repulsive potential created using a blue detuned laser is rotated, controls the flow velocity $\Omega$ and the strength of the potential $U$ controls and drives transitions (via phase slips) \cite{wright13} between different circulation states. Comparing to the Hamiltonian~(\ref{eq:SFringham}) the repulsive potential actuates two of the terms namely the rotation frequency $\Phi = \Omega/\Omega_0$ and the ``impurity'' term $\lambda$ whose strength is controlled by $U$.

In order to observe this hysteresis loop in the experiment, the BEC is prepared initially in either the $n=0$ or $n=1$ state in the trap and then this repulsive trap potential with a chosen strength $U_2$ and different rotation velocity $\Omega_2$ is applied for a fixed time of $2$ s, followed by a time of flight image to determine the final rotational state $n$. Due to the swallowtail loop structure, at a given strength $U_2$, as shown in upper panel of Figure \ref{fig:eckelexp13} right, the transition from $0$ to $1$ and \emph{vice-versa} clearly happen at different angular momenta $\Omega_{c\pm}$. Moreover as the strength of the applied potential $U_2$ is increased, $\Omega_{c \pm} \rightarrow \Omega_0/2$ and the swallowtail loop size decreases and eventually vanishes. The size of the loop as a function of $U_2$ was determined from the experiment and is plotted in the bottom panel of Figure \ref{fig:eckelexp13} right. The discrepancy from theoretical calculations that include relaxation effects required to accomplish the non-adiabatic transitions indicate that further work may be required to understand some quantitative aspects of the experiment. A detailed treatment of modeling the relevant excitations leading to the dissipation by vortex-antivortex pairs was already provided by the authors in \cite{eckel14}.

\subsection{Other Extensions}

Owing to the general nature of swallowtail shaped dispersions resulting from the interplay of atom-atom interactions and periodicity, they have been predicted to occur in a variety of systems different from the setting mainly considered in this review---namely that of BECs in an optical lattice with effectively 1D dynamics. In this subsection we catalogue these developments without going into the details owing to the restricted scope of the review.

Lin {\it et al.} \cite{dipswallow} study BECs with magnetic dipole-dipole interactions in optical lattices. In this case the effective atom-atom interaction can be controlled by changing the alignment of the atomic dipoles to the optical lattice axis by applying magnetic fields. At strong enough interaction, they observe swallowtail loops whose sizes and stability may be controlled by modifying the magnetic dipole orientations. In \cite{cavloops}, Venkatesh {\it et al.} study band-structures of atoms confined in optical lattices formed inside optical cavities that are continuously driven by an external laser. In the limit of very dilute gases, $s$-wave contact interactions do not play a role but the cavity-induced atomic interactions in the strong coupling regime of cavity quantum electrodynamics (QED) can lead to swallowtail loop structures in the atomic bands. Moreover this also corresponds to bistable solutions for the steady state photon number in the cavity. In the work of Watanabe and colleagues \cite{swallowtail}, swallowtail band structures of superfluid Fermi gases in optical lattices in the BEC-BCS crossover are investigated. They find that typically the width of the swallowtail is largest at unitarity. In~addition, they find that the microscopic mechanism of the emergence of the swallowtail in the BCS side of the crossover is very different in nature from that of the BEC case: a narrow band in the quasiparticle energy spectrum close to the chemical potential plays a crucial role for the appearance of the swallowtail in the BCS~side. It is also pointed out that, as a consequence, the incompressibility experiences a profound dip.
Chen and Wu extend the study of interplay between interactions and band structures of superfluid systems in optical lattices to two dimensions in \cite{diracpts} considering BECs in honeycomb shaped optical lattices. Such optical lattices serve as analogues to the structure of graphene and support Dirac points in their band close to which the energy dispersion is linear in 2D having characteristic conical shape. In \cite{diracpts} the authors show that even for arbitrarily small interaction strength, the Dirac point is extended into a closed curve and a tube like structure, a 2D version of the 1D swallowtail loop, arises around the original Dirac point. Moreover in work that followed closely thereafter Hui {\it et al.} showed that even in the case of 2D optical lattice with double-well superlattice like geometry along one direction, swallowtail loop structures emerge for any interaction strength \cite{twodloop}. Thus, the possibility of having swallowtail loops structures or analogues thereof for arbitrary small interactions seems to be a more ubiquitous feature in 2D as opposed to 1D where a the nonlinearity given by interactions has to be comparable to the lattice potential. {In \cite{impurity}, a BEC trapped in a double-well potential with an additional degree of freedom given by a single bosonic impurity atom that interacts with the condensate is considered. In this setup, as the impurity-BEC interaction strength is tuned above tunneling energy of the bosons, swallowtail loops appear in the adiabatic energy dispersions as a function of the tilt of one of the wells relative to the other. The relation between swallowtail loops and self-trapping of the condensate in one well or the other as well as relation to the Dicke model are explored.}

\subsection{Future Prospects}

Presently, experiments showing direct evidence for looped band structures for BECs in optical lattices have not been performed. One of the impediments preventing the experimental observation of loops in optical lattices is the requirement of large atom-atom interactions so that a large part of the loop solution is energetically stable \cite{danshita07,sc07,seaman05a}. The simplest way to ascertain breakdown of adiabaticity caused by the loop is to study Bloch oscillations in optical lattices. In this regard it is quite important to control and characterize other sources of loss of adiabaticity such as LZ tunneling to higher bands and distinguish them from the effect of the loops. Clearly for this a control of atom-atom interaction from very small to large enough to obtain loops is required. In this context, some recent experiments in the group of N\"{a}gerl with the ability to tune interactions using Feshbach resonances is promising \cite{haller10,meinert14}. Also, the extended theoretical schemes for 2D optical lattices \cite{diracpts,twodloop} may be easier to implement in an experiment as they do not require large atom-atom interactions to have looped energy dispersions. On the theoretical side, a clear understanding of the quantum mechanical underpinnings of the mean-field loops is already available for the case of repulsive and attractive interacting BECs in double-well potentials \cite{quantloops,chen11}. An extension of such a study to optical lattices in two dimensions or for fermionic atoms can be interesting. Another interesting theoretical consideration would be examine the idea of shortcuts to adiabaticity \cite{shortcuts} that has received a lot of attention of late to systems where the underlying evolution equation is not linear and understand if one may conceive of protocols where the loss of adiabaticity predicted due to the loops could be~avoided.


\section{Multiple Period States in Cold Atomic Gases in Optical lattices\label{sec:multiperiod}}

Density structures and patterns caused by the interplay of competing effects are ubiquitous in nature. In the case of superfluids flowing in a periodic potential, non-trivial density patterns can emerge due to the interplay of spatial periodicity imposed by the external potential and the nonlinearity due to the superfluid order parameter.

According to the conventional wisdom of the Bloch theorem, in the linear system described by the Schr\"odinger equation, the density pattern of the stationary solution in a lattice is periodic with periodicity coinciding with that of the lattice potential. However, nonlinearity can cause non-trivial density patterns with different periodicity. For BECs in a periodic potential, it has been found that nonlinearity of the interaction term can cause the appearance of stationary states whose period does not coincide with that of the lattice; instead, a multiple of it \cite{machholm04,pethick_smith}. Such states are called multiple (or~$n$-tuple) period states.

In this section, we discuss multiple period states of superfluid atomic gases in optical lattices. In the Section \ref{subsec:basics}, we provide the basic physical idea of the emergence of the multiple period states due to nonlinearity. To provide a physical picture concisely, we take the BEC case as an simple example. In the Section \ref{subsec:pdbec}, we present an account of existing results of the multiple period states in BECs. In the Section \ref{subsec:pdfermi}, we present some theoretical results of the multiple period states in superfluid Fermi gases along the BCS-BEC crossover focusing on their unique features in contrast to the multiple period states in BECs.

\subsection{Basic Physical Idea: A Simple Explanation of the Emergence of Multiple Period States by a Discrete~Model \label{subsec:basics}}

The emergence of the multiple period states in BECs can be explained by the discrete model (Equation (\ref{hamilt})) in a simple manner \cite{machholm04}. In the following exposition, we follow the discussion given in the above cited paper. For clarity, here we focus on the period-2 states: states whose period is equal to twice of the lattice constant $d$.

The stationary states with a fixed total number of particles $N$ can be obtained by the variation of $H' \equiv H - \mu N$ with respect to the amplitude $\psi_j^*$ at site $j$, where $\mu$ is the chemical potential and $N= \sum_j |\psi_j|^2$,
\begin{equation}
  \frac{\delta H'}{\delta \psi_j^*} = -K (\psi_{j+1} + \psi_{j-1}) + U |\psi_j|^2 \psi_j - \mu \psi_j = 0\, .
\label{eq:dhdpsi}
\end{equation}
We then separate from $\psi_j$ a plane wave part at site $j$, $e^{ikjd}$, as $\psi_j = e^{ikjd} g_j$ with $\hbar k$ being quasimomentum of the bulk superflow flowing in the same direction of the periodic potential and $g_j$ being the complex amplitude at site $j$. Equation (\ref{eq:dhdpsi}) becomes
\begin{equation}
  -K (g_{j+1} e^{ikd} + g_{j-1} e^{-ikd}) + U |g_j|^2 g_j - \mu g_j = 0\, .
\label{eq:statsolj}
\end{equation}

Due to the boundary conditions of the period-2 states, we have $g_0 = g_2$ and $g_1 = g_3$. We solve combined two Equations (\ref{eq:statsolj}) for $j=1$ and $2$ with these boundary conditions. Subtracting these two equations, 
we obtain
\begin{equation}
  -2K \cos{kd} \left(\frac{|g_2|}{|g_1|}\, e^{i(\phi_2-\phi_1)} - \frac{|g_1|}{|g_2|}\, e^{i(\phi_1-\phi_2)}\right) + U (|g_1|^2 - |g_2|^2) = 0\, ,
\label{eq:statsolper2}
\end{equation}
with $g_j \equiv |g_j| e^{\phi_j}$.

For the linear case ($U=0$), we can readily see that $|g_1| \ne |g_2|$ cannot satisfy Equation (\ref{eq:statsolper2}) except at $kd = \pi/2$, which corresponds to a trivial solution of $g_1=g_2=0$. On the other hand, solutions with $|g_1| = |g_2|$ exist provided $\phi_2-\phi_1 = 0$ (modulus of $2\pi$): thus these solutions are normal period-1~states.

For the nonlinear case ($U \neq 0$), nonzero contribution from the kinetic energy part (the first term in the left-hand side of Equation (\ref{eq:statsolper2})) can be compensated by that from the interaction energy part (the second term in the left-hand side of Equation (\ref{eq:statsolper2})) so that this equation can be satisfied. Therefore, the emergence of the period-2 states is a purely nonlinear phenomenon. Since the second term in the left-hand side is real, the phase difference $\phi_2 - \phi_1$ should be $0$ or $\pi$, namely:
\begin{equation}
  \pm 2 K \cos{kd}\, \left(\frac{|g_2|}{|g_1|} - \frac{|g_1|}{|g_2|}\right) = U (|g_1|^2 - |g_2|^2)\, .
\end{equation}
Thus we obtain
\begin{equation}
  \pm \frac{\cos{kd}}{\frac{U\nu}{2K}} = \frac{|g_1| |g_2|}{\nu}\, ,
\label{eq:statsolfin}
\end{equation}
where the filling factor $\nu \equiv (|g_1|^2 + |g_2|^2)/2$ is the average number of particles per cell (in the present case of the period-2 states, the cell consists of two lattice sites). Since the right-hand side of Equation~(\ref{eq:statsolfin}) takes $0 \le |g_1| |g_2|/\nu \le 1$, solutions with period $2d$ exist when $|\cos{kd}| \le U\nu/2K$ \cite{machholm04}.

\begin{figure}[H]
\center{
\begin{tabular}{c  c  c}
\includegraphics[width=0.3\textwidth]{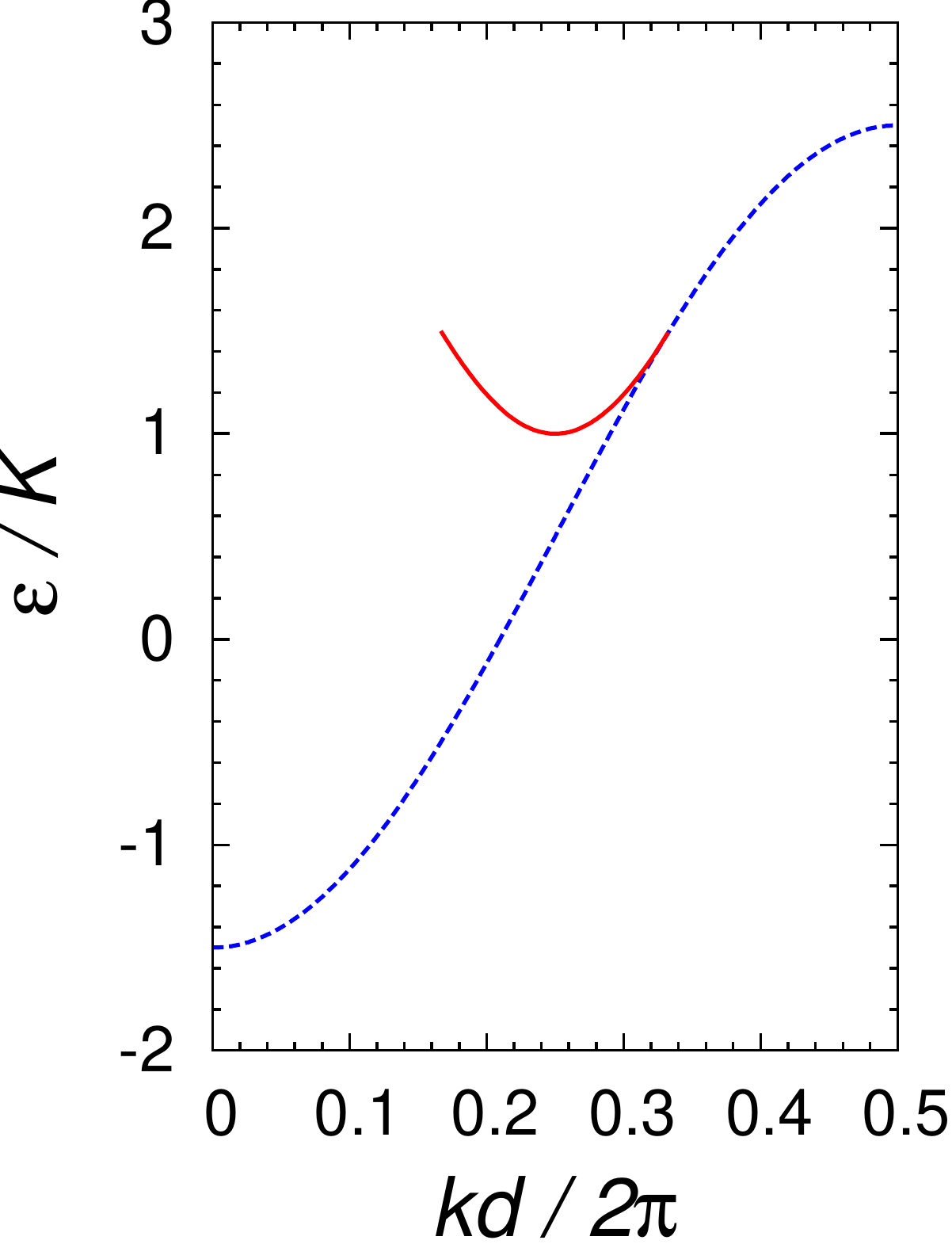}&
\includegraphics[width=0.3\textwidth]{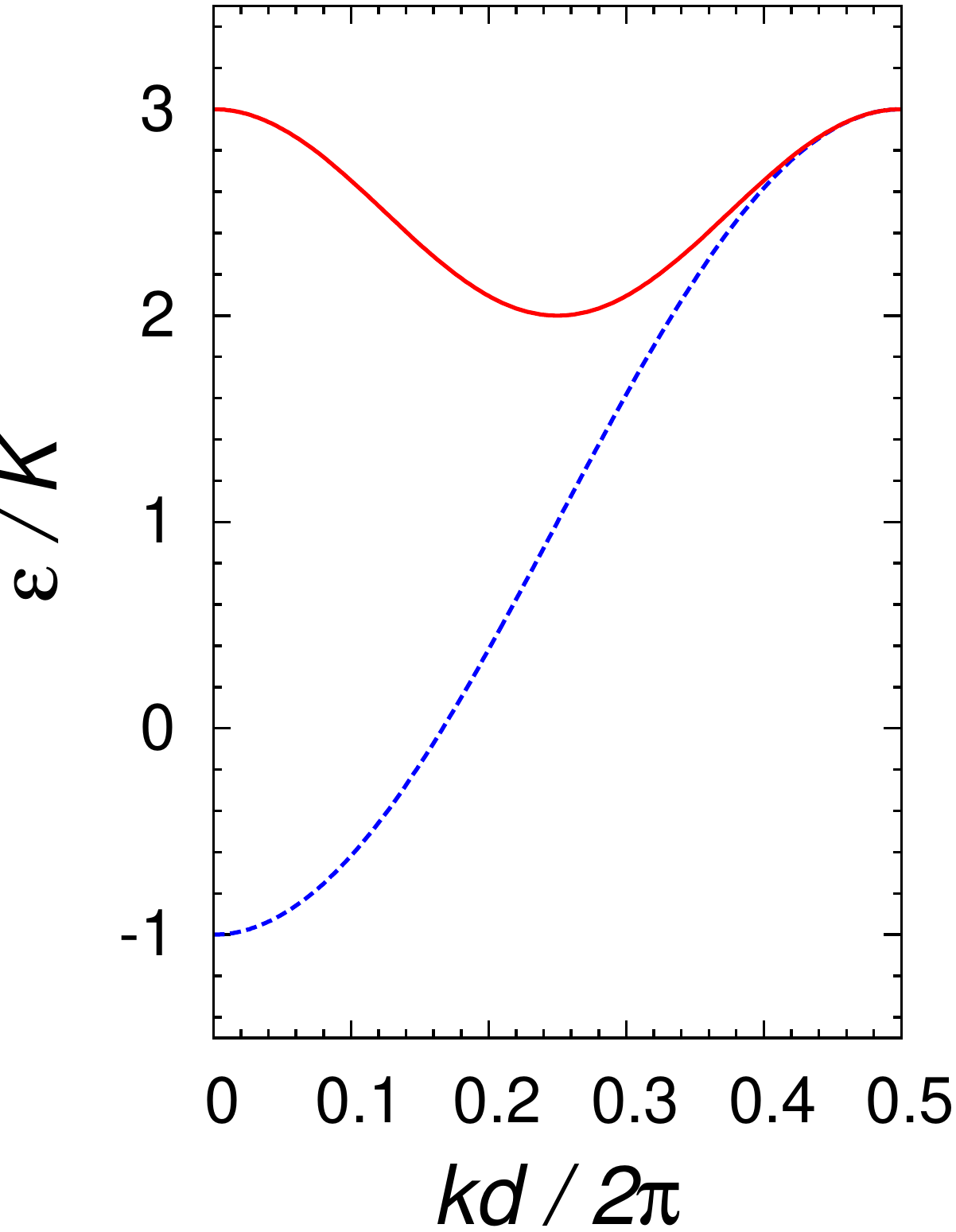}&
\includegraphics[width=0.3\textwidth]{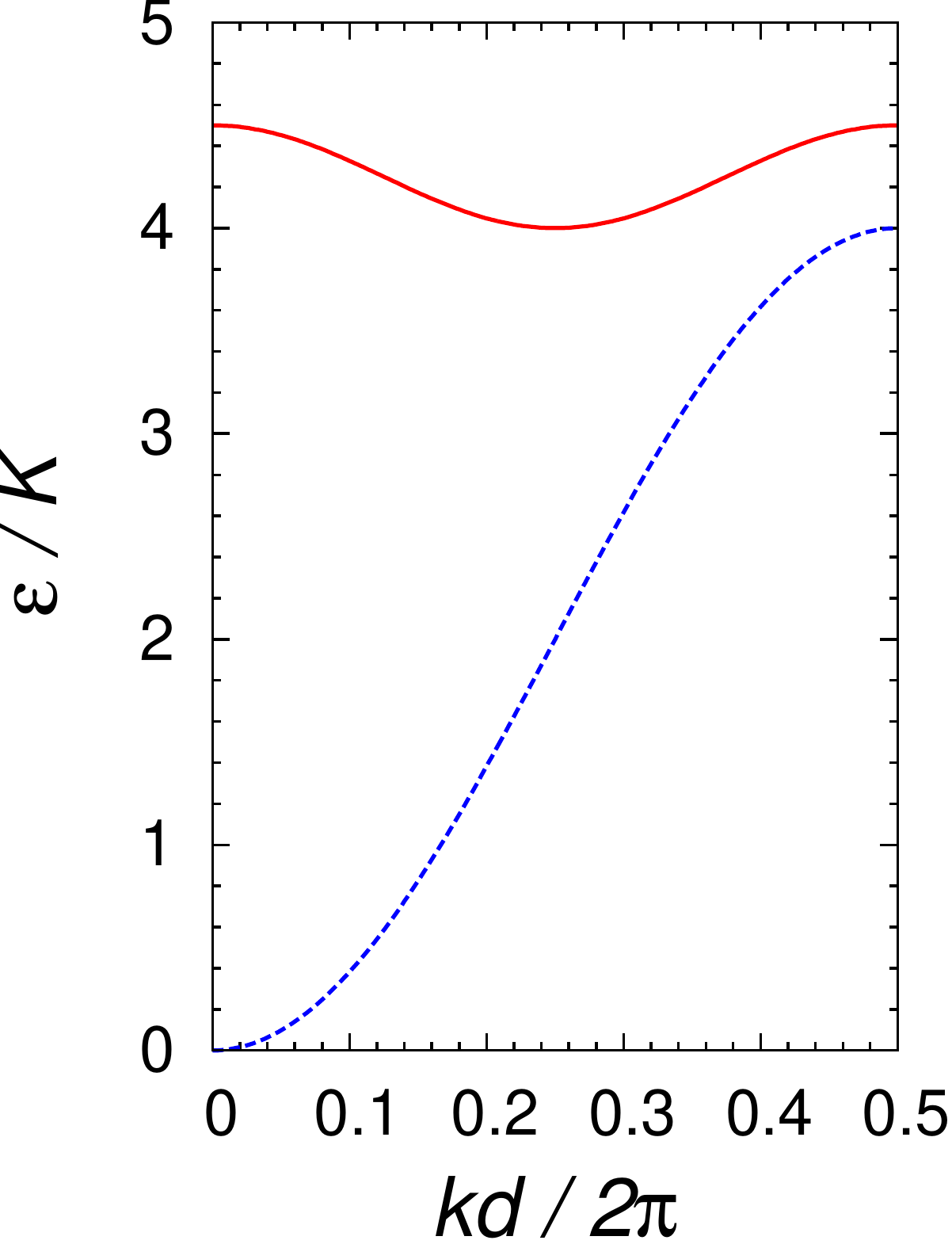}\\
       ({\bf a}) & ({\bf b}) & ({\bf c}) 
\end{tabular}
}
\caption{Energy per particle $\epsilon$ as a function of $k$ for period-1 (blue dashed lines) and period-2 (red solid lines) states of BECs in a periodic potential obtained from the discrete model, for ({\bf a}) $U\nu/2K=1/2$; ({\bf b}) $U\nu/2K=1$; ({\bf c}) $U\nu/2K=2$. In the case of the ({\bf a}), period-2 states exist in the limited region of $1/6 \le kd/2\pi \le 1/3$.}
\label{fig:pdtb}
\end{figure}

In Figure \ref{fig:pdtb}, we show the energy bands of the period-1 and period-2 states for $U\nu/2K=1/2$, $1$, and $2$. Note that, in the case of the Figure \ref{fig:pdtb}a for $U\nu/2K=1/2$, the period-2 states exist in the limited region of $1/6 \le kd/2\pi \le 1/3$. At infinitesimally small $U\nu/2K$, the period-2 states exist only at $kd/2\pi=1/4$. 
As $U\nu/2K$ increases, the region in which the period-2 states exist increases and it finally extends over the whole Brillouin zone for $U\nu/2K \ge 1$ (see Figures \ref{fig:pdtb}b and \ref{fig:pdtb}c).

Note that there is another class of period-2 states called the phase states \cite{machholm04}. From Equation~(\ref{eq:statsolper2}), we see that, at $|k|d = \pi/2$, $|g_1|=|g_2|$ is a solution for arbitrary phase difference $\phi_2-\phi_1$.\linebreak The periodicity of the density distribution and the energy of the phase states are the same as the normal Bloch state at $|k|d = \pi/2$, but only the phase profile has the period $2d$ \cite{machholm03}.

\subsection{Multiple Period States in BECs\label{subsec:pdbec}}

Multiple period states in BECs in optical lattices were first predicted by Machholm {\it et al.} \cite{machholm04}. Using both (1) the simple discrete model within the tight-binding approximation to the mean-field GP equation and (2) the more general continuum GP equation, they studied BECs flowing along the 1-dimensional external periodic potential of the form given by Equation (\ref{eq:lat}) (multiple period states of BECs in a Kronig--Penney potential (a periodic delta-function potential) were studied in \cite{li04}).\linebreak They have shown the existence of the multiple period states as stationary states and have clarified that they emerge due to nonlinearity originating from superfluidity.

\begin{figure}[H]
\centering
\rotatebox{0}{
\resizebox{!}{6cm}
{\includegraphics{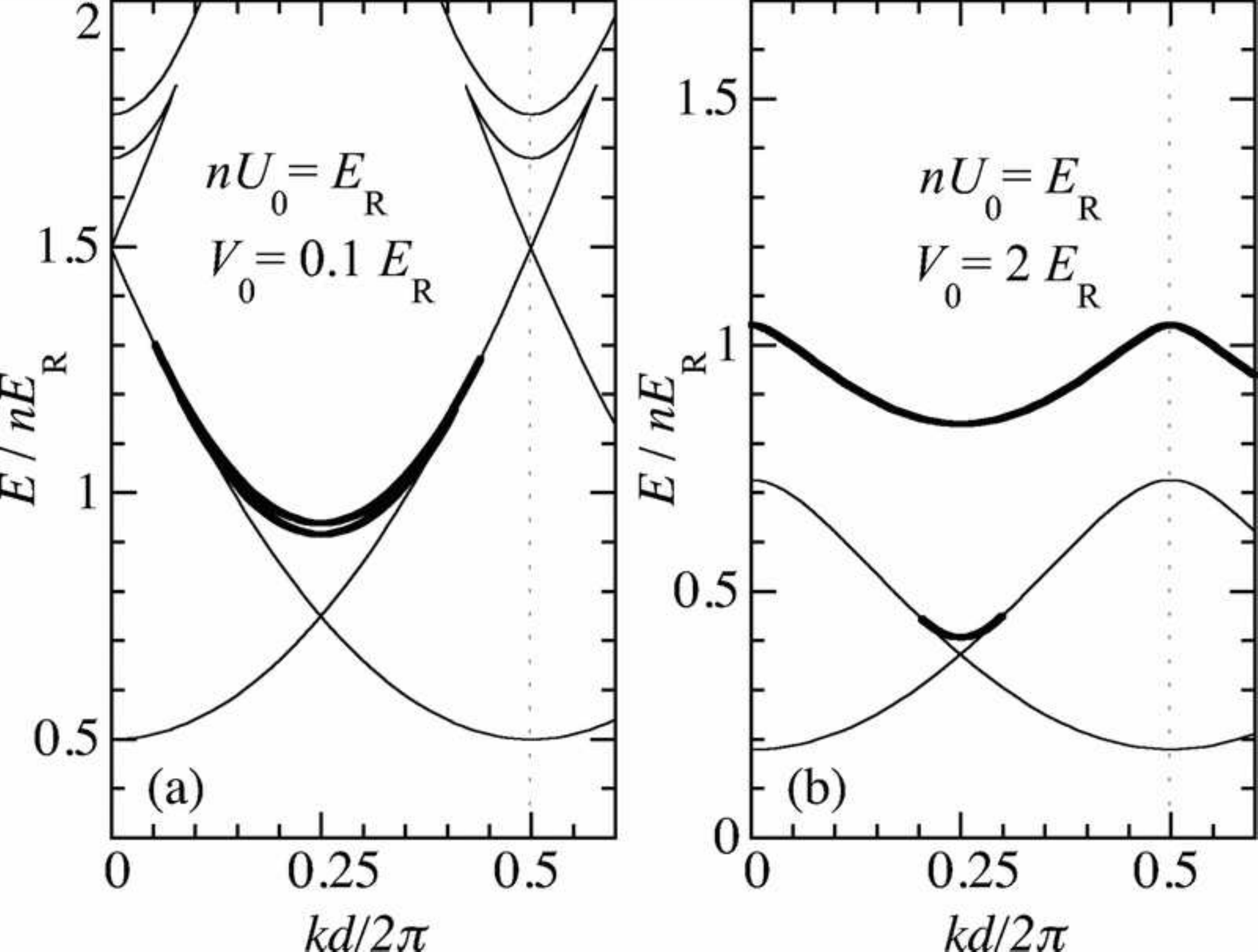}}}
\caption{Energy per particle $E/n$ of BECs in a periodic potential as a function of $k$ for the lowest bands obtained from the GP equation for the continuum model. The bands of the period-2 states are shown by the thick solid lines. In the notations of the present article, $U_0 = g$ and $V_0 = sE_R/2$. \mbox{This figure is taken from} \cite{machholm04}.}
\label{fig:pdcont}
\end{figure}

Figure \ref{fig:pdcont} shows the lowest energy bands obtained by solving the GP equation for the continuum model. A striking difference from the energy bands obtained from the discrete model discussed in Section \ref{subsec:basics} is that the phase states form a band in the continuum model (see the lower thick solid lines in Figure \ref{fig:pdcont}) and their density profiles have period $2d$.

\begin{figure}[H]
\centering
\rotatebox{0}{
\resizebox{!}{4.2cm}
{\includegraphics{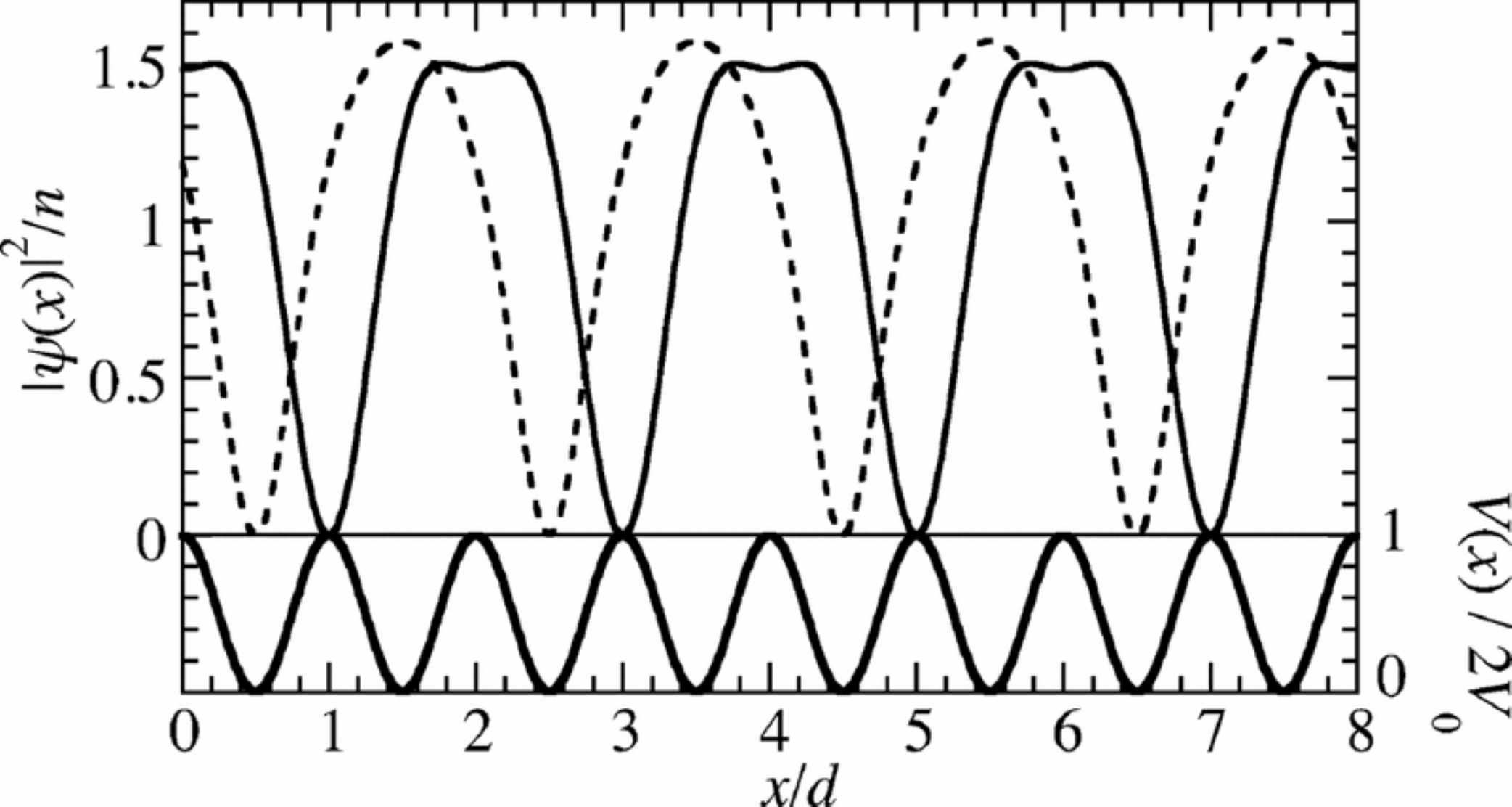}}}
\caption{Density profiles $|\psi(x)|^2$ of period-2 states (upper panel) of BECs and the periodic external potential $V(x)$ (lower panel). The solid (dashed) line in the upper panel shows $|\psi(x)|^2$ for the lower-energy (higher-energy) period-2 state of Figure \ref{fig:pdcont}a at $kd/2\pi = 1/4$, $gn = E_R$, and $s=0.2$. This figure is taken from \cite{machholm04}.}
\label{fig:profiles}
\end{figure}

Figure \ref{fig:profiles} shows the density profiles of the lower-energy and higher-energy period-2 states. By~comparing with the external potential $V(x)$ shown in the lower panel, we can see that the periodicity of these states is indeed $2d$. Here, the quasi-wave number $k$ of the superflow is at $k=\pi/2d$ corresponding to the first Brillouin zone edge of the system with period $2d$, and thus the condensate wave function $\psi$ has a node in each period. The density profile shown by the solid (dashed) line, which has nodes at the potential maxima (minima), is that of the period-2 state in the lower (higher) energy branch (see the lower (upper) thick solid line in Figure \ref{fig:pdcont}). According to the energy bands shown in Figure \ref{fig:pdcont}, we can also see that the lowest band of the period-2 states appears as an upper edge of the swallowtail for the period $2d$ system. Connection between the period-2 states and the swallowtail was studied in depth in \cite{seaman05b}.

It was pointed out that the lowest band of the period-2 states is closely connected with the dynamical instability \cite{machholm04}. The linear stability analysis has shown that, with increasing $k$, the dynamical instability of the normal Bloch state sets on at the quasimomentum where the band of the normal Bloch states merges with the lowest band of the period-2 states. There, the dynamical instability is caused by the growing perturbation mode with wavelength $2d$ \cite{wu01,machholm03,wu03,modugno04}.\linebreak The growth of the mode with wavelength $2d$ accompanying the dynamical instability has been observed experimentally as well \cite{gemelke05}. Since the lowest band of the period-2 states appears as the saddle of the swallowtail for the period $2d$ system and it forms the upper edge of the swallowtail, these period-2 states are dynamically unstable \cite{machholm03} while the upper branch of the period-2 states can be dynamically stable in some region of $k$ \cite{machholm04}. The lowest multiple period states can be dynamically stable by introducing long-range interactions. 
For example, it was demonstrated that, in dipolar BECs, multiple period states with period $2d$ and $3d$ can be dynamically stable even at $k=0$ provided the dipole-dipole interactions are repulsive and sufficiently strong \cite{maluckov12a,maluckov12}.

\subsection{Multiple Period States in Superfluid Fermi Gases\label{subsec:pdfermi}}

The emergence of the multiple period states in BECs in optical lattices is one of the novel nonlinear phenomena caused by the presence of the superfluid order parameter. However, in the normal repulsively interacting BECs without long-range interaction, the lowest multiple period states are higher in energy than the normal Bloch states and are dynamically unstable. Yoon {\it et al.}\cite{period_double} have shown that the situation is very different for the multiple period states of superfluid Fermi gases in the BCS regime.

\begin{figure}[H]
\centering
\rotatebox{270}{
\resizebox{!}{12cm}
{\includegraphics{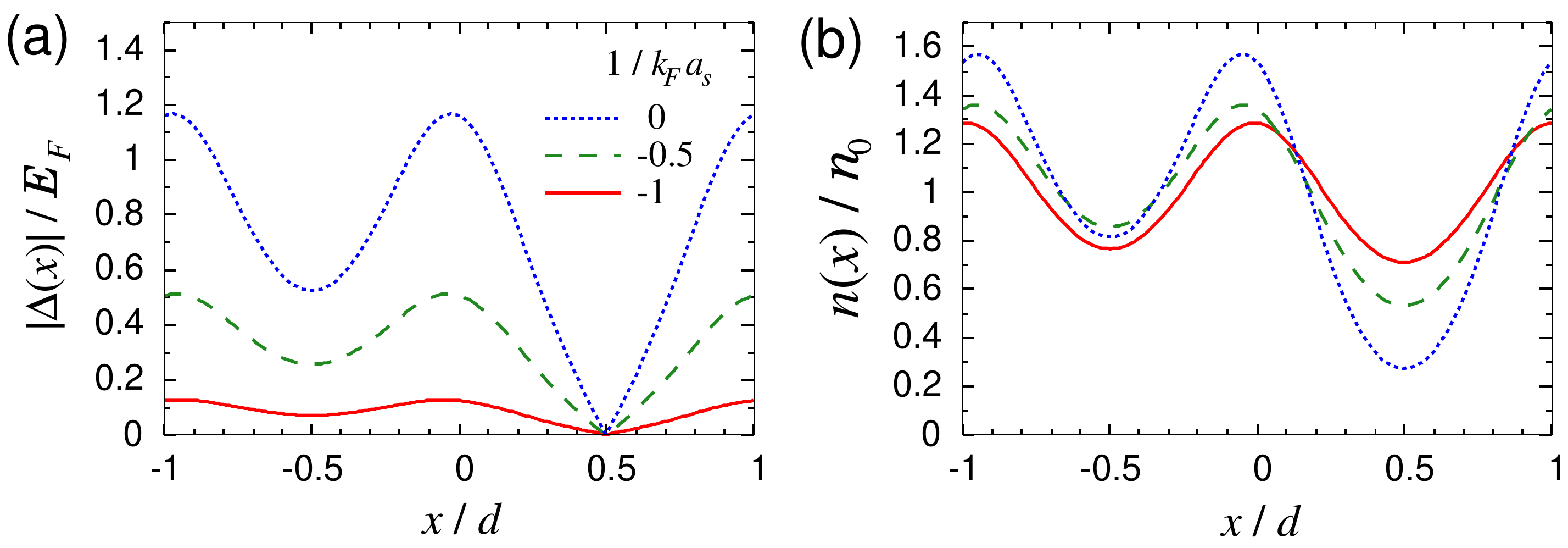}}}
\caption{Profiles of (\textbf{a}) the magnitude of the pairing field $|\Delta(x)|$ and (\textbf{b}) the density $n(x)$ of the lowest period-2 states of superfluid Fermi gases in the BCS-BEC crossover: $1/k_{F}a_s =-1$ (red solid~line), $-0.5$ (green dashed line), and $0$ (blue dotted line). The quasimomentum $P$ of the superflow is set at the Brillouin zone edge $P=P_{\rm edge}/2=\hbar q_{B}/4$ of the period-2 states, and other parameters are set at $s=1$ and $E_{F}/E_{R}=0.25$. This figure is adapted from \cite{period_double}.}\label{fig:profilesfermi}
\end{figure}

Figure~\ref{fig:profilesfermi} shows the profiles of the magnitude of the pairing field $|\Delta(x)|$ and the density $n(x)$ of the lowest period-2 states of superfluid Fermi gases in the BCS-BEC crossover obtained by solving the BdG equations for the continuum model (\ref{eq:bdg}). The striking difference from the BEC case shown in Figure \ref{fig:profiles} is that the feature of the period doubling shows up in the pairing field rather than the density. The difference between the regions of $-1<x/d\le0$ and $0<x/d\le1$ can be clearly seen in $|\Delta(x)|$ at any value of $1/k_F a_s$. On the other hand, the difference in $n(x)$ between these two regions is small in the deeper BCS side ($1/k_F a_s =-1$) (see the red line in Figure~\ref{fig:profilesfermi}b) and finally disappears in the BCS limit.

\begin{figure}[H]
\begin{center}
\rotatebox{0}{
\resizebox{7cm}{!}
{\includegraphics{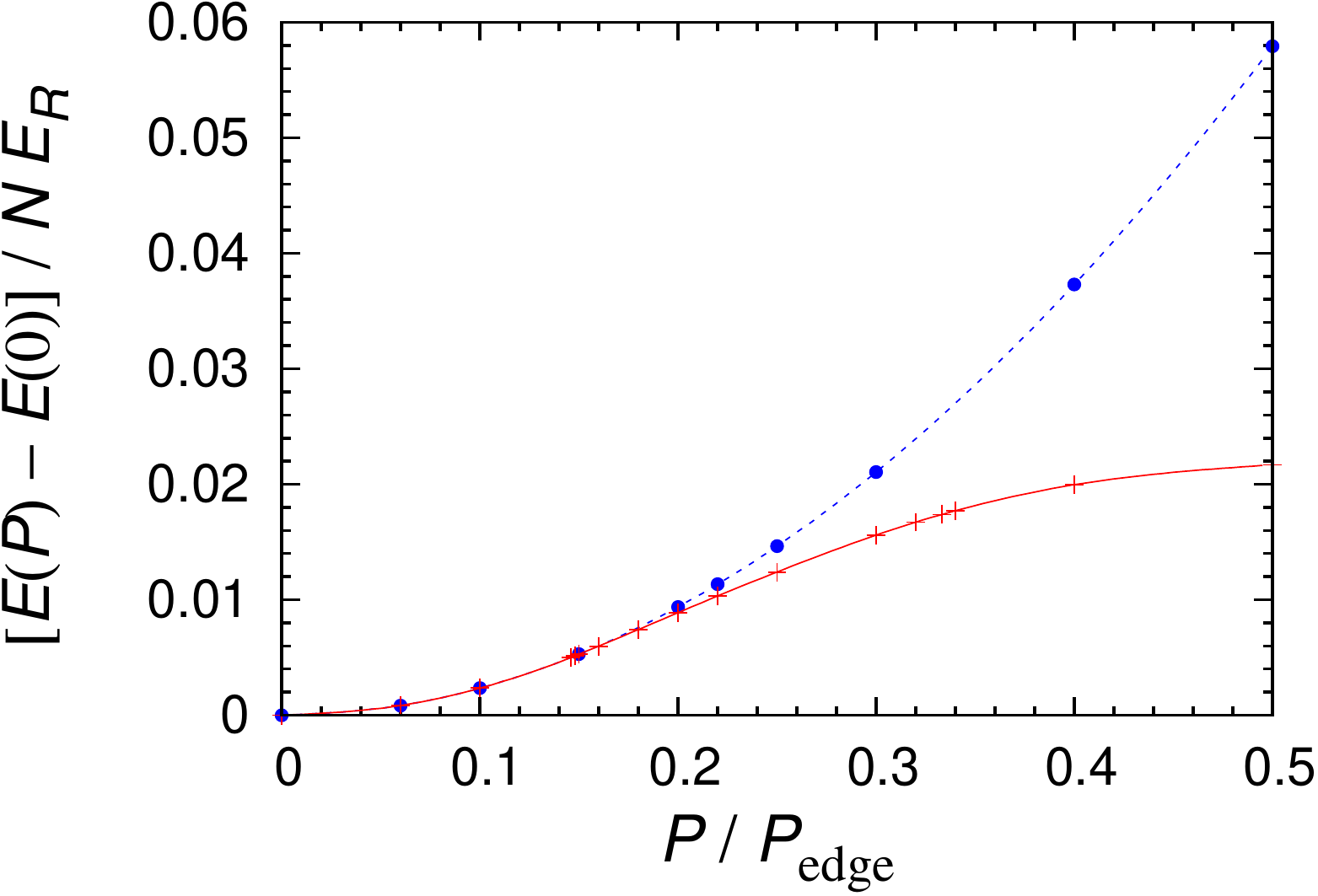}}}
\caption{Energy $E$ per particle of superfluid Fermi gases in a periodic potential as a function of the quasimomentum $P$. Parameter values are $s=1$, $E_F/E_R =0.25$, and $1/k_F a_s =-1$. The normal Bloch states with period $d$ are shown by the blue dotted line with $\bullet$ symbols, and the period-2 states are shown by the red solid line with $+$. Note that the period-2 states are energetically more stable than the normal Bloch states in the region of $0.2 \lesssim P/P_{\rm edge} \le 0.5$.}
\label{fig:efermi}
\end{center}
\end{figure}

Figure~\ref{fig:efermi} shows the energy bands of the lowest period-2 states in the BCS regime ($1/k_Fa_s=-1$) together with that of the normal Bloch states. In the region of small $P$, the line of the period-2 states coincides with that of the normal Bloch states, as they are equivalent in this region, the states with period 1 being just a subset of any multiple period states with integer periods. Unlike the period-2 states in BECs, which form the concave upper edge of the swallowtail (see Figure \ref{fig:pdcont}), here the band of the period-2 states is convex upward. Remarkably, the lowest period-2 states are energetically more stable compared to the normal Bloch states around the Brillouin zone edge of the period-2 states ($P/P_{\rm edge} \sim 0.5$).

\begin{figure}[H]
\begin{center}
\rotatebox{0}{
\resizebox{7cm}{!}
{\includegraphics{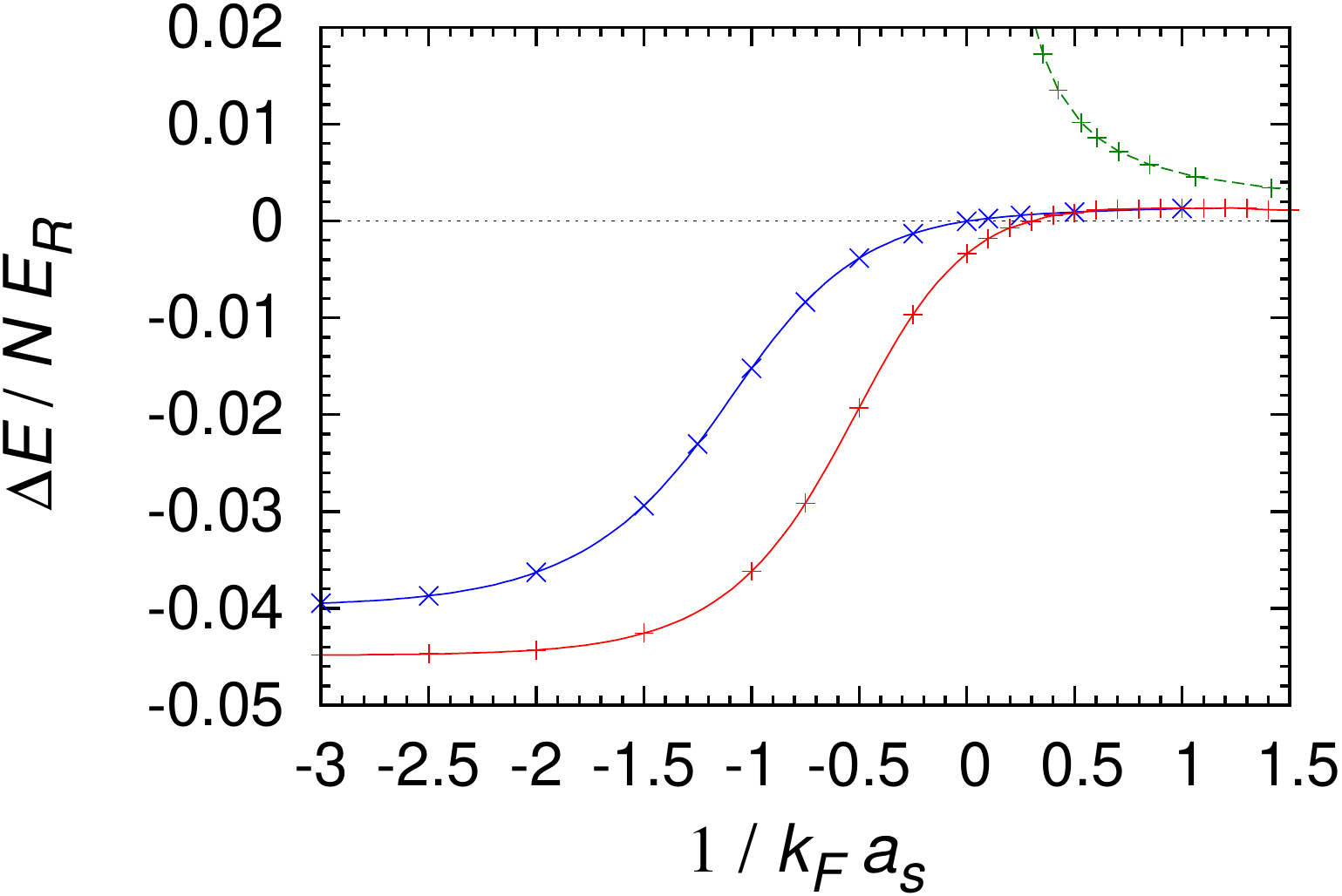}}}
\caption{Difference $\Delta E \equiv E_2-E_1$ of the total energy per particle between the period-2 states ($E_2$) and the normal Bloch states ($E_1$) at $P=P_{\rm edge}/2$ along the BCS-BEC crossover. The red solid line with $+$ is for $s=1$ and the blue solid line with $\times$ is for $s=2$; $E_F/E_R=0.25$. The green dashed line shows the results by the GP equation for parameters corresponding to $s=1$ and $E_F/E_R=0.25$. This figure is taken from \cite{period_double}.
}\label{fig:defermi}
\end{center}
\end{figure}

The manner in which the energy of the lowest period-2 states relative to that of the normal Bloch states changes along the BCS-BEC crossover can be seen in Figure~\ref{fig:defermi}.
As we have already seen, the period-2 states are energetically more stable (\textit{i.e.}, $\Delta E<0$) in the deep BCS regime, where the band of the period-2 states is convex upward. With $1/k_{F}a_s$ increasing from the deep BCS regime, $\Delta E$ increases from a negative value and finally period-doubled states become higher in energy than the normal Bloch states (\textit{i.e.}, $\Delta E>0$) in the BEC side, where the band of the period-2 states forms the concave upper edge of the swallowtail.

The energetic stability of the period-2 states in the BCS regime can be physically understood as follows. Let us consider the different behavior of $\Delta(x)$ and $n(x)$ for a period-2 state and a normal Bloch state at $P=P_{\rm edge}/2$. In the case of the normal Bloch state, since $|\Delta(x)|$ is exponentially small in the BCS regime, we can distort the order parameter $\Delta(x)$ to produce a node, like the one in the period-2 state, with a small energy cost (per particle) up to the condensation energy $|E_{\rm cond}|/N \ll E_F$, where $E_{\rm cond} \equiv g^{-1} \int d^3r\, |\Delta({\bf r})|^2$. However, making a node in $\Delta(x)$ kills the supercurrent $j=V^{-1}\partial_P E$, which yields a large gain of kinetic energy (per particle) of the superfluid flow of order $\sim P_{\rm edge}^2/m \sim E_R$. Even if $\Delta(x)$ is distorted substantially to have a node, the original density distribution of the normal Bloch state is almost intact so that the increase of the kinetic energy and the potential energy due to the density variation is small. Therefore, the period-2 state is energetically more stable than the normal Bloch state in the BCS regime.
In the above discussion, the key point is that $\Delta(x)$ and $n(x)$ can behave in a different way in the BCS regime.
On the other hand, in the BEC limit, the density is directly connected to the order parameter as $n(x)\propto |\Delta(x)|^2$, and distorting the order parameter accompanies an increase of the kinetic and potential energies due to a large density variation.

\begin{figure}[H]
\begin{center}
\rotatebox{0}{
\resizebox{8cm}{!}
{\includegraphics{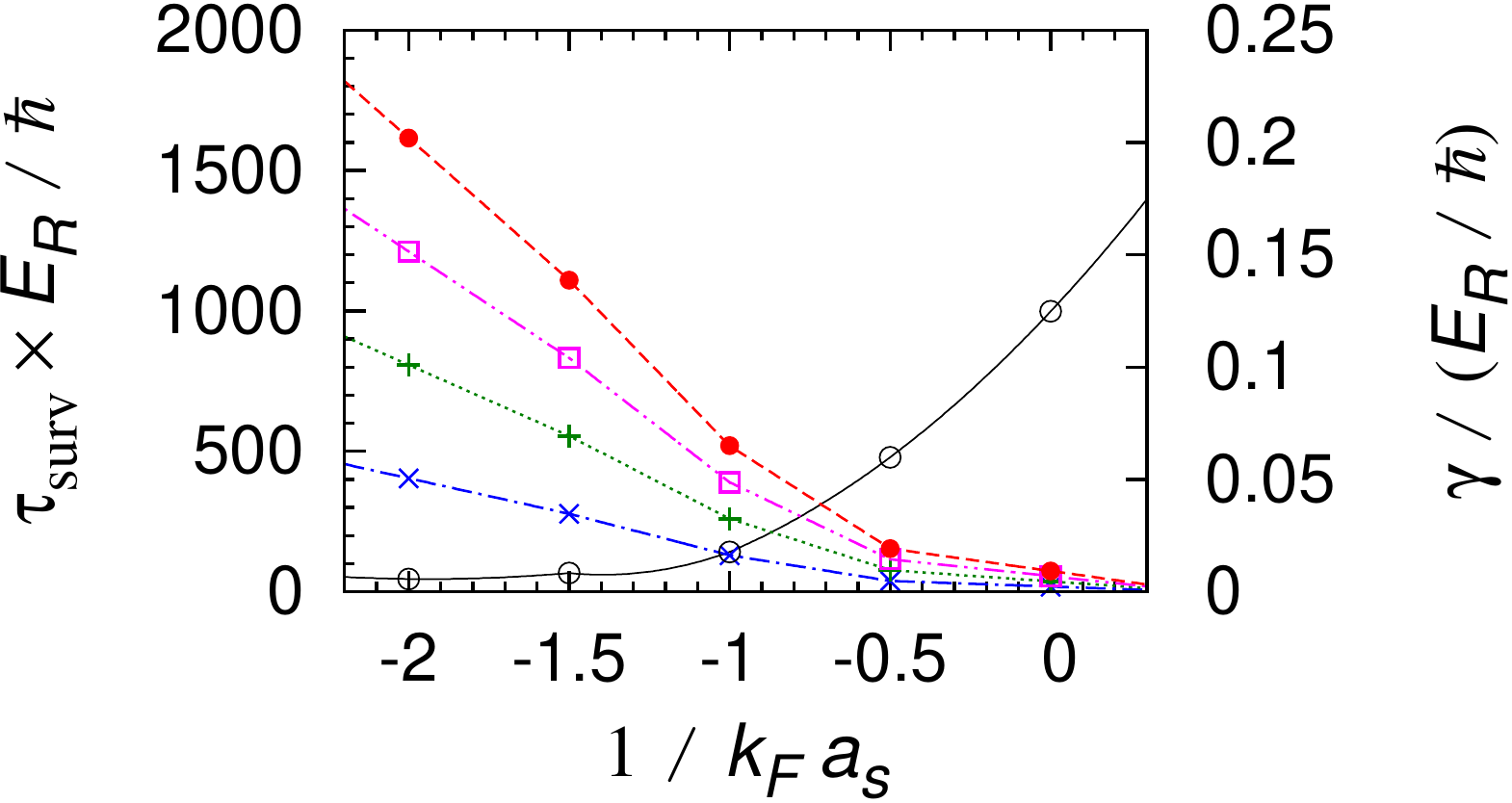}}}
\caption{Growth rate $\gamma$ of the fastest growing mode (black solid line) and survival time $\tau_{\rm surv}$ of the period-2 state at $P=P_{\rm edge}/2$ ($s=1$ and $E_F/E_R=0.25$). Blue dashed-dotted, green dotted, magenta dashed double-dotted, and red dashed lines show $\tau_{\rm surv}$ for relative amplitude $\tilde{\eta}(0)$ of the initial perturbation of $10\%$, $1\%$, $0.1\%$, and $0.01\%$, respectively. This figure is taken from \cite{period_double}.}
\label{fig:survival}
\end{center}
\end{figure}

In the deep BCS regime, the period-2 states are not only energetically stable, but also they can be long-lived although dynamically unstable. The black solid line in Figure \ref{fig:survival} shows the growth rate $\gamma$ of the fastest exponentially growing mode $|\eta(t)|=|\eta(0)|\, e^{\gamma t}$ of the deviation $|\Delta(x,t)|-|\Delta_0(x)|$ from the true stationary state $\Delta_0(x)$ for the period-2 states.
We see that $\gamma$ is suppressed with decreasing $1/k_Fa_s$, which makes the period-2 states long-lived in the BCS regime.
The growth rate $\gamma$ corresponds to the imaginary part of the complex eigenvalue for the fastest growing mode obtained by the linear stability analysis \cite{ring,pethick_smith}, which is an intrinsic property of the initial stationary state independent of the magnitude of the perturbation.

On the other hand, the actual survival time $\tau_{\rm surv}$, the timescale for which the initial state is destroyed by the large-amplitude oscillations, depends on the accuracy of their initial preparation. The survival time can be estimated by $\tilde{\eta}(0) e^{\gamma t} \sim 1$, where $\tilde{\eta}(0)$ is the relative amplitude of the initial perturbation with respect to $|\Delta_0|$ for the fastest growing mode. In Figure~\ref{fig:survival}, we show $\tau_{\rm surv}$ for various values of $\tilde{\eta}(0)$. This result suggests that if the initial stationary state is prepared within an accuracy of 10\% or smaller, this state safely survives for time scales of the order of $100\hbar/E_R$ or more in the BCS side,  corresponding to $\tau_{\rm surv}$ of more than the order of a few milliseconds for typical experimental parameters \cite{miller07}: for $E_{R, b}=2\pi\times 7.3 \mathrm{kHz}\times \hbar$ used in the experiment of \cite{miller07}, $1 \hbar/E_R = 0.0109$ msec. In the deep BCS regime ($1/k_Fa_s \ll -1$), $\tau_{\rm surv}$ increases further and may become larger than the time scale of the experiments, so that the period-doubled states can be regarded as long-lived states and, in addition, they have lower energy than the normal Bloch states in a finite range of quasimomenta. Therefore, by (quasi-)adiabatically increasing the quasimomentum $P$ of the superflow starting from the ground state at $P=0$, multiple-period states such as the period-doubled states could be realized experimentally in the deep BCS regime.

\section{Nonlinear Lattices\label{sec:nonlinlat}}

This section deals with a special kind of optical lattices, called ``nonlinear lattices''. Here the coupling constant of the nonlinear term itself (\textit{i.e.}, the interatomic interaction strength or the scattering length) has a space-periodic dependence. This is quite different from the systems which we have discussed so far: unlike cold atomic gases in a linear external periodic potential, those in a nonlinear lattice with periodically modulated interaction in space can be designed to have no linear periodic counterpart at all. While, in the former, properties are determined by the competition between the linear periodic potential and a nonlinear term, in the latter, periodicity and nonlinearity are generated by a single term. This leads to unique stability properties of the superfluid, and imposes additional conditions on its survival \cite{zhang13,nonlinlat}.

In the first subsection, we provide a basic sketch of how the sustenance of superfluidity depends on the geometry (homogeneous$/$in a linear lattice$/$in a nonlinear lattice) of the BEC system. In the second subsection, we explain the stability properties of BECs in a nonlinear lattice in terms of a simple discrete model. The third subsection presents the results of studies on ultracold bosons and fermions in nonlinear lattices for various parameter regimes. 
Finally, in the last subsection, we present a short outline of the experimental setup that constructs such a space-periodic dependence of the interatomic interaction strength.

\subsection{Dynamical Stability of the Superfluid: Special Properties of Nonlinear Lattices}

For uniform and homogeneous BECs, the dynamical stability of the superfluid is determined by the nature of the interatomic interaction. If it is repulsive (\textit{i.e.}, the scattering length is positive), the superfluid remains dynamically stable for any value of the momentum $\hbar k$ of the superflow. On the other hand, if the interaction is attractive (\textit{i.e.}, the scattering length is negative), the long-wavelength modes with 
\begin{equation}
  q^2 < \frac{4mng}{\hbar^2}
\end{equation}
grow or decay in time exponentially for any value of $\hbar k$, thus invoking an instability in the system (e.g., \cite{pethick_smith}). However, shorter-wavelength modes are stable because for them, the kinetic energy dominates over the interaction energy.

The dynamical stability properties of the superfluid changes in the presence of an external periodic potential. The periodic nature of the system may lead to Bloch solutions of the form $\psi(x)=e^{ikx} \phi(x)$, where $\phi(x)$ is a periodic function with the same periodicity as that of the lattice. The quasimomentum of the superflow is given by $\hbar k$, $k$ being the corresponding Bloch wave number. Here, for simplicity, we have assumed that the superfluid flows in the same $x$ direction as the periodic potential. Unlike the homogeneous system, the system has a nonzero critical value of $k$ above which the Bloch states are dynamically unstable \cite{wu01} as has been seen in Section \ref{sec:loopstability}: in other words, the $k=0$ state is always dynamically stable.

For a nonlinear lattice, however, this picture changes still further. The coupling constant $g$ in the GP equation (\ref{GP1}) for bosons and the BdG equations (\ref{eq:bdg}) and (\ref{eq:gap}) for fermions now depends on the space coordinate $x$. It can be thought of having a form as
\begin{equation}
\label{gperiodic}
g(x) = V_1 + V_2 \cos{2k_0x}\, ,
\end{equation}
\textit{i.e.}, $g(x)$ consists of one constant part and one sinusoidal component. $k_0$ is related to the period $d$ of the modulation by $k_0 = \pi/d$. If the nonlinear lattice is realized by an optical Feshbach resonance (details are given in the last part of this section), $k_0$ is equivalent to the wave number of the laser beam. This special type of periodic nonlinearity gives rise to a dynamical instability for the $k=0$ state \cite{zhang13}, which is in contrast with the linear lattice case (\textit{i.e.}, the case of an external periodic potential). We~shall explain this in the following subsection.

\subsection{Basic Physical Idea: The Dynamical Stability of Nonlinear Lattices}

Zhang \textit{et al.} \cite{zhang13} and Dasgupta \textit{et al.} \cite{nonlinlat} studied extended states of BECs in quasi-one-dimension with a periodically modulated interaction in space, \textit{i.e.},~a nonlinear lattice with no periodic linear potential. It was observed that when the coefficient of the nonlinear term is purely sinusoidal (\textit{i.e.},~$V_1=0$ and $V_2 \neq 0$), Bloch states at $k=0$ are dynamically (and energetically) unstable \cite{zhang13}.\linebreak In addition, even though $k=0$ state is dynamically unstable, states for nonzero $k$ could be dynamically stable in some region in $0.25\le k/k_0\le 0.5$ (this point will be seen in more detail later in Figure \ref{dyn_p1}) \cite{zhang13}.


Why is the $k=0$ state unstable in the nonlinear lattice, in contrast to the stable $k=0$ state in the linear periodic potential? Why are the states with higher values of $k$ possibly dynamically stable even though the $k=0$ state is unstable? To explain these, we take resort to a simple discrete model~\cite{nonlinlat}.\linebreak We map the 1D nonlinear lattice with $V_1=0$ and $V_2 \neq 0$ to a discrete model with the on-site interaction alternating between $U$ and $-U$ (with $U>0$):
\begin{equation}
  H = -K \sum_j (\psi_j^* \psi_{j+1} + \psi_{j+1}^* \psi_j) 
+ \frac{U}{2} \left[\, \sum_{j={\rm even}}|\psi_j|^4 - \sum_{j={\rm odd}} |\psi_j|^4\, \right]\, .
\end{equation}
So, if we denote the distance between two adjacent sites in the discrete model as $\tilde{d}$, the actual lattice constant $d$ of the unit cell in the original system is $2\tilde{d}$ (\textit{i.e.}, the unit cell in the original system corresponds to a ``supercell'' with two sites in the discrete model).

\begin{figure}[H]
\begin{center}
\includegraphics[scale=.6]{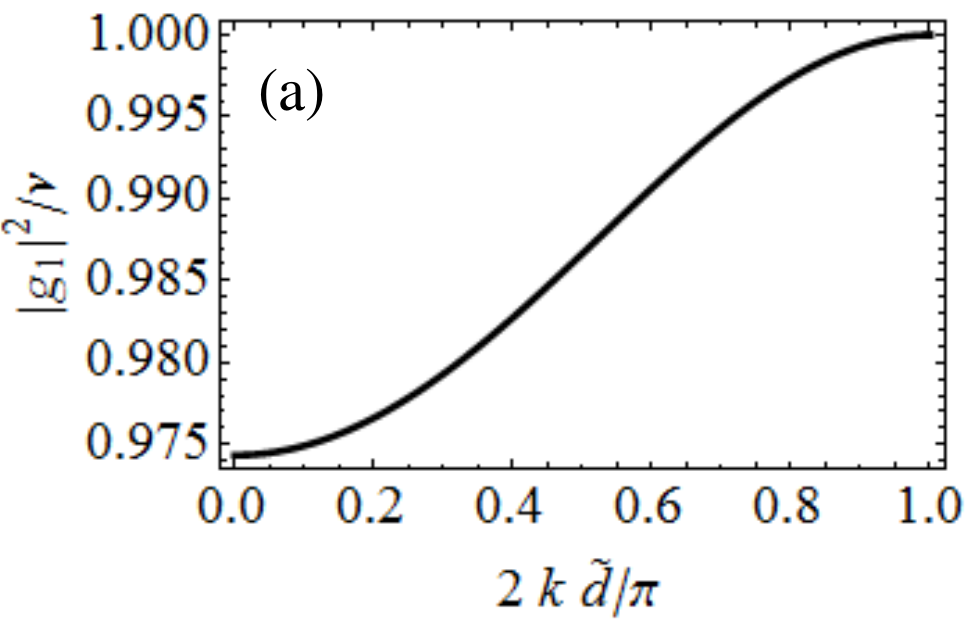}
\includegraphics[scale=.6]{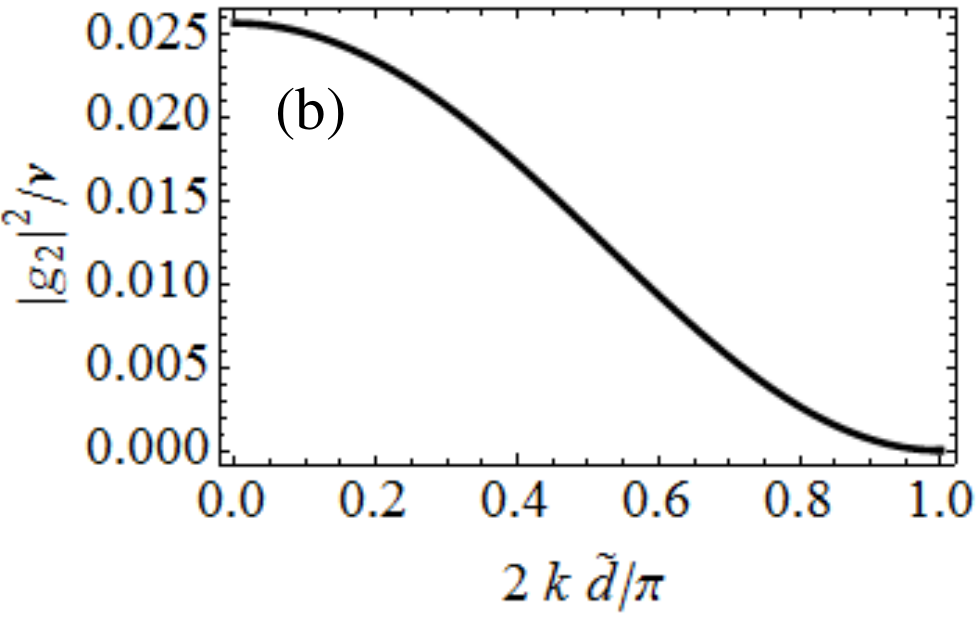}\\
\includegraphics[scale=.6]{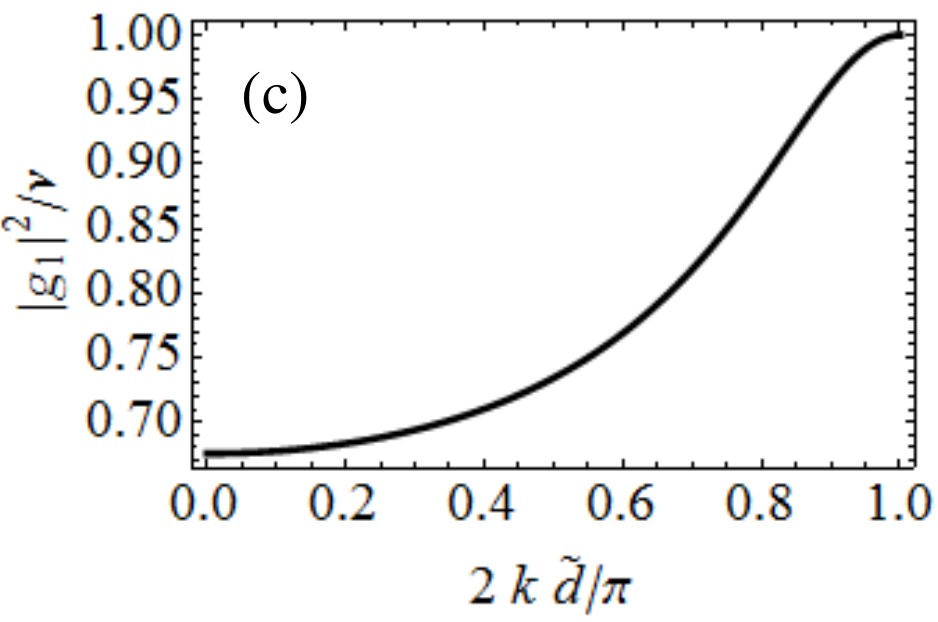}
\hspace{0.2cm}
\includegraphics[scale=.6]{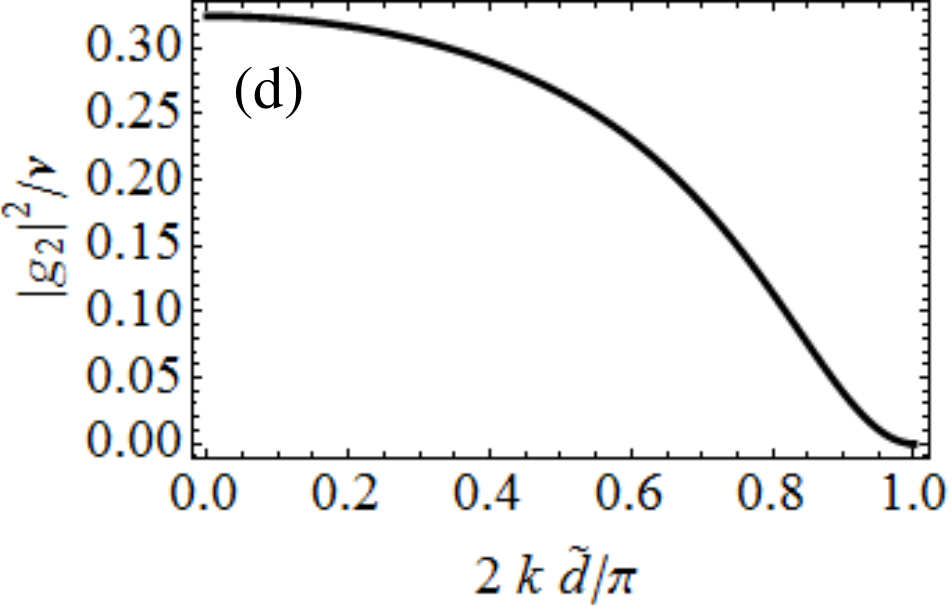}
\caption{Density distributions in the lowest band of the normal Bloch states (\textit{i.e.}, period-1 states whose period is one supercell) as functions of $k$ for different values of $U\nu/2K$. Panels (\textbf{a}) and (\textbf{b}): Populations of $|g_1|^2$ (attractive site) and $|g_2|^2$ (repulsive site) for $U\nu/2K=6$, respectively. Panels (\textbf{c}) and (\textbf{d}): Populations of $|g_1|^2$ and $|g_2|^2$ for $U\nu/2K=0.75$, respectively. This figure is taken from \cite{nonlinlat}.}
\label{p1den}
\end{center}
\end{figure}

Assuming the state is in the Bloch form, we write the amplitude $\psi_j$ at site $j$ as $\psi_j=g_j e^{i kj\tilde{d}}$, where $\hbar k$ is quasimomentum of the superflow and $g_j$ is the complex amplitude at site $j$ with the periodic boundary conditions, $g_j = g_{j+2}$ (Note that the unit cell contains 2 sites.). In addition, the amplitudes $g_j$'s are subject to the normalization condition $|g_1|^2 + |g_2|^2 = \nu$, where $\nu$ is the total number of particles in a unit cell with two sites. One can obtain the stationary solutions of $g_1$ and $g_2$ by solving the combined equations of $\delta H/\delta \psi_1^* =0$ and $\delta H/\delta \psi_2^* =0$ in almost the same manner as in Section \ref{subsec:basics}. Resulting populations $|g_1|^2$ and $|g_2|^2$ in the attractive and the repulsive sites, respectively, for the lowest Bloch band are
\begin{equation}
  \frac{|g_1|^2}{\nu} = n_+ \quad\mbox{and}\quad \frac{|g_2|^2}{\nu} = n_-
\label{eq:pop}
\end{equation}
with
\begin{equation}
  n_{\pm} = \frac{1}{2} \left\{1 \pm \left[ \left(\frac{\cos{k\tilde{d}}}{U\nu/2K}\right)^2+1 \right]^{-1/2} \right\}\, .
\label{eq:npm}
\end{equation}
The populations $|g_1|^2$ and $|g_2|^2$ for two different values of $U\nu/2K$ are shown in Figure \ref{p1den} as functions of $k$ within the first Brillouin zone.

As seen from Figure \ref{p1den}, the population difference between the adjacent sites is the smallest at the zone center $k=0$ while it increases as going toward the zone edge at which all the particles are accumulated in the attractive sites. To understand why $k=0$ state is dynamically unstable, it is instructive to see the interaction energy averaged over one unit cell (with 2 sites) \cite{zhang13}.\linebreak From Equations~(\ref{eq:pop}) and (\ref{eq:npm}), we obtain the average interaction energy per particle for the lowest Bloch band as
\begin{equation}
  \frac{E_{\rm int}/K}{N} = -\frac{U}{2K\nu} (|g_1|^4 + |g_2|^4) = -\frac{U\nu}{2K} \left[\left(\frac{\cos{k\tilde{d}}}{U\nu/2K}\right)^2 +1 \right]^{-1/2} < 0\, .
\end{equation}
Note that, in the lowest Bloch band, the average interaction energy per particle is negative for any value of $k$. So, roughly speaking, this situation resembles a BEC with attractive interparticle interaction, which is dynamically unstable as we have mentioned in the last subsection, and the dynamical instability of BECs in nonlinear lattices at $k=0$ could be understood as a consequence of the net attractive interaction energy \cite{zhang13}.

\begin{figure}[H]
\begin{center}
\includegraphics[scale=.55]{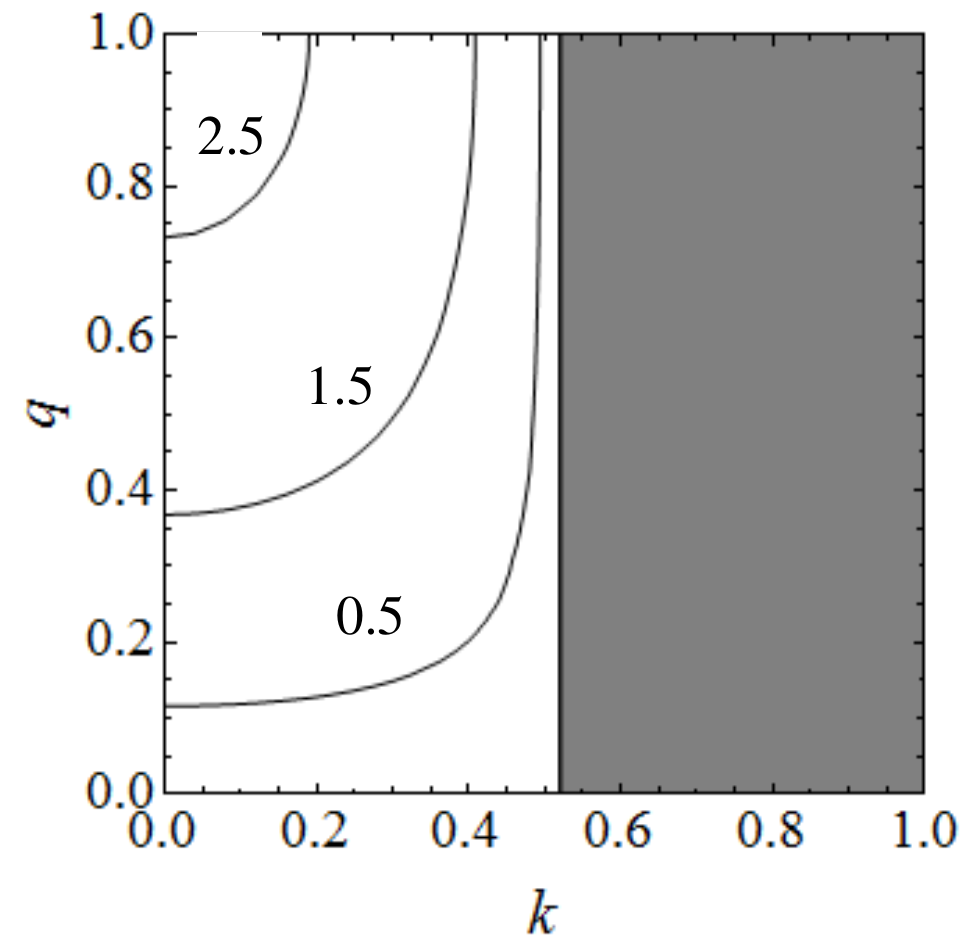}
\includegraphics[scale=.55]{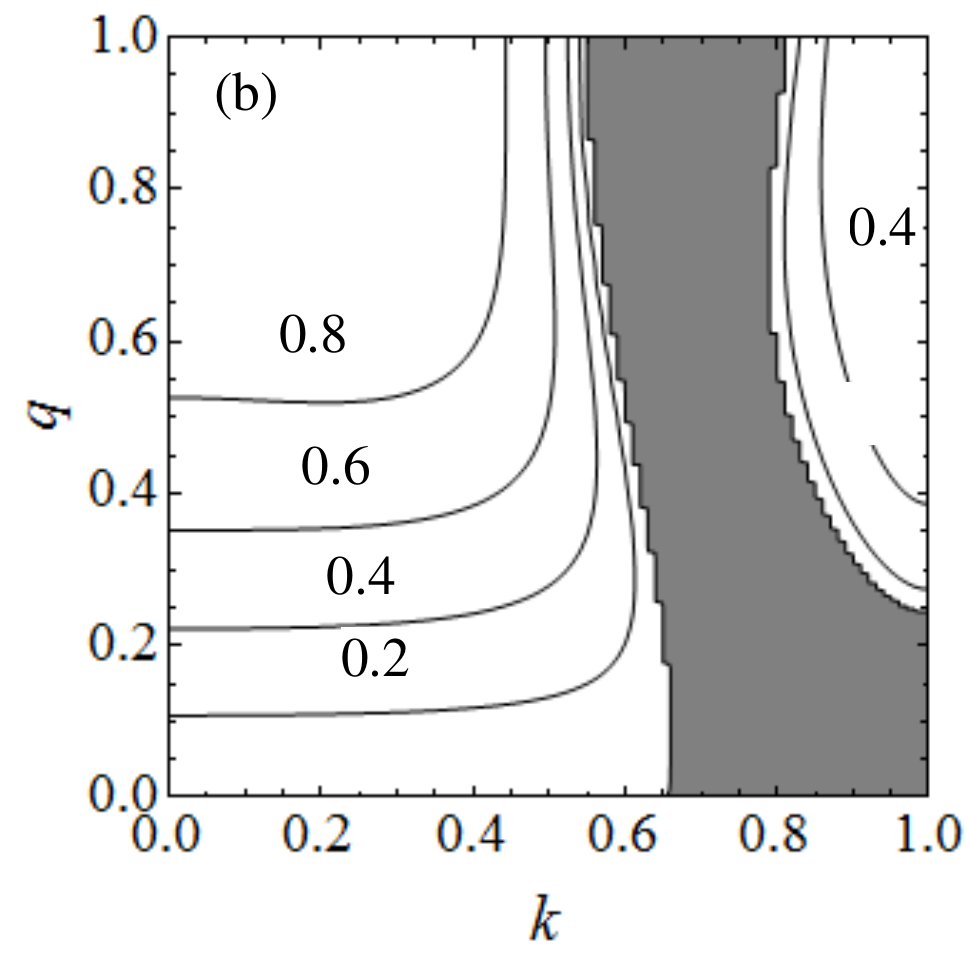}
\caption{Dynamical stability diagrams for the normal Bloch states (\textit{i.e.}, period-1 states) for \mbox{$U\nu/2K= 6$} (\textbf{a}) and $0.75$ (\textbf{b}). Quasi-wave numbers $k$ and $q$ are in units of $\pi/2\tilde{d}$. \mbox{The white regions }are the dynamically unstable regions and the gray-shaded regions are the dynamically stable regions. \mbox{The contours show the} growth rate of the fastest growing mode, \textit{i.e.}, the largest maximum absolute value of the imaginary part of the eigenvalues of matrix $\sigma_z M(q)$ in Equation (\ref{dynmat}) in units of $K$. \mbox{This figure is taken from} \cite{nonlinlat}.
}
\label{dyn_p1}
\end{center}
\end{figure}

Figure \ref{dyn_p1} shows the dynamical stability diagram of the stationary states in the lowest Bloch band in the $k$--$q$ plane, where $q$ is the quasi-wave number of the perturbation on the stationary states.
It is noted that there is a region at larger values of $k$ in which the lowest Bloch states are dynamically stable (e.g., the gray-shaded region of $0.5 \lesssim 2k\tilde{d}/\pi \le 1$ in Figure~\ref{dyn_p1}a) even though they are dynamically unstable at smaller values of $k$ including the zone center at $k=0$. To understand this somewhat counterintuitive fact, we shall take a closer look at the populations $|g_1|^2$ and $|g_2|^2$ shown in Figure~\ref{p1den}. As we have briefly mentioned before, we note that almost all the particles are accumulated in the attractive sites and thus the repulsive sites are almost empty near the zone edge at $2k\tilde{d}/\pi = 1$: at~the zone edge, only alternate sites are occupied and fragments of the BEC in these alternate sites are isolated. Since the transition amplitude between the states with populations $\{|g_1|^2, |g_2|^2\}$ and $\{|g_1|^2\pm 1, |g_2|^2\mp 1\}$ is $\sim \sqrt{|g_1| |g_2|} K$, tunneling of particles between neighboring sites is suppressed (\textit{i.e.}, the inter-site dynamics is frozen), and thus the dynamical instability is suppressed near the zone edge. Therefore, in nonlinear lattices, the lowest Bloch states can be dynamically stable at higher values of $k$ near the zone edge \cite{nonlinlat}. On the other hand, at the zone center $k=0$, the difference between the respective populations in adjacent sites is the smallest: No sites are empty and inter-site tunneling is non-negligible. Thus the suppression of the dynamical instability does not work around the zone center, rendering the system dynamically unstable due to the net attractive interaction mentioned before.
Since the isolation of fragments of the BEC is a result of the attractive interaction in alternate sites and this new mechanism of dynamical stability is more effective for larger $U\nu/2K$ (see wider gray-shaded area in Figures \ref{dyn_p1}a than that in \ref{dyn_p1}b) resulting in larger net attractive interaction energy, this mechanism can be called ``attraction-induced dynamical stability'' \cite{nonlinlat}.

We note that, in addition to the net attractive interaction energy and the suppression of the tunneling near the zone edge, there would be other factors to determine the dynamical stability of the nonlinear lattice system. When $U\nu/2K$ is sufficiently small, we observe that a dynamically unstable region appears near the zone edge (see, e.g., Figure \ref{dyn_p1}b) and the dynamically stable region is located at the intermediate values of $k$ ($0.65 \lesssim 2k\tilde{d}/\pi \lesssim 0.8$ in the case of Figure \ref{dyn_p1}b ). This non-trivial reentrant behavior suggests that there are several other factors that affect the stability, which are collectively responsible for the complicated stability diagram like Figure \ref{dyn_p1}b.

As a final comment in this subsection, we mention that the discussion here is based on the discrete model, but the attraction-induced dynamical stability has been confirmed in the continuum model as well \cite{nonlinlat}. The main difference is that, in the continuum model, if the value of $V_2$ is increased beyond a certain point, the attractive interaction between intra-site particles becomes dominant and eventually leads to the collapse of fragments of the BEC. This intra-site dynamics cannot be accounted for by the discrete model, which does not include the intra-site degrees of freedom.

\subsection{Superfluid Cold Atomic Gases in Nonlinear Lattices}

\begin{figure}[H]
\centering
\resizebox{!}{7cm}
{\includegraphics{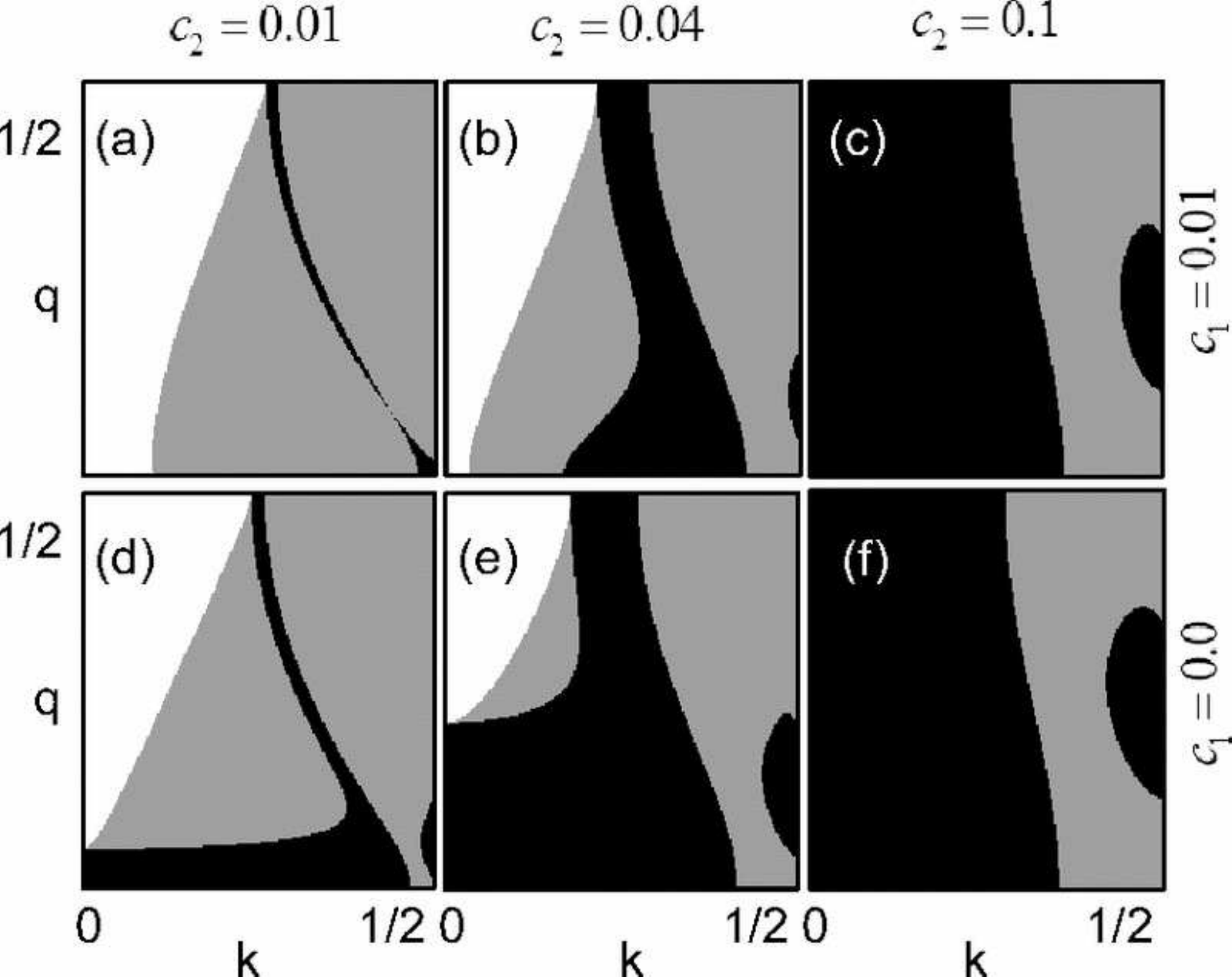}}
\caption{Stability diagrams in nonlinear lattices for various values of $c_1\equiv mnV_1/(4\hbar^2k_0^2)$ and $c_2\equiv mnV_2/(4\hbar^2k_0^2)$, where $n$ is the average number density. The quasi-wave numbers $k$ and $q$ are in units of $2k_0 = 2\pi/d$ with $d$ being the lattice constant of the unit cell in the notation of this review. The Bloch states are stable in the white area. In the gray area the Bloch states are energetically unstable but dynamically stable while, in the black area, they are unstable both energetically and dynamically.
This figure is taken from \cite{zhang13}.
}
\label{fig:zhang}
\end{figure}

Extended states of BECs in nonlinear lattices were first studied by Zhang \textit{et al.} in \cite{zhang13} (Localized states such as solitons in nonlinear lattices were studied earlier in, e.g., \cite{sakaguchi05,abdullaev07} (see also \cite{malomed_review} and references therein).). In this work, they considered quasi-1D BECs in nonlinear lattices described by the 1D version of the GP equation (\ref{GP1}) with the periodically modulated interaction strength in space given by Equation (\ref{gperiodic}): $g(x) = V_1 + V_2 \cos{2k_0x}$ with $V_1$ and $V_2 \ge 0$. 
They studied stationary Bloch states in nonlinear lattices and their energetic and dynamical stability summarized in Figure \ref{fig:zhang}.\linebreak As we have discussed in the previous subsection, the key result is that, when $V_1=0$, $k=0$ state is dynamically (and energetically) unstable for any value of $V_2\ne 0$ (see black regions in the lower panels of Figure \ref{fig:zhang} for $c_1=0$). This is in contrast to BECs in a linear external periodic potential whose dynamically unstable region is restricted in the domain of $1/4 < k/2k_0 \le 1/2$ (\textit{i.e.}, the right half of each panel in Figure \ref{fig:niustab1}b). 
In addition, states at higher values of $k$ near the zone edge can be dynamically stable even though $k=0$ state is dynamically unstable. It was pointed out that the dynamical instability of the $k=0$ state can be partially explained by the net attractive average interaction energy as we have discussed in the previous subsection. They also discussed the stability of the superfluidity due to the competition between the spatially modulated part ($V_2$ term) and the uniform, repulsive component ($V_1$ term); the latter tends superfluids to be stable. With increasing $V_1$ from zero for a fixed nonzero $V_2$, state at $k=0$ becomes dynamically and energetically stable\linebreak (see Figure \ref{fig:zhang}a,b), and by further increasing $V_1$ a swallowtail loop starts to appear at the zone edge when $V_1 > V_2$.

\begin{figure}[H]
\begin{center}
\includegraphics[scale=.6]{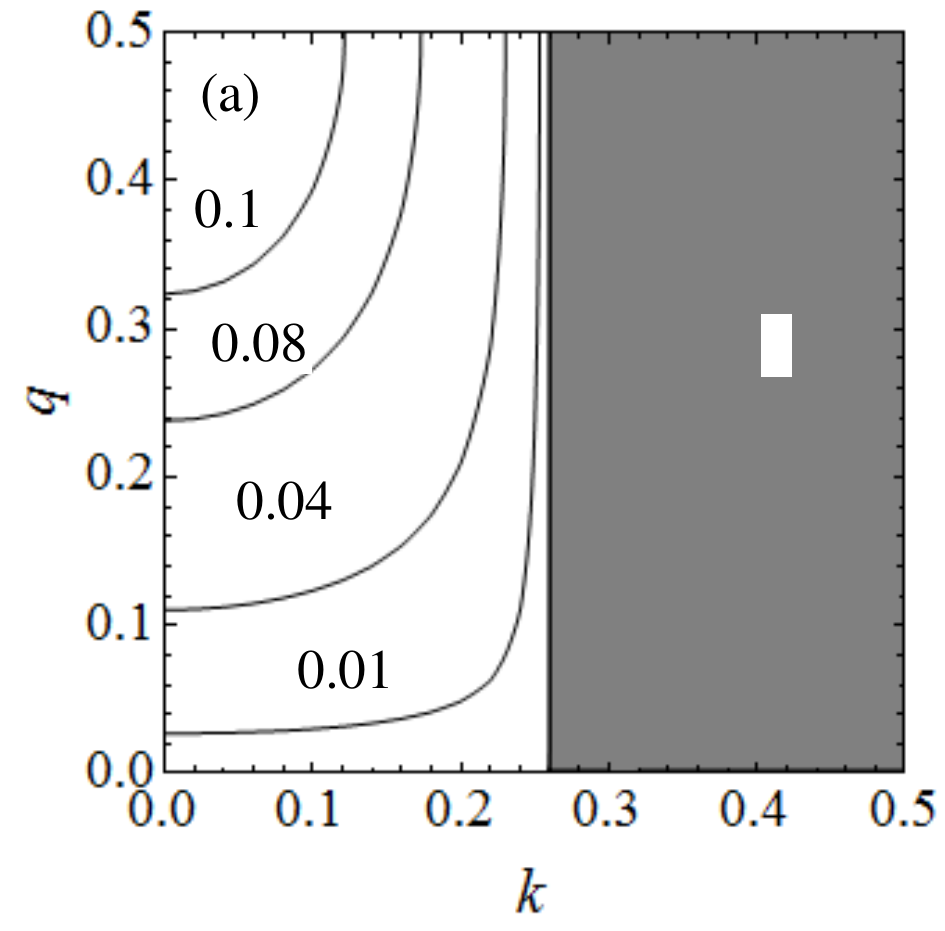}
\includegraphics[scale=.6]{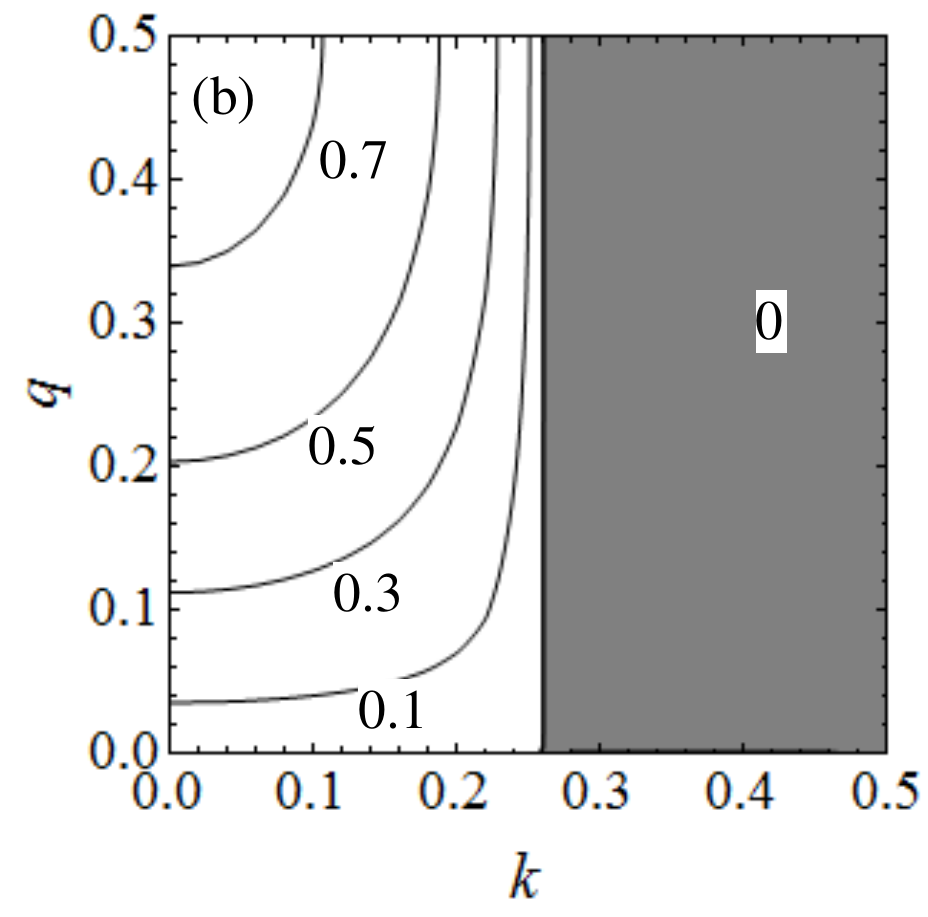}
\vspace{-6pt}
\caption{The same as Figure \ref{dyn_p1}, but for period-2 Bloch states. Taken from \cite{nonlinlat}.
}
\label{dyn_p2}
\end{center}
\end{figure}

Dasgupta {\it et al.} \cite{nonlinlat} studied multiple-period states of BECs in nonlinear lattices. They discussed stationary states with larger integer periodicity and their energetic and dynamical stability using the GP equation for both the discrete and continuum models. 
The main result of this work is that they found a new mechanism of dynamical stability called ``attraction-induced dynamical stability'' and provided the understanding of the dynamical stability around the Brillouin zone edge due to the isolation of fragments of the BEC in each attractive domain in the nonlinear lattice as discussed in the previous subsection.
This attraction-induced dynamical stability is even better manifested for the period-2 case because the majority of the particles are stored in every second attractive domain (\textit{i.e.}, every fourth site in the discrete model) making the fragments of the BEC more firmly separated, while for period-1 case they are stored in every attractive domain (\textit{i.e.}, every second site in the discrete model). 
Figure \ref{dyn_p2} shows the dynamical stability diagrams for period-2 Bloch states calculated for the discrete model at the same values of $U\nu/2K$ as Figure \ref{dyn_p1}. It is noted that the growth rate of the fastest growing mode in Figure \ref{dyn_p2}a is much smaller than the period-1 counterpart shown in Figure~\ref{dyn_p1}a; the dynamically stable region in Figure~\ref{dyn_p2}b is larger than the period-1 counterpart shown in Figure~\ref{dyn_p1}b.

\begin{figure}[H]
\centering
\resizebox{!}{5.5cm}
{\includegraphics{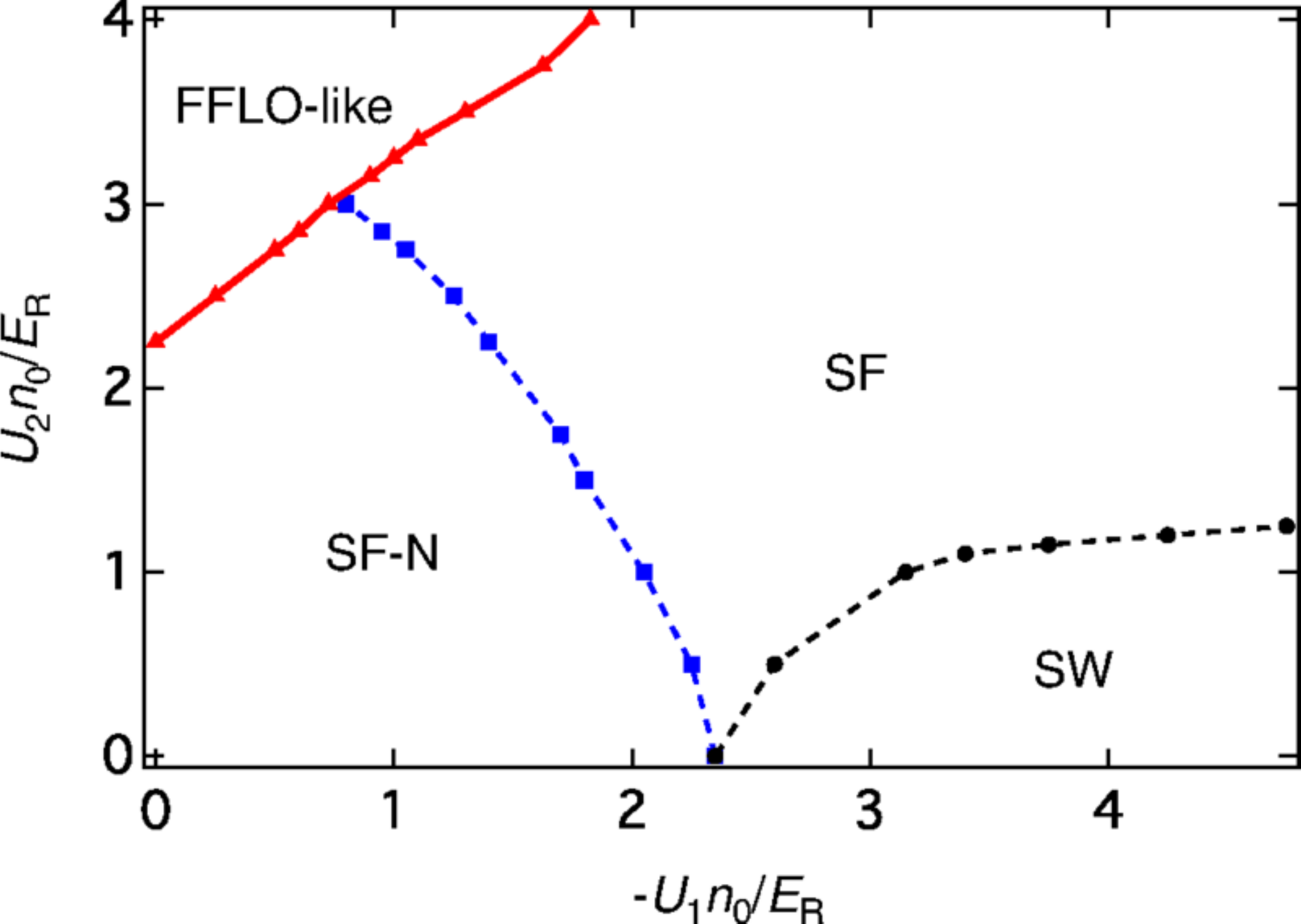}}
\caption{Phase diagram of the Bloch states of superfluid Fermi gases in nonlinear lattices in the $|U_1|$--$U_2$ plane, where $U_1=V_1$ and $U_2=V_2$ in the notation of this review. In the parameter region denoted by ``SF-N'', the system has a critical quasimomentum of the superflow above which the normal state has lower energy compared to the superfluid state while, in the regions denoted by ``SF'' and ``SW'', there is no such critical value and the superfluid state is always lower in energy than the normal state in the whole Brillouin zone. Particularly, in the region of ``SW'' at larger $|U_1|/U_2$, the energy band has a swallowtail loop around the zone edge. In the region denoted by ``FFLO-like'' (Fulde--Ferrell--Larkin--Ovchinnikov) at smaller $|U_1|/U_2$, the ground state has a nonzero quasimomentum of the superflow. Taken from \cite{yu15}.
}
\label{fig:yu}
\end{figure}

Superfluid Fermi gases in nonlinear lattices were studied by Yu {\it et al.} \cite{yu15}. They considered quasi-1D 2-component superfluid Fermi gases based on the 1D version of the BdG equation (\ref{eq:bdg}) with spatially modulated interaction strength of the form $g(x) = V_1 + V_2 \cos{2k_0x}$ with $V_1<0$ and $V_2\ge 0$. Note that it has been assumed that the uniform component is attractive ($V_1 <0$) so that the system is in the superfluid phase. The properties of the Bloch states in this system for various parameter values of $V_1$ and $V_2$ are summarized in Figure \ref{fig:yu}.

The key point is the competition between the effects of the nonlinear uniform interaction by $V_1$ and the periodicity of the system induced by $V_2$. It was found that the former dominates over the latter for lager $|V_1|/V_2$ (the region denoted by ``SW'' in Figure~\ref{fig:yu}), the Bloch band has a swallowtail loop around the Brillouin zone edge. This situation is similar to superfluid Fermi gases in a periodic potential, where the swallowtail appears due to the effect of the nonlinear interaction dominating over the periodicity of the system induced by the external potential~\cite{swallowtail}. On the other hand, for smaller $|V_1|/V_2$ (the region denoted by ``FFLO-like'' in Figure~\ref{fig:yu}), it was predicted that the state at the Brillouin zone edge with nonzero quasimomentum of the superflow (quasimomentum per atom $P=\hbar k_0/2$ in the notation of this review article) becomes the ground state of the system. They call this state as ``FFLO-like state'' because of the nonzero value of the quasimomentum of the superflow (Note that the current of this state, however, is zero.).\linebreak The stability of the Bloch states of superfluid Fermi gases in nonlinear lattices has yet to be studied.

\subsection{Experimental Setup}

Finally we give a brief sketch how nonlinear lattices can be created experimentally. With the aid of modern experimental techniques, it is now indeed possible to construct systems where the nonlinear term has an explicit spatial dependence. A very efficient way to do this is to employ optical Feshbach resonances (OFRs) \cite{fatemi,chin}.

Ultracold atoms offer an immense controllability over physical quantities like scattering length. It all started with magnetic Feshbach resonances, that use Zeeman shifts to make the scattering states resonate with a bound molecular state. It was later suggested \cite{fedichev96} that lasers can be used to induce the resonance optically. This Feshbach resonance via optical fields offers additional advantages over magnetic ones: the intensity as well as the frequency of the laser beams can be rapidly and precisely controlled. In addition, high resolution (submicron level) spatial control of the scattering length is possible by creating specially structured laser fields.

In OFRs, the basic scheme is to use a laser beam (tuned near the photoassociation resonance) that couples an initial state of free atoms to a molecular bound state. The scattering length is accordingly modified. If the OFR is driven by a standing wave with a certain periodicity, the interparticle interaction derived from it has the same periodic nature. Thus, a nonlinear lattice is generated. \linebreak
Using OFRs, Yamazaki {\it et al.} \cite{yamazaki10} demonstrated rapid, spatial modulation of the scattering length periodically at the submicron level. In this particular setup, a pulsed optical standing wave was applied to a BEC of Ytterbium atoms. The resonant wavelength was 556 nm, and the optical pulse could generate a scattering length periodically modulated in space with wavelength 278 nm.

\section{Conclusions \label{sec:conclusion}}

Over the last two decades ultracold atoms in optical lattices \cite{bloch_review,lattice,yukalov_review,jkps_review} have emerged as a key paradigm to study the \emph{ideal} realizations of many important problems in highly controllable settings, ranging from single particle quantum mechanical effects such as Bloch oscillations to strongly correlated many-body effects such as the superfluid-Mott insulator quantum phase transition. In the recent past the focus in the field has shifted more towards observing equilibrium and non-equilibrium quantum many-body effects and topological phenomena \cite{lewenstbook}. In this light the subject of this review, namely, the interplay of mean-field atomic interactions and the periodicity of the applied external potential serves to re-emphasize the fact that there are unique and interesting phenomena even in a much simpler setting. Our hope is that such a review can rekindle some interest in this area especially from the experimental side as to date there has not been a clear observation of swallowtail loop structures or period-doubled solutions for optical lattices. Moreover, in striving to control complex quantum systems such as ultracold atoms to an ever greater degree in order to realize complex many-body ground states \cite{lewenstbook} it becomes very important to be aware of and understand fundamental limitations on state preparation due to unavoidable adiabaticity breaking imposed by phenomena such as swallowtail loop dispersions.

In this review we focused on some interesting phenomena originating from the nonlinear, mean-field interactions of superfluid atomic gases in periodic potentials. We began with a summary of the basic theoretical description of Bose--Einstein condensates (BECs) and superfluid Fermi gases within the mean-field framework in Section \ref{sec:framework}. In Section \ref{sec:swallowtail} we provided a comprehensive overview of the phenomenon of swallowtail loops in the band-structure of superfluid atomic gases in an optical lattice, followed by the the discussion of Bloch states with multiple periods of the applied optical lattice potential in Section \ref{sec:multiperiod}, and Bloch states in nonlinear lattices, \textit{i.e.}, situations in which the nonlinear interaction term is itself a periodic function in space in Section \ref{sec:nonlinlat}. While we have covered a substantial portion of the various interesting phenomena that can arise due to the nonlinearity in the mean-field theory of superfluids, this is by no means complete. For instance, as mentioned in the introduction, we have made no attempt to describe localized soliton solutions to the Gross--Pitaevskii equation and suggest the excellent review \cite{malomed_review} on this topic for the interested reader.

Although considerable amount of research work has already been accomplished in this field, it is still relatively young and has flourished only over the last two decades beginning with the experimental discovery of BECs. As a result we believe there are still a range of open problems that can be investigated. Some examples include, the stability of Bloch and swallowtail loop states of superfluid Fermi gases in nonlinear lattices, Bloch oscillation dynamics for both bosons and fermions in regimes with swallowtail loops, the study of quantum equivalents of mean-field nonlinear phenomena including solitons and loops, and nonlinear phenomena in more exotic systems such as BECs with spin-orbit interactions or spinor BECs \textit{etc}.

\vspace{6pt}
\acknowledgments{We acknowledge Mauro Antezza, Franco Dalfovo, Elisabetta Furlan, Jonas Larson, Takashi~Nakatsukasa, Duncan O'Dell, Giuliano Orso, Francesco Piazza, Lev P. Pitaevskii, Sandro Stringari, and Sukjin Yoon for collaborations.
This work was supported by IBS through Project Code (IBS-R024-D1);  {by the Zhejiang University 100 Plan; by the Junior 1000 Talents Plan of China;} by the Max Planck Society, MEST (Ministry of Education, Science and Technology) of Korea, Gyeongsangbuk-Do, and Pohang City for the support of the JRG at APCTP; and by the Basic Science Research Program through NRF by MEST (Grant No. 2012R1A1A2008028). Raka Dasgupta would like to acknowledge Department of Science and Technology, Government of India for Inspire Faculty Award [04/2014/002342].}



\conflictofinterests{The authors declare no conflict of interest.}



\bibliographystyle{mdpi}
\renewcommand\bibname{References}



%


%

\end{document}